%% file: FireSalesSIFINVersionRevised.tex
\documentclass[11pt,a4paper]{article}

\usepackage{amsmath,amssymb}

\usepackage{amsthm}
\usepackage{hyperref}
\usepackage{mathtools}
\usepackage[shortlabels]{enumitem}
\usepackage{bbm}
\usepackage{graphicx,graphics,float}
\usepackage{subfigure}
\usepackage[left=1in, right=1in, top=1in, bottom=1in, includefoot, headheight=15pt]{geometry}
\usepackage{stmaryrd}

\usepackage{color}
\definecolor{darkgreen}{rgb}{0, .5, 0}
\definecolor{darkred}{rgb}{.5, 0, 0}
\definecolor{orange}{rgb}{1, 0.3, 0.1}

\usepackage{setspace}

\theoremstyle{plain}
\newtheorem{theorem}{Theorem}[section]
\newtheorem{proposition}[theorem]{Proposition}
\newtheorem{corollary}[theorem]{Corollary} %%
\newtheorem{assumption}[theorem]{Assumption} 
\newtheorem{lemma}[theorem]{Lemma} %%
\newtheorem{example}[theorem]{Example}
\theoremstyle{definition} %%
\newtheorem{definition}[theorem]{Definition}

\numberwithin{equation}{section}
\usepackage{multicol}

\input{definitions.tex}

\DeclareRobustCommand*\circFSuper[1]{\accentset{\circ}{f}^#1}
\newcommand{\circFSuperSub}[2]{\accentset{\circ}{f}^#1_#2}
\newcommand{\circG}{\accentset{\circ}{g}}

\newcommand{\circRho}{\accentset{\circ}{\rho}}

\usepackage{relsize}
\usepackage{accents}
\usepackage{array}
\usepackage{bm}

\begin{document}
\singlespacing
\title{\LARGE \bf
%An Asymptotic Model for Fire Sales \\ A Stochastic Model for Fire Sales \\ Fire Sales: a Stochastic Model and Resilience Criteria \\ Fire Sales: Modelling and Risk Management \\ 
%Modeling and Managing the Risk of Fire Sales
%Fire Sales: Modelling and Gaining Control \\
Suf{}focating Fire Sales}
\author{Nils Detering\thanks{Department of Statistics and Applied Probability, University of California, Santa Barbara, CA 93106, USA. Email: detering@pstat.ucsb.edu}, Thilo Meyer-Brandis\thanks{Department of Mathematics, University of Munich, Theresienstra\ss{}e 39, 80333 Munich, Germany. Emails: meyerbra@math.lmu.de, kpanagio@math.lmu.de and ritter@math.lmu.de}, Konstantinos Panagiotou\footnotemark[2], Daniel Ritter\footnotemark[2]  }
\maketitle

\begin{abstract}
\noindent
Fire sales are among the major drivers of market instability in modern financial systems. Due to iterated distressed selling and the associated price impact, initial shocks to some institutions can be amplified dramatically through the network induced by portfolio overlaps. In this paper, we develop a mathematical framework that allows us to investigate central characteristics that drive or hinder the propagation of distress. We investigate single systems as well as ensembles of systems that are alike, where similarity is measured in terms of the empirical distribution of all defining properties of a system. This asymptotic approach  ensures a great deal of robustness to statistical uncertainty and temporal fluctuations. A characterization of those systems that are resilient to small shocks emerges, and we provide criteria that regulators might exploit in order to assess the stability of a financial system. 
 We illustrate the application of these criteria for some exemplary configurations in the context of capital requirements and test the applicability of our results for systems of moderate size by Monte Carlo simulations.
\end{abstract}

\medskip
%\noindent\textit{Keywords:} systemic risk, financial and price-mediated contagion, asset fire sales, capital requirements

\section{Introduction}
Modern financial systems exhibit an inevitably complex structure of dependencies. This complexity is visible on multiple levels, including for example the intricate distribution of corporate obligations and also the -- sometimes significant -- overlap in the asset holdings of the participating institutions. This interlocking and pervasive structures can make the system fragile to possible initial local shock events. The problem of modeling, understanding, and managing this notion of \emph{systemic risk} has been an important research topic, and it has become particularly prominent after the strike of the global financial crisis in the years 2007/08.

One of the first quantitative papers to address systematically the effect of -- what they termed -- cyclical dependencies among the market participants was the seminal work \cite{Eisenberg2001}. They considered default contagion, that is, the successive default on liabilities across financial institutions, and showed, among other results, that under mild assumptions a unique clearing vector exists that clears the obligations of all members. From today's viewpoint,  however, it is  well understood that there are various other important channels of contagion. For instance, in the book \cite{Hurd2016}, \emph{Asset Correlation}, \emph{Default Contagion}, \emph{Liquidity Contagion} and \emph{Asset Fire Sales} are listed as the four main channels of direct and indirect default propagation.

\begin{figure}
\centering
\includegraphics[width=0.63\textwidth]{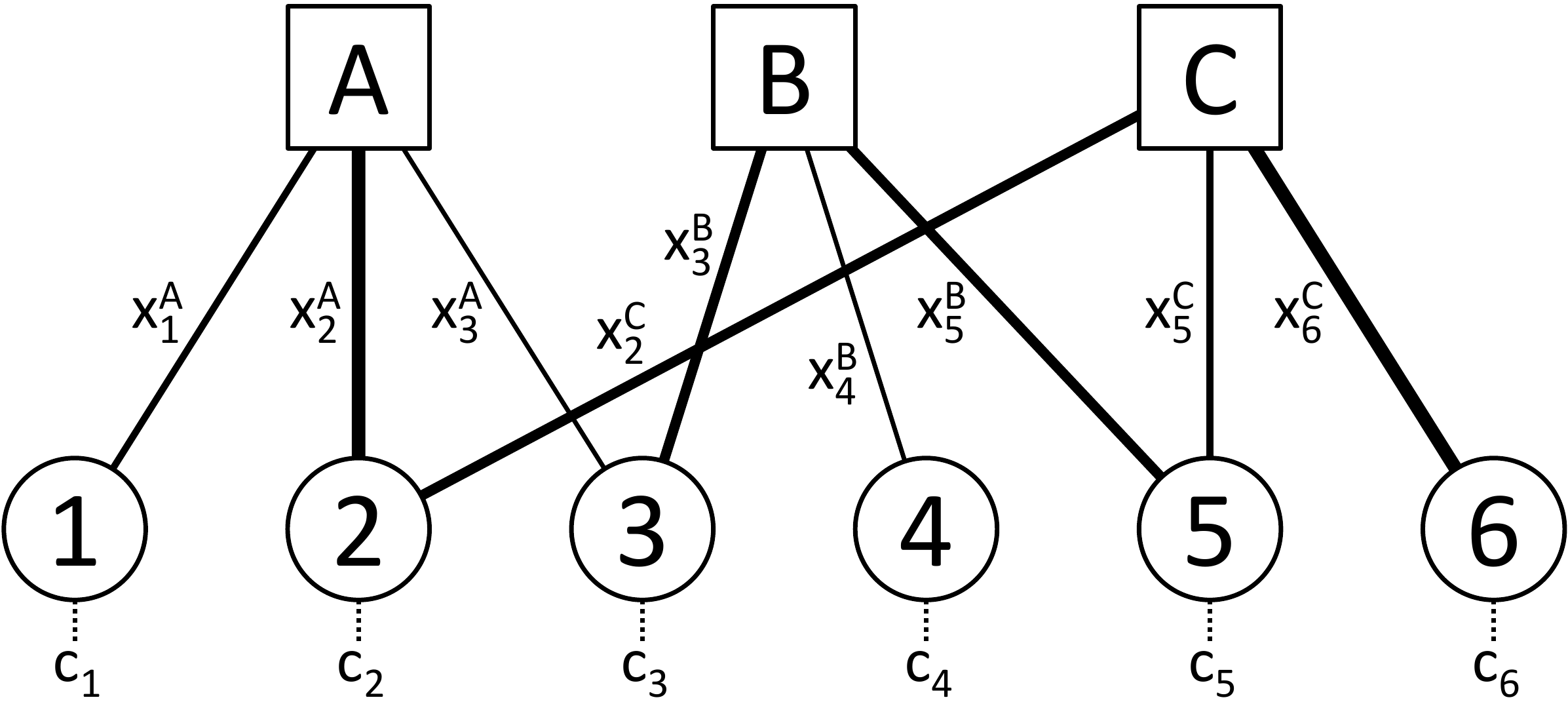} 
\caption{An illustration of a system with $\bm{n=6}$ financial institutions (circles) and $\bm{M=3}$ assets (squares). Edges represent investments of the institutions in the assets and their thickness indicates the investment volume $\bm{x_i^m}$. Furthermore, capitals $\bm{c_i}$ are attached to the institutions.}\label{fig:bipartite}
\end{figure}

Here we focus on fire sales, which is widely accepted as one of the main drivers of market instability (\cite{ELLUL2011596},\cite{Khandani2011},\cite{Cont2013a},\cite{RePEc:ecb:ecbwps:20202373},\cite{RePEc:bca:bocawp:20-41}). A recent example of a fire sales event is the selloff that followed the decline in the stock price of ViacomCBS on March 24, 2021. It was particularly amplified due to the leveraged bets by the hedge fund Archegos Capital, which as a result of the price drop was forced to unwind its positions in ViacomCBS.\footnote{https://www.reuters.com/article/usa-markets-blocktrades-timeline-idUSL1N2LS332} The underlying dynamics that we consider are defined abstractly as follows. Consider a system of institutions that are invested in certain assets, where each institution $i$ is equipped with some initial capital (equity) $c_i$ and holds some number~$x_i^A$ of shares of an asset $A$ (see also Figure~\ref{fig:bipartite}). The portfolios of the institutions may therefore overlap, and the dependencies are quantified by the collection of the $x_i^A$'s. As a reaction to some initial shock event, one or more of the institutions sell according to an individual strategy a non-negligible number of shares of their assets. These sales may cause a decline of the assets' share prices and all investors in the assets that were sold incur losses in their portfolio values. This may start another round of asset sales, possibly on an extended set of assets, again reducing share prices and so on. By this iterative process the initial stress is propagated through the system and can be amplified considerably. In particular, institutions that were spared from the initial shock can get into trouble, if their portfolios overlap with those of distressed investors.

\vspace{2pt}
\noindent {\bf Literature}
Various approaches have been developed to model and understand the impact of fire sales. The works \cite{Cifuentes2005,Rogers2013,Weber2016,RePEc:chf:rpseri:rp1520,Chen2016AnOV} extend the classical setting from~\cite{Eisenberg2001} for given financial networks so that various aspects of fire sales are incorporated. \cite{Caccioli2014} use a branching
process approximation to model fire sales, and they find that the system is stable when certain market parameters are below a critical value that they specify. In a two-period model \cite{RePEc:eee:jfinec:v:115:y:2015:i:3:p:471-485} show how contagion effects propagate across the banking sector. \cite{Ibragimov2011} study the benefits and disadvantages of overlapping portfolios not only from the perspective of single institutions, but also for the market as a whole. An important consequence of their model is that for some specifications a divergence between private and social welfare may arise. Alike and related considerations are also made in \cite{Beale2011}, where a small number of assets is considered, and in \cite{Wagner2010}, where using a microfounded model it is shown that the risk of joint liquidation motivates institutions to create heterogeneous portfolios. A related setting for reinsurance markets with overlapping insured objects is analyzed by \cite{Kley2016}. \cite{Cont2013a} analyze the impact of fire sales on asset price dynamics and correlation in continuous time. \cite{FEINSTEIN2020449} propose a continuous time model for price-mediated contagion and derive conditions based on the risk-weights of assets that ensure that a solution to the model exists. In~\cite{Cont2016b}, extending and building upon work of \cite{Khandani2011}, they further describe the impact of the liquidation of large portfolios on the covariance structure of asset returns and provide a quantitative explanation for spikes in volatility and correlations observed.

Apart from the theoretical work there is a significant amount of empirical studies that consider the effect of fire sales, develop viable models and propose measures to capture quantitatively their effects. In \cite{Guo2016} the topology of the network of common asset holdings is analyzed and \cite{Braverman2014}  proposes a network representation to quantify the interrelations induced by common asset holdings. \cite{Cont2017} develops a stress testing framework for fire sales that is extended by indices of centrality for institutions participating in a fire sales process in \cite{Cont2018}. There are several works that develop exposure- and market-based measures to depict the effect of fire sales. For example, \cite{girardi18} uses the scalar product of two portfolios' weights, and \cite{Kritzman2011} uses the so-called absorption ratio, the degree of variation induced by the first principal components of asset returns, to measure dependence between different sources of risk in a portfolio. Additional measures are constructed from a statistical analysis of the equity returns of institutions, as in \cite{Acharya2017,Brownlees2017,Adrian2016}.

\vspace{2pt}
\noindent {\bf Our Contribution}
We study a fairly general fire sales process here that accounts for the \emph{actual} losses incurred in each round, which makes the model complex and its dynamics quite involved. 
%To the best of our knowledge, this is the first model for a financial system that considers the actual losses based on the realized price history and thus accounts for the fact that shares sold earlier during a fire sales cascade are sold at a more favorable price. 
To analyze our model we resort to a simpler related auxiliary model that is based on one clearing price for all shares. The final outcome of this simpler process can then be determined by a fixed-point equation. The dimensionality of this fixed-point equation only depends on the number of assets and not on the size of the financial system. This is in contrast to previous works (\cite{Cifuentes2005,Rogers2013,Weber2016}) in which the clearing vector for a financial system has been derived by solving a fixed-point equation with dimensionality equal to the size of the system. The auxiliary process only serves as a tool for us and we establish coupling results that tie the auxiliary process to the original process in a novel way that provides both, lower and upper bounds for the outcome of the original process. These results enable us to determine the final state of the system at the end of the original fire sales process. 

However, we are also interested in a more fundamental question: what are the important underlying structures that propel the process of fire sales? In other words, which system characteristics favor the emergence of large fire sales cascades, and which prohibit them?
In order to address these questions, we establish quite involved continuity properties for the fixed-point of the auxiliary system with respect to changes in the empirical distribution of the system. Facilitating these results, instead of restricting our attention only to a single system, we can then consider an \emph{ensemble} of systems that share common structural characteristics. By reducing any particular system to its bare bone characteristics, we obtain robustness and flexibility. Moderate uncertainties of the precise system configuration are absorbed and we can study large financial systems ($n\rightarrow \infty$, $n$ the number of institutions) given that their empirical distribution converges. This asymptotic approach then allows us to detect unstable system configurations and provides a \emph{parameter-free} notion of resilience  of a given system. This parameter free notion of (non)-resilience inherently captures a large system phenomenon, depending only on the tails of the system's empirical distribution. It thus focuses on fire sales in large systems due to its asymptotic nature. Establishing these results is possible again by continuity arguments where we study whether the fixed-point is converging to zero as the initial shock becomes small (resilient system). If the system is not resilient on the contrary, small shocks may have disastrous outcome. 

More than a mere description of the fire sales process, a central objective can then be to devise measures to prevent disastrous fire sales cascades. We show for a stylized system how our mathematical framework may be used to derive explicit capital requirements that ensure the stability of the system. For this example setting, it turns out that for each institution the capital requirement is solely based on its own asset holdings, thus ensuring full transparency and fairness, which is an essential question to address when determining the systemic riskiness of the involved institutions. 
As a further application, our results allow us to quantify the effects of portfolio diversification and similarity. Similarly as in (\cite{Ibragimov2011,Beale2011,Wagner2010,Capponi}) we find that diversification, when it comes at the cost of portfolio similarity, leads to larger fire sales cascades. However, heterogeneous, diversified portfolios are favorable from a systemic risk perspective. We also find that for well-capitalized systems, diversification is beneficial, even if it comes at the cost of portfolio similarity. We hope that some of these and additional findings that can be derived from our model will contribute to ongoing discussions on financial market regulation. 

Currently the main burden to apply our results might be the availability of data. Although we show in simulations that the different regimes (resilience/non-resilience) are already visible for systems of moderate size ($n=10^4$), current regulatory stress tests are applied to only the most systemically important banks. In fact, the 2021 stress test of the European Banking Authority (EBA) is only based on 50 banks from EU and EEA countries.\footnote{https://www.eba.europa.eu/risk-analysis-and-data/eu-wide-stress-testing} Applying our asymptotic results to such a small set of banks will most likely not provide useful information about the resilience of the actual system. In this regard our method calls for more data collection and an analysis on a larger data set, which might become available at some point. In fact, the financial systems of many economies are quite large. For instance, Germany alone is home to more than $1500$ banks. Adding other financial institutions and possibly even assessing the risk on a European level, one easily reaches a size of the financial system for which the asymptotic effects become relevant, as shown in the simulations. Without the availability of larger data sets, we think that one can still obtain some useful general structural conclusions from the results, as shown in some first examples.

%the regulatory discussions in the literature, see \cite{Beale2011,Frey2018,Ibragimov2011,Wagner2010} for example.
%This approach was also used recently to study default contagion in the financial mathematics community, see the papers~\cite{Cont2016,Detering2016,Detering2018}. Some of our results  are similar in nature, but the process that we consider has entirely different dynamics.
%Moreover, we demonstrate that they can be combined with the classical single firm risk management policies defined by the \cite{BaselCommittee2011} in terms of value-at-risk and we can expose their positive effects.

\vspace{2pt}
\noindent {\bf Outline} 
In Section \ref{sec:fire:sales} we present our model of fire sales, we introduce the aforementioned notion of resilience and outline our main results. Section \ref{sec:aux:model} deals with the analysis of the simpler auxiliary model. Then, in Section~\ref{sec:fsp} we build the bridge between the two models, and in particular we derive abstract and explicit criteria that guarantee resilience for the original fire sales system.  Section \ref{sec:applications} deals with applications of the developed theory and provides simulations, as outlined in the previous paragraph. The paper closes with a conclusion that summarizes future research directions, in particular the problem of applying our results to systems of moderate size. All proofs that are omitted in the main part and some generalized results can be found in Section~\ref{sec:proofs}.

\section{ {Model, Resilience and Overview of the Results}}\label{sec:fire:sales}

In this section we present our model of fire sales and outline our results.  

\subsection{{The Fire Sales Process}}

% We first describe the parameters and assumptions and then determine the final state of the system after the fire sales cascade has completed both in the deterministic as well as in the stochastic setting.

%\vspace{-7pt}

{\bf Model Parameters}  We consider a financial system consisting of $n\in\N$ institutions that can invest in $M\in\N$ different (not perfectly liquid) assets or asset classes. To each institution $i\in[n]:=\{1,\ldots,n\}$ we assign the number of shares $x_i^m\in\R_{+}$ that institution $i$ holds of asset $m\in[M]$. Here, by $\R_{+}$ we denote the set of positive real numbers including zero. See Figure \ref{fig:bipartite} for an illustration. Further, we denote by $c_i\in\R_+$ the initial capital of institution $i$. This could for example be the equity for a leveraged institutions, such as a bank, or the portfolio value for an institution that exclusively invests in assets, such as a fund. We assume that the capitals incur exogenous losses $\ell_i\in\R_{+}$ due to some shock event.
In the case of a market crash for instance, it could be that $\ell_i=\sum_{1\le m\le M} x_i^m\delta^mp^m$, where $p^m$ denotes the initial price of one share of asset $m\in[M]$ and $\delta^m\in(0,1]$ is the relative price shock on the asset. Furthermore, let the empirical distribution $F_n:\R_{+}^{M}\times\R_{+}\times\R_{+}\to[0,1]$ of the institutions' parameters be denoted by
\begin{equation}
\label{eq:empDistFixedSize}
F_n(\bm{x},c,\ell) = n^{-1}\sum_{i\in[n]}\1\{x_i^1\leq x^1,\ldots,x_i^M\leq x^M, c_i\leq c, \ell_i\leq \ell\}
\end{equation}
and let in the following $(\bm{X}_n,C_n,L_n)$ be a random vector with distribution $F_n$.

\noindent
{\bf  Asset Sales} We assume that due to the exogenous losses some of the institutions are forced to liquidate parts of their asset holdings in order to comply with regulatory or market-imposed constraints (e.\,g.~leverage constraints), self-imposed risk preferences and policies to adjust the portfolio size, or to react to investor redemptions. These sales are described by a non-decreasing
function $\rho:\R_{+}\to[0,1]$ such that each institution $i\in[n]$ incurring a loss of $\Lambda$ sells $x_i^m\rho(\Lambda/c_i)$ of its shares of asset $m$. 
The fraction $\Lambda/c_i$ describes the relative loss of institution $i$ measured against its initial equity. It is hence sensible to assume that
$$\rho(0)=0, \quad \rho(u)\leq1  \text{ and } \rho(u)=\rho(1) \text{ for all }  u\ge 1.$$
If at default of an institution the whole portfolio is to be liquidated, then $\rho(1)=1$. In general, however, the remaining assets at default may be frozen by the insolvency administrator and only be sold to the market on a longer time scale.
In this case,  $\rho(1)=\lim_{u\to1-}\rho(u)\in[0,1)$. Our assumptions on $\rho$ are rather mild and allow for a flexible description of various scenarios. A simple example is given by $\rho(u)=\1\{u\geq1\}$, which describes complete liquidation of the portfolio at default (if the institution is leveraged) resp.~dissolution. More realistic examples, as for instance derived from leverage constraints, can be found in Section~\ref{eq:salesfctex}.

Some remarks are in order here. First, 
%The actual reasoning behind asset sales is in general more complex than the presented examples, and this is why we consider a general \daniel{(non-decreasing)} sale function $\rho$ in this article. 
to simplify the exposition and the notation in the main part, we restrict our analysis to continuous $\rho$, and we refer to Section~\ref{sec:proofs}, where we provide proofs of all our results in full generality.
%A natural assumption is that $\rho$ is right-continuous. By replacing $\rho(u)$ with its right-continuous modification $\overline{\rho}(u):=\lim_{\epsilon\to0+}\rho((1+\epsilon)u)$ throughout, our results become applicable also for arbitrary (not right-continuous) sale functions $\rho$. Finally, denote by $\accentset{\circ}{\rho}(u):=\lim_{\epsilon\to0+}\rho((1-\epsilon)u)$ the left-continuous modification of~$\rho$.
Second, it is possible to consider a more general model in which we choose different sales functions $\rho^m$ for the assets $m\in[M]$ to account for different asset specific aspects. It is then possible to replace the scalar function $\rho(u)$ by the diagonal matrix $\mathrm{diag}(\rho^1(u),\ldots,\rho^M(u))$ in all forthcoming considerations. Further, we may partition the set of institutions into different types (banks, insurance companies, hedge funds, \dots) and choose different $\rho$ or $\rho^m$ for each type. Finally, our proofs in this article also work for  arguments (of $\rho$) other than $ \Lambda/c_i$ %for $\rho$ 
(where $\Lambda$ are the losses), but for the sake of simplicity we stick to this particular form. %\nils{Need to address reviewer question here whether liquidation amounts can depend on liquidated amounts of other institutions. I would say that this happens only implicitly through the losses but not explicitly as they might have in mind.}

\noindent
{\bf  Price Impact} Since the assets are not perfectly liquid (the limit order book has finite depth), the sales of shares triggered by the exogenous shock cause prices to decline. This on the other hand causes further losses for all the institutions invested in the assets due to mark-to-market accounting.
We model the price loss of asset $m\in[M]$ by a continuous function $h^m:\R_{+}^M\to[0,1]$ which is non-decreasing in each coordinate. That is, if $\bm{y}=(y^1,\ldots,y^M) \in \R_{+}^M$ and $ny^m$ shares of asset $m$ have been sold in total, then we assume that the price of  $m$ drops by $h^m(\bm{y})$.
%,  \kosta{and crucially, that}  each institution $i\in[n]$ suffers losses of $\bm{x}_i\cdot h(\bm{y})$, where $\bm{x}_i:=(x_i^1,\ldots,x_i^M)$ and $h(\bm{y})=(h^1(\bm{y}),\ldots,h^M(\bm{y}))$.

%ome comments on the choice of the model are appropriate here. First,
Note the relative parametrization with the number of institutions $n$, where we assumed that $ny^m$ (instead of $y^m$) shares of asset $m$ are sold.
For fixed $n$ this is arbitrary; however, in due course we will consider an ensemble of systems (see Assumption \ref{ass:regularity:fire:sales} below), and then this parametrization will turn out to be rather convenient to state our results. In particular, it reflects the fact that the market depth scales with the number of institutions involved.
%Further, note that $\bm{x}_i\cdot h(\bm{y})$ is an upper bound on institution $i$'s losses at the time that $n\bm{y}$ shares were sold. This upper bound is usually strict, since $i$ might have sold part of its shares already at an earlier time and thus at a higher price. However, as we shall see later, this upper bound will turn out to be very handy when studying fire sales, see Section~\ref{sec:aux:model}  and the discussion at the end of this section. 

\noindent
{\bf Fire Sales} %The fire sales process that we consider combines all previous  ingredients. 
Triggered by some exogenous event the institutions start selling a portion of their assets and drive down the prices. Due to mark-to-market effects, this means that institutions experience further losses and are forced into further sales. This iterative process continues until the system stabilizes and no further sales, losses and price changes  occur.
More specifically, let us denote by $\bm{\tau}_{(k)} = (\tau_{(k)}^j)_{1\le j \le M}$ the vector of cumulatively sold shares at the beginning of round $k\in \mathbb{N}$, that is, the number of actually sold shares in round $k$ is $\bm{\tau}_{(k+1)} - \bm{\tau}_{(k)}$ and, obviously, $\bm{\tau}_{(1)} = 0$. Moreover, for each bank $i \in [n]$ and any round $k$ let
\begin{itemize}
	\item $\bm{x}_{i,k} = (x_{i,k}^j)_{1\le j \le M}$ be the number of shares that $i$ holds at the beginning of round $k$;
 	\item $\ell_{i,k}$ be the total loss of $i$ accumulated until the beginning of round $k$.
\end{itemize}
From these definitions we readily obtain that the number of shares held at the beginning of round~$k$ is
\begin{equation}
\label{eq:xik}
	\bm{x}_{i,1} = \bm{x}_i, \quad \bm{x}_{i,k} = \bm{x}_{i}\big(1 - \rho(\ell_{i,k-1}/c_i)\big)\quad i\in[n], k \ge 2, 
\end{equation}
and the total number of sold shares is given by
\begin{equation}
\label{eq:tau}
	\bm{\tau}_{(1)} = \mathbf{0},
	\quad
	\bm{\tau}_{(k)} = \sum_{i \in [n]} \bm{x}_i \rho\big(\ell_{i,k-1}/c_i\big),~~k\ge 2.
\end{equation}
Moreover, from the specification  of the fires sales process we obtain that
\begin{equation}
\label{eq:lik}
	%L_{i,0} = \ell_i, \quad
	\ell_{i,k} = \ell_i + \bm{x}_{i,k}h\big(\bm{\tau}_{(k)}/n\big) + \sum_{j=1}^{k-1} (\bm{x}_{i,j} - \bm{x}_{i,j+1})h\big(\bm{\tau}_{(j)}/n\big) , \quad i\in[n], k \in \mathbb{N},
\end{equation}
since at the beginning of round $k$ there remain   $\bm{x}_{i,k}$ shares which are affected with a price impact of $h(\bm{\tau}_{(k)}/n)$, and in any  preceding round $1 \le j \le k-1$ the number of sold shares is $\bm{x}_{i,j} - \bm{x}_{i,j+1}$, sold with a loss of $h(\bm{\tau}_{(j)}/n)$. We remark that the realized price for units sold during round $k$, which is $\bm{x}_{i,k} - \bm{x}_{i,k+1}$ is based on the losses triggered by $\bm{\tau}_{(k)}$, the total cumulative assets sold up to (and including) round $k-1$. The losses due to the asset sales in round $k$ itself are not endogenized for this round. 

By induction $(\bm{\tau}_{(k)})_{k\in\N}$ is non-decreasing componentwise and bounded by $ \sum_{i \in [n]} \bm{x}_i$. The limit 
\begin{equation}
\label{eq:finallysoldshares}
n\bm{\psi}_n := \lim_{k\to\infty}\bm{\tau}_{(k)},
\end{equation}
  that is, the vector of finally sold shares, exists. In addition to $\bm{\psi}_n$ we will also be interested in the number of defaulted institutions given by $\mathcal{D}_n:=\{i\in[n]\,:\,\lim_{k\rightarrow \infty }\ell_{i,k}  \geq c_i\}$ and hence
\begin{equation}
\label{eq:finalDefFixedSizeReal}
n^{-1}\vert\mathcal{D}_n\vert = n^{-1}\sum_{i\in[n]}\1\{\lim_{k\rightarrow \infty }\ell_{i,k}  \geq c_i\} .
\end{equation}
The set $\mathcal{D}_n$ provides additional information about the vulnerability of the financial system and the impact of fire sales on leveraged institutions such as banks or hedge funds. This completes the description of the model for fire sales that we study here. We would like to mention that within the already existing work, the model used in \cite{RePEc:eee:jfinec:v:115:y:2015:i:3:p:471-485} is possibly closest to the setup presented above. In \cite{RePEc:eee:jfinec:v:115:y:2015:i:3:p:471-485} the authors consider a fire sales process in a financial system of institutions that try  to maintain a fixed leverage ratio. In contrast to our setting, the price impact is assumed to be linear and the asset sales in one asset do not have a direct effect on the price of other assets. In addition, the authors focus on first order effects while we are particularly interested in higher order effects.
%As remarked previously, $\bm{x}_i\cdot h(\bm{\tau}_{(k)})$ is an upper bound for the losses of institution $i$ in round $k$.
%Here we work with the \emph{actual} losses incurred in each round, which makes the model complex and its dynamics quite involved. To the best of our knowledge and in direct comparison with previous works, this is the the first model that considers the actual losses instead of bounding them from above only. 

\subsection{Ensembles of Systems}
\label{ssec:stores}

Up to now we have described the fire sales process in any specific (finite) system. Our aim is, however, to understand qualitatively the characteristics of a system that promotes or hinders the spread of fire sales. As portrayed in the introduction, in the following we thus consider an ensemble of systems that are similar in the sense that they all share some (observed) statistical characteristics. This similarity is measured in terms of the most natural parameters, namely the joint empirical distribution~\eqref{eq:empDistFixedSize} of the asset holdings, the capital/equity, and the initial losses. In particular, we assume that we have a collection of systems with a varying number $n$ of institutions with the property that the sequence $(F_n)_{n\in \mathbb{N}}$ stabilizes, i.e., has a limit. Additionally, we assume convergence of the average asset holdings to a finite value; this is a standard assumption avoiding condensation of the distribution of the asset holdings. %We collect our assumptions:
\begin{assumption}\label{ass:regularity:fire:sales}
Let $M \in \N$ be a fixed number of assets. For each $n\in\N$ consider a system with $n$ institutions and with asset holdings $\bm{x}(n)=(\bm{x}_i(n))_{1\le i\le n}$, capitals $\bm{c}(n)=(c_i(n))_{1\le i\le n}$, and exogenous losses $\bm{\ell}(n)=(\ell_i(n))_{1\le i\le n}$. Let $F_n$ be the empirical distribution function of these parameters for $n\in\N$ (as in~\eqref{eq:empDistFixedSize}) and let $$(\bm{X}_n,C_n,L_n) = \big((X_n^1, \dots, X_n^M),C_n,L_n)\big) \sim F_n.$$ Then assume the following.
\begin{enumerate}[(a)]
\itemsep-1pt
\item \emph{\bf Convergence in distribution:} There is a distribution function $F$ such that as $n \to \infty$, $F_n(\bm{x},y,z)\to F(\bm{x},y,z)$ at all continuity points of $F$.
\item \emph{\bf Convergence of means:} Let $(\bm{X},C,L) = ((X^1, \dots, X^M),C,L)\sim F$. Then as $n\to\infty$,
\[ \E[X_n^m]\to\E[X^m]<\infty, \quad m\in[M]. \]
\end{enumerate}
\end{assumption}
An ensemble of systems satisfying Assumption~\ref{ass:regularity:fire:sales} will be called an \emph{$(\bm{X},C)$-system with initial shock $L$} in the sequel. A particular and probably the most relevant example is as follows. Suppose that a distribution $F$ is specified, possibly obtained by an empirical analysis of a real system. Then, for each $n\in \N$ we construct a system by assigning to each institution $i\in [n]$ independently asset holdings, capital and losses sampled from $F$. By the strong law of large numbers, with probability 1, the sequence of systems that we obtain satisfies Assumption~\ref{ass:regularity:fire:sales}. As in the (deterministic) model our aim is to describe in this broader setting the final state of the system. As institutions without any asset holdings play no role in the process and can be removed, we shall assume that $\P (X^1=0,\dots, X^m=0)=0$.

In order to simplify the presentation in the main part of this article we assume that the sales function is strictly increasing around 0. It can be waived and the results in full generality are presented the Section~\ref{discont:description} and proved in Section~\ref{proofs:general:case}.  
\begin{assumption}\label{ass:main:part}
The sales function $\rho:\R_{+}\to[0,1]$ is continuous and strictly increasing in a neighbourhood of $0$.
%\begin{enumerate}[(a)]
%\item The distribution of $(\bm{X},C,L)$ has no atoms \kosta{is continuous?} and
%\item the sales function $\rho:\R_{+,0}\to[0,1]$ is continuous and strictly increasing.
%\end{enumerate}
\end{assumption}

\noindent
{\bf (Non-)Resilience} In this part of the section we approach the heart of the matter and develop a notion of how \emph{resilient} or \emph{non-resilient} a given ~$(\bm{X},C)$-system is with respect to fire sales, triggered by some initial shock $L$. Note that all information about the system itself comes from $(\bm{X},C)$, whereas an initial shock is specified by the random variable $L$. We can then consider shocks of different magnitudes on the same, crucially, a priori unshocked system.
%In the following, if we use the notation~$g$, $f^m$, %$\circG$, $\circFSuper{m}$, $\hat{\bm{\chi}}$  and $\bm{\chi}^*$, we mean the quantities from the previous section (see Equations~\eqref{eqn:fire_sales:fm} and \eqref{eqn:def:g}) for the \daniel{unshocked} $(\bm{X},C)$-system, that is, with initial shock $L\equiv0$. It turns out that these quantities contain valuable information not only about resilience under the auxiliary process but also under the real process. 
%If instead we index these quantities by $\cdot_L$, we mean the shocked  $(\bm{X},C,L)$-system.
 From a regulator's perspective, for example, a desirable property of an $(\bm{X},C)$-system is the ability to absorb small local shocks $L$ without larger parts of the system being harmed. In our model we can even vary the statistical properties of $L$ arbitrarily. The following way of defining resilience thus  emerges  naturally: we let the shock $L$ `become small' in the sense that $\E[L/C]\to0$, and we call the system \emph{resilient} if the asymptotic number of sold shares  $\psi_{n}^m$ (that now -- explicitly -- depends on $L$) also tends to $0$. We would like to stress that we do not pose any restrictions on the joint distribution of $(\bm{X},C,L)$. In particular, we do not assume that the shock $L$ is independent of $(\bm{X},C)$.
 
\begin{definition}[Resilience]\label{def:resilience}
An $(\bm{X},C)$-system  is said to be \emph{resilient},
%under the auxiliary, respectively the real fire sales process
 if for any $\epsilon>0$ there exists $\delta>0$ such that for all $L$ with $\E[L/C]<\delta$ it holds %$\limsup_{n\to\infty}\chi_n^m  \leq \epsilon$, respectively
 $\limsup_{n\to\infty}\psi_n^m  \leq \epsilon$ for all $m\in [M]$. 
\end{definition} 
Our definition of resilience is in terms of the final number of sold shares. There are also other options, for example instead we could have based it on the final fraction of institutions that defaulted, which is the `standard' choice in the default contagion literature, see i.e.~\cite{Detering2016,Detering2018}. However, we selected the final number of sold shares, as the economic cost of fire sales can be large even when the number of defaults is rather small. In any case, all our results can easily be adapted to a resilience condition based on the default fraction.

We now move on to define non-resilient systems. 
%Note, however, that in our model description we made the conservative assumption that each institution $i\in[n]$ in the system is exposed to the final price impact $h(\bm{\chi}_n)$ with its total initial asset holdings $\bm{x}_i$. One can argue that institutions sell off their assets gradually and are hence not exposed to the total price change. The following result considers non-resilience under this conservative assumption. For other scenarios it can serve as a first indication of non-resilience.
%
% In this subsection, we restrict ourselves to initial shocks of the form $\ell_i\in\{0,2c_i\}$ for all $i\in[n]$, where $\P(L=2C)>0$ and $L$ is independent of $(\bm{X},C)$. That is, each institution $i$ defaults initially with positive probability.
%Rather than $\ell_i=2c_i$, the first natural choice to model the default of institution $i$ would be $\ell_i=c_i$. Note, however, that in the setting of Section~\ref{sec:fire:sales}, even if $\P(L=C)>0$, it is possible that no initial defaults occur since $(L,C)$ is defined as the weak limit of a sequence $(L_n,C_n)$ and it is possible that $L_n<C_n$ for all $n\in\N$ and still $L=C$ almost surely. In order to derive meaningful results one therefore has to choose $\ell_i=2c_i$ (or any other multiple larger than $1$). Note that this does not change the outcome of the fire sales process since $\rho(u)=\rho(1)$ for all $u\geq1$.
In complete analogy to the previous considerations, we call a financial system \emph{non-resilient} if the fraction of eventually sold shares is lower bounded by some positive constant which -- crucially -- does not dependent on the specification of the shock $L$, as long as this is positive. In order to ensure that the process starts at all, the shock has to affect some banks, so we require $L$ to be such that $\P (L>0)>0$.

\begin{definition}[Non-Resilience]\label{def:non:resilience}
An $(\bm{X},C)$-system is \emph{non-resilient}, if there exists $\Delta>0$ and an asset $m\in [M]$ such that  $\liminf_{n\to\infty} \psi_{n}^m \geq \Delta$ for any initial shock $L$ with $\P (L>0)>0$.
\end{definition}
The definition of non-resilience is rather strong, as we only require $\P (L>0)>0$, and apart from that, it is not really the complement of the notion of resilience in Definition~\ref{def:resilience}. Rather surprisingly, we will see later that under Assumption~\ref{ass:main:part} there is essentially no gap between Definition~\ref{def:resilience} and Definition~\ref{def:non:resilience}, that is, in most cases a system is either resilient or non-resilient. However, when $\rho$ is not  strictly increasing, the situation may be different. This is a particularly interesting and important setting, as it describes a situation where shock driven asset sales only occur if some institutions have been affected to some larger extent. It thus ensures that minor losses (e.g.~resulting from mark-to-market accounting of naturally volatile assets) and major shocks as the default of Lehman Brothers in 2008 are treated differently. Since in Section~\ref{sec:proofs} we consider a sales function $\rho$ that is not strictly increasing close to $0$, we provide the required notion of weak non-resilience here. 
\begin{definition}[Weak Non-Resilience]\label{def:weak:non:resilience}
An $(\bm{X},C)$-system is \emph{weakly non-resilient}, if it is not resilient, that is there exists $\Delta >0$ and an asset $m\in [M]$ such that  for every $\varepsilon >0$ there exists an initial shock $L$ such that $0 < \mathbb{E}[L/C] \le \varepsilon$  but $\liminf_{n\to\infty} \psi_{n}^m \geq \Delta$. 
\end{definition}
We will later see that, with the exception of some pathological cases, $\lim_{n\to\infty} \psi_{n}^m $ exists for all $m\in M$ and thus $\liminf_{n\to\infty} \psi_{n}^m=\limsup_{n\to\infty} \psi_{n}^m$, which shows that Definition~\ref{def:weak:non:resilience} complements Definition~\ref{def:resilience}.

\noindent
  {\bf Overview of the Results}
      Having defined the notions of (non-)resilience we can give an informal overview of our main findings.  Given an $(\bm{X},C)$-system with initial shock $L$ we first derive explicit bounds
 %-- depending on the system parameters $\bm{X},C,L$ only --
 for the final number of sold shares $\psi_n^m$, $m\in [M]$ (Section~\ref{ssec:aux:stoch}), and for the default fraction $n^{-1}|{\cal D}_n|$ in Section~\ref{proofs:general:case}). To this end, we resort to a different \emph{auxiliary} process described in the next section, which makes an important simplification to the rather involved fire-sales process given by~\eqref{eq:xik}--\eqref{eq:finalDefFixedSizeReal}: instead of considering the actual losses incurred in some round $k$,  we \emph{overestimate} them and replace them by the simple upper bound $\bm{x}_i\cdot h(\bm{\tau}_{(k)}/n)$, where $\bm{\tau}_{(k)}$ is the vector of the total number of sold shares in the first $k$ rounds. The advantage is that this auxiliary process can be handled analytically by solving a fixed-point equation. We then derive continuity properties of the fixed-point as a function of the system's parameters, which allow us to describe the auxiliary process for an ensemble of systems by a collection of functions $(f^m)_{m \in [M]}$. These functions essentially describe the evolution of the price of the corresponding asset during the execution of the process. The joint zero of these functions coincides with the final number of sold shares in the auxiliary process. The disadvantage of turning to the auxiliary process, however, is that it is not clear how much this simplification distorts from the behavior of the original process. Here we provide an explicit connection of the two processes: we manage to `sandwich' the fire sales process between two carefully crafted auxiliary processes, so that (non-)resilience properties are not or minimally affected (see Section~\ref{sec:real:process}).
 
With this coupling at hand we manage to derive explicit (non-)resilience criteria for $(\bm{X},C)$-systems. As it turns out (see Section~\ref{sec:real:process} and~\ref{ssec:resilience:real}), the resilience of such a system is linked to the behavior of the functions $(f^m)_{m \in [M]}$ near zero. We are then able to derive qualitative criteria that depend on the parameters of system -- $\bm{X}$,$C$ -- and of the process -- the sales function $\rho$ and the price impact $h$ -- only and characterize resilience. For example for a system with one asset ($M=1$), linear price impact and $X,C$ such that $\mathbb{E}[X^2/C]<\infty$, we show that if $\mathbb{E}[X^2/C] \rho'(0) > 1$, then the system is non-resilient: the capital is too small in direct comparison to the asset holdings, and selling begins immediately as a response to a price drop. On the other hand, if $\mathbb{E}[X^2/C] \rho'(0) <1$ and $\rho'(0) > 0$, then the system is resilient. We also consider cases in which $\rho'(0) = 0$ and $\mathbb{E}[X^2/C]=\infty $; there we show that the actual interplay between $\rho$ and $h$ (price drop) determines if the system is resilient or not. Several results of this kind and their interpretations are presented in Section~\ref{ssec:resilience:real}.
 
\section{The Auxiliary Model}\label{sec:aux:model}

\subsection{The Auxiliary  Process}

As already described previously, the aim of this section is to formulate an auxiliary process that will turn out useful for an analytic treatment of the fire sales process defined by~\eqref{eq:xik}--\eqref{eq:finalDefFixedSizeReal}. Note that   the essence of the fire sales process lies in Equation~\eqref{eq:lik}:  the behavior is rather complex, as we need to keep track of the asset sales in all  rounds. In order to make the analysis tractable, we will substitute~\eqref{eq:lik} in the auxiliary process as follows: we forget the actual prices at which assets were sold in previous rounds, and calculate the loss by using the price impact in the current round -- which is of course at least as large as the impact in previous rounds. More specifically, it holds that
\begin{equation}
\label{eq:likbound}
	%L_{i,0} = \ell_i, \quad
	\ell_{i,k} \le  \ell_i + \bm{x}_{i}h(\bm{\tau}_{(k)}/n), \quad i\in[n], k \in \mathbb{N}.
\end{equation}
In the \emph{auxiliary} fire sales process we thus replace the actual loss $\ell_{i,k}$ by this simple upper bound. %(As we already said, this process will serve as a tool to gain understanding of the real process.) 

We now provide a description of the final state of the system after this auxiliary fire sales process is completed; in particular, we are interested in the vector $\bm{\chi}_n$ of the number of finally sold shares divided by $n$ and the final price impact $h^m(\bm{\chi}_n)$ on any asset $m\in[M]$. 

\begin{figure}
\centering
    \begin{tabular}{ c || c | c }
  
    & Fire Sales Process & Auxiliary Process \\ \hline\hline
    vector of finally sold shares  & $\bm{\psi_n} = (\psi_n^1, \dots, \psi_n^m)$ & $\bm{\chi_n} = (\chi_n^1, \dots, \chi_n^m)$ \\ 
    divided by $n$ & & \\ \hline 
     vector of cumulatively sold shares & $\bm{\tau}_{(k)} = (\tau_{(k)}^1, \dots, \tau_{(k)}^M)$ & $\bm{\sigma}_{(k)} = (\sigma_{(k)}^1, \dots, \sigma_{(k)}^M)$ \\
     in rounds $1, \dots, k$ & & 
    \end{tabular}
\caption{Notation for the fire sales and the auxiliary process.}
\label{eq:tablenotation}
\end{figure}

In order to describe $\bm{\chi}_n$ we again consider the process in rounds, where in each round institutions react to the price changes from the previous round. Denote by $\bm{\sigma}_{(k)}=(\sigma_{(k)}^1,\ldots,\sigma_{(k)}^M)$ the vector of cumulatively sold shares in round $k$. We readily obtain
\begin{equation}
\label{eq:selling:1}
\bm{\sigma}_{(1)} = \sum_{i\in[n]}\bm{x}_i \rho\left(\frac{\ell_i}{c_i}\right) = n\E\left[\bm{X}_n\rho\left(\frac{L_n}{C_n}\right)\right]. 
\end{equation}
Similarly, in round $k\geq2$
\begin{equation}
\label{eq;selling}
\bm{\sigma}_{(k)} = \sum_{i\in[n]} \bm{x}_i\rho\left(\frac{\ell_i + \bm{x}_i\cdot h(n^{-1}\bm{\sigma}_{(k-1)})}{c_i}\right) = n\E\left[\bm{X}_n\rho\left(\frac{L_n + \bm{X}_n\cdot h(n^{-1}\bm{\sigma}_{(k-1)})}{C_n}\right)\right].
\end{equation}
Thus, by induction $(\bm{\sigma}_{(k)})_{k\in\N}$ is non-decreasing componentwise and bounded by $n\E[\bm{X}_n]$. The limit $n\bm{\chi}_n := \lim_{k\to\infty}\bm{\sigma}_{(k)}$ -- the vector of finally sold shares in the auxiliary process -- exists.

Table~\ref{eq:tablenotation} contrasts our notation for the fire sales process and the auxiliary process which will be used throughout the rest of the paper. We want to remind the reader that we work with a given financial system at this point and that the distribution of $(\bm{X}_n,C_n,L_n)$ is chosen such that we can rewrite the deterministic sums appearing in (\ref{eq:selling:1}) and (\ref{eq;selling}) in terms of expectations. This  turns out to be convenient later when we consider ensembles satisfying Assumption~\ref{ass:regularity:fire:sales}, as it avoids long expressions involving multivariate integrals. 

In contrast to the original fire sales process, it turns out that for the auxiliary process, the number of sold shares $\bm{\chi}_n$ can be derived by solving a rather simple equation. %\nils{Here we need a comment regarding uniqueness and relate to the condition used in \cite{RePEc:chf:rpseri:rp1520}.}
\begin{proposition}\label{Prop:det:setting}
Consider the auxiliary fire sales process as described above. Then  $\bm{\chi}_n = n^{-1}\lim_{k\to\infty}\bm{\sigma}_{(k)}$, the number of sold shares divided by $n$ at the end of the auxiliary process, is the smallest (componentwise) solution of
\begin{equation}\label{eqn:fixed:point:sigma}
{
\E\left[\bm{X}_n\rho\left(\frac{L_n+\bm{X}_n\cdot h(\bm{\chi})}{C_n}\right)\right] - \bm{\chi} = 0.}
\end{equation}
\end{proposition}
\proof
By continuity of $\rho$ and the dominated convergence theorem 
\begin{align*}
\bm{\chi}_n &= n^{-1}\lim_{k\to\infty}\bm{\sigma}_{(k)} = \lim_{k\to\infty}\E\left[\bm{X}_n\rho\left(\frac{L_n + \bm{X}_n\cdot h(n^{-1}\bm{\sigma}_{(k-1)})}{C_n}\right)\right]\\
&= \E\left[\bm{X}_n\rho\left(\frac{L_n + \bm{X}_n\cdot h(n^{-1}\lim_{k\to\infty}\bm{\sigma}_{(k-1)})}{C_n}\right)\right] = \E\left[\bm{X}_n\rho\left(\frac{L_n + \bm{X}_n\cdot h(\bm{\chi}_n)}{C_n}\right)\right]
\end{align*}
and $\bm{\chi}_n$ is thus a solution of \eqref{eqn:fixed:point:sigma}.

By the Knaster-Tarski theorem there exists a smallest fixed point $\hat{\bm{\chi}}_n$. Clearly, $\bm{\sigma}_{(0)}:=\bm{0}\leq n\hat{\bm{\chi}}_n$. Hence assume inductively that $\bm{\sigma}_{(k)}\leq n\hat{\bm{\chi}}_n$ for $k\geq1$. Then
\begin{equation}\label{eqn:induction:sigma}
\bm{\sigma}_{(k+1)} = \sum_{i\in[n]}\bm{x}_i \rho\left(\frac{\ell_i + \bm{x}_i\cdot h(n^{-1}\bm{\sigma}_{(k)})}{c_i}\right) \leq \sum_{i\in[n]}\bm{x}_i \rho\left(\frac{\ell_i + \bm{x}_i\cdot h(\hat{\bm{\chi}}_n)}{c_i}\right) = n\hat{\bm{\chi}}_n
\end{equation}
by monotonicity of $\rho$, and hence $\bm{\chi}_n=n^{-1}\lim_{k\to\infty}\bm{\sigma}_{(k)}\leq\hat{\bm{\chi}}_n$. By definition of $\hat{\bm{\chi}}_n$ it thus holds that $\bm{\chi}_n=\hat{\bm{\chi}}_n$.
\hfill\hfill
\endproof
Further, given $\bm{\chi}_n$, we readily obtain that under the auxiliary process the set of finally defaulted institutions is $\mathcal{D}_n:=\{i\in[n]\,:\,\ell_i+\bm{x}_i\cdot h(\bm{\chi}_n) \geq c_i\}$  and hence
\begin{equation}
\label{eq:finalDefFixedSize}
n^{-1}\vert\mathcal{D}_n\vert = n^{-1}\sum_{i\in[n]}\1\{\ell_i+\bm{x}_i\cdot h(\bm{\chi}_n)\geq c_i\} = \P\left(L_n+\bm{X}_n\cdot h(\bm{\chi}_n)\geq C_n\right).
\end{equation}
Note that we slightly `overload' the notation, as we use $\mathcal{D}_n$ for both the fires sales and the auxiliary process. Since it will be always clear from the context which processes is studied, this should cause no confusion.

Let us close this section with a remark about the case of non-continuous sales function $\rho$. The following example shows that also in this case it may be possible to determine the final state of the system by the smallest solution of \eqref{eqn:fixed:point:sigma}. Consider $\rho(u)=\1\{u\geq1\}$, that is, institutions sell their portfolio as they go bankrupt. Then $\bm{\sigma}_{(k)}\neq\bm{\sigma}_{(k-1)}$ only if in round $k$ at least one institution defaults that was solvent in round $k-1$. Since there are only $n$ institutions, the fire sales process stops after at most $n-1$ rounds and the vector $\bm{\chi}_n$ of finally sold shares divided by $n$ solves \eqref{eqn:fixed:point:sigma}. Again by \eqref{eqn:induction:sigma} we then obtain $\bm{\chi}_n=\hat{\bm{\chi}}_n$ is the smallest solution of \eqref{eqn:fixed:point:sigma}. By a similar reasoning we derive more generally for any sales function $\rho$ with finitely many discontinuities that $\bm{\chi}_n=\hat{\bm{\chi}}_n$. The case of arbitrary non-decreasing sales function is more difficult and treated in Section~\ref{discont:description}.

\subsection{The Auxiliary Process as $n\rightarrow \infty$}\label{ssec:aux:stoch}
 
Proposition~\ref{Prop:det:setting} describes the final state of any (finite) system when we consider the auxiliary process. Here we extend our considerations to an ensemble of systems. We are given an $(\bm{X},C)$-system with initial shock $L$, that is, an ensemble of systems, where for each $n\in\N$ a system with $n$ institutions and~$M$ assets is specified by sequences of asset holdings, capitals and exogenous losses with joint distribution $(\bm{X}_n,C_n,L_n)$. We also assume that the system satisfies Assumption~\ref{ass:regularity:fire:sales}, that is, the distribution of $(\bm{X}_n,C_n,L_n)$ converges to that of $(\bm{X},C,L)$ and the asset holdings are in $\mathcal{L}^1$.

\begin{figure}[t]
    \hfill \subfigure[]{\includegraphics[width=0.3\textwidth]{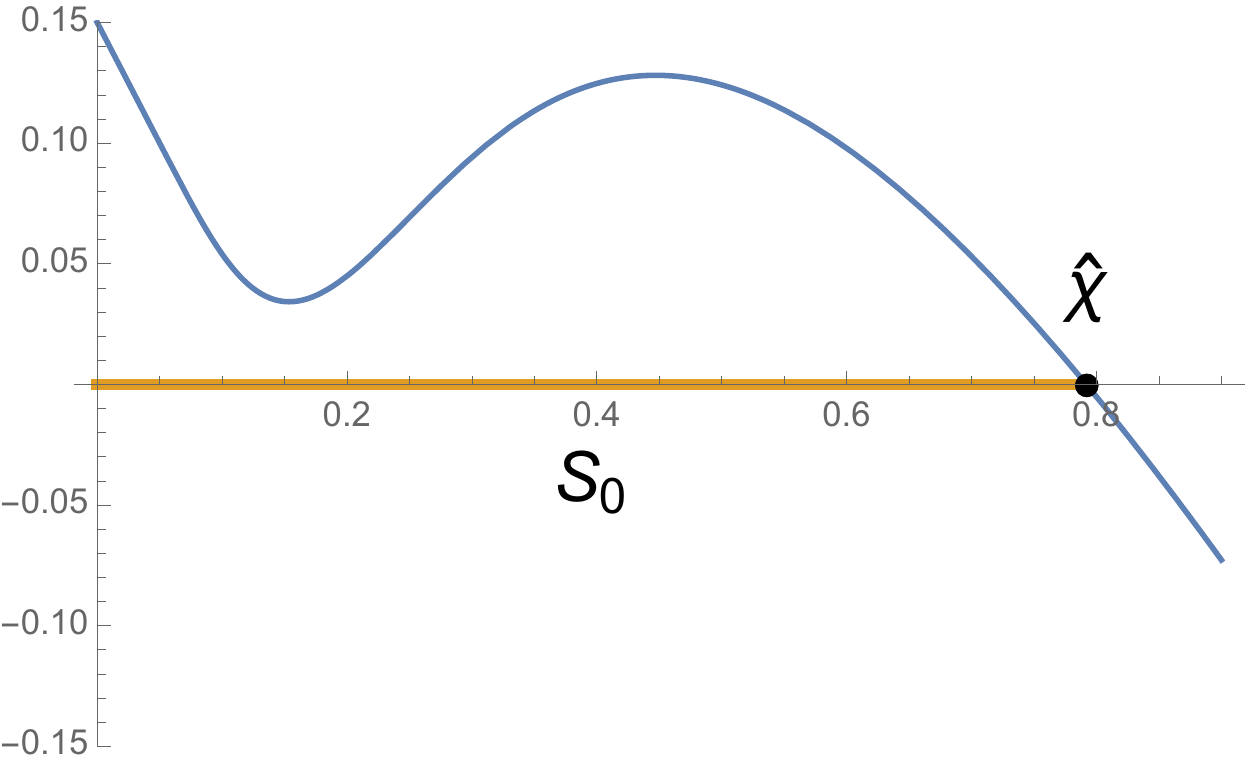}\label{fig:one:joint:root:large}} \hfill\subfigure[]{\includegraphics[width=0.3\textwidth]{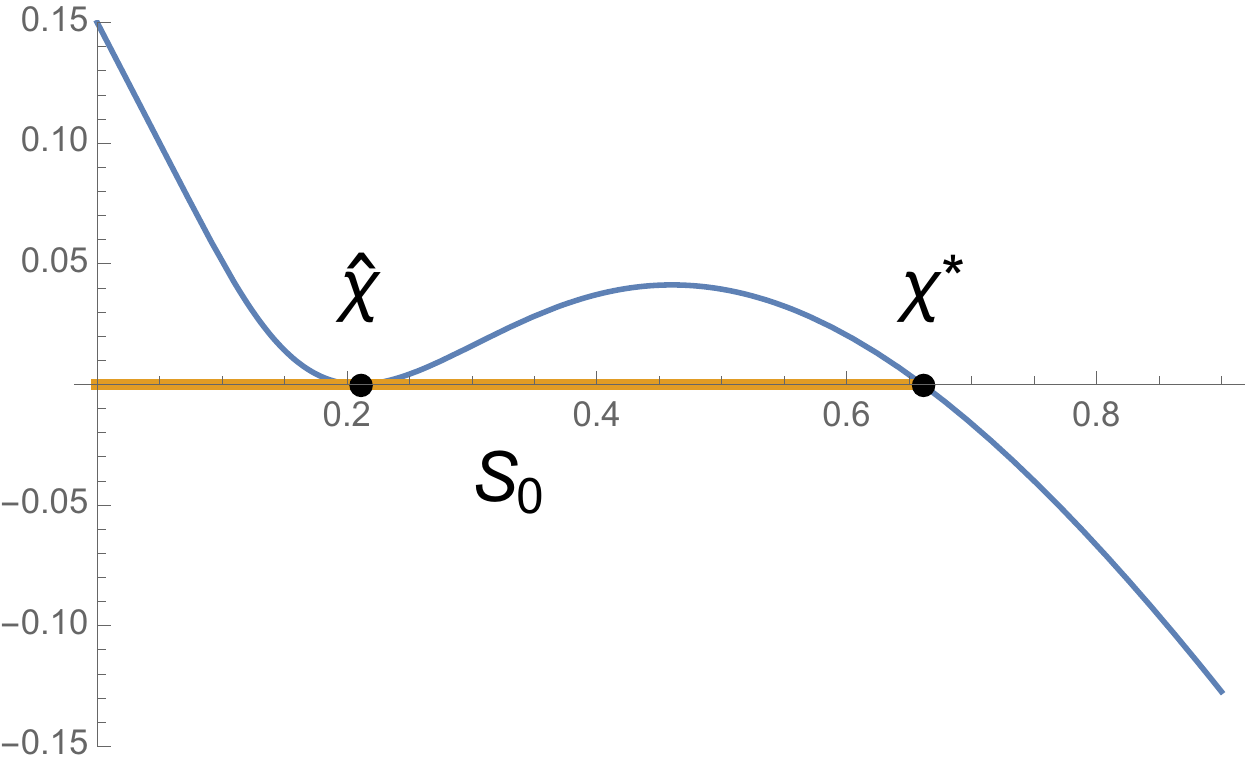}\label{fig:two:joint:roots}} \hfill \subfigure[]{\includegraphics[width=0.3\textwidth]{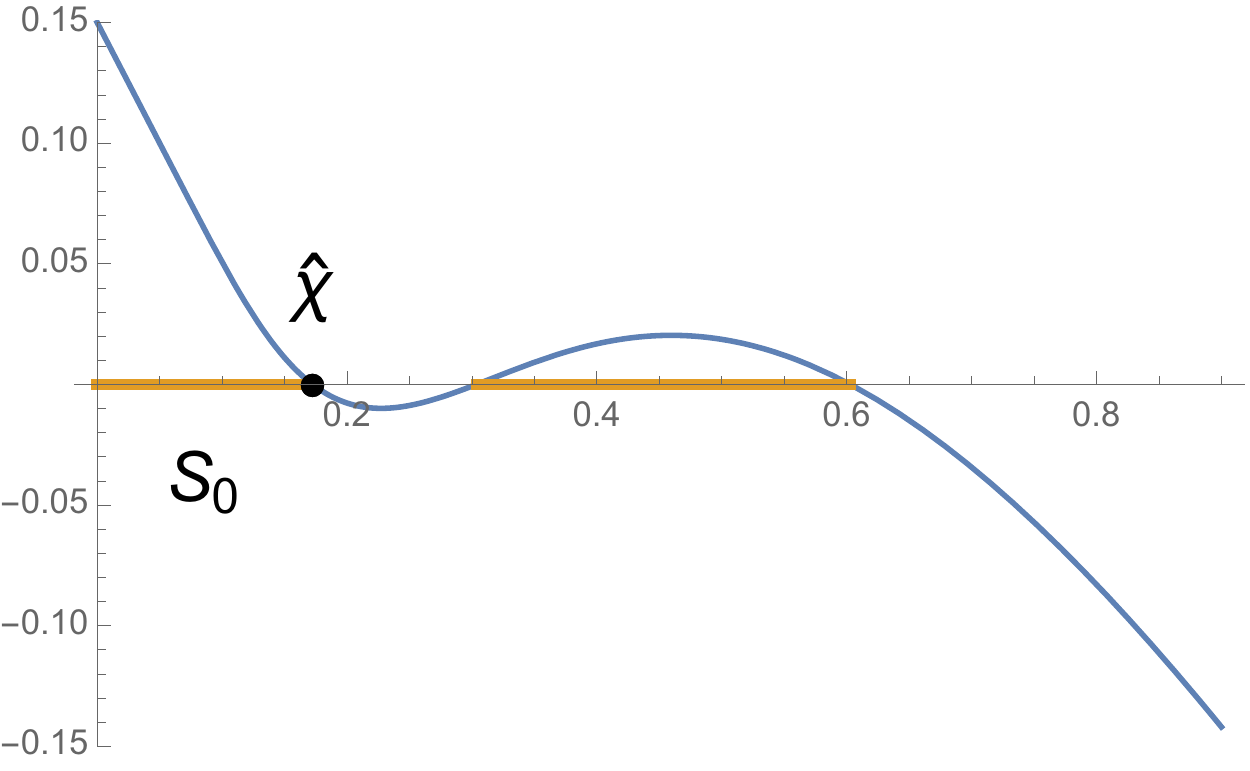}\label{fig:one:joint:root:small}}
    \hfill
\caption{Figure corresponding to Example \ref{example:1}. Plot of function $\bm{f(\chi)}$ (blue) for three different example systems. The set $\bm{S}$ is depicted in orange. }\label{fig:examples:chi:hat}
\end{figure}

Applying Proposition~\ref{Prop:det:setting} to each one of the systems in the ensemble, we obtain that for each $n \in\mathbb{N}$, the number of sold shares divided by $n$ at the end of the auxiliary process is the smallest solution of~\eqref{eqn:fixed:point:sigma}. Since $(\bm{X}_n,C_n,L_n) \to (\bm{X},C,L)$ it is thus natural to believe that for large $n$, the number of finally sold shares $\bm{\chi}_n$ is close to the smallest (componentwise) solution of the `limiting equation'
\begin{equation}
\label{eq:limitingeq}
\E\left[\bm{X}\rho\left(\frac{L+\bm{X}\cdot h(\bm{\chi})}{C}\right)\right] - \bm{\chi} = 0.
\end{equation}
As it will turn out, except for few pathological situations, this intuition is indeed correct and motivates the following definitions. Let 
%Before we do so, let us give some definitions that are handy in the forthcoming description. First, recall that \daniel{in the deterministic model} the eventual number of sold shares is characterized by the smallest solution to~\eqref{eqn:fixed:point:sigma}. \daniel{This \kosta{what?} can be explained as follows. Assume that asset sales occur continuously in time such that $n\bm{\chi}(\tau)$ denotes the vector of sold shares at time $\tau\geq0$ for some \nils{increasing} continuous map $\bm{\chi}:\R_{+,0}\to\R_{+,0}^M$. Then $n\E\left[X_n^m\rho\left(\frac{L_n+\bm{X}_n\cdot h(\bm{\chi}(\tau))}{C_n}\right) \right]$ describes the number of shares of asset $m$ the system as a whole needs to sell at time $\tau$ and the fire sales process stops \kosta{it doesn't necessarily stop} as soon as the number of sold shares of each asset $m\in[M]$ reaches this required number\nils{revise!}. Moreover, at every time $\tau$ during the fire sales process it must hold that the required number $n\E\left[X_n^m\rho\left(\frac{L_n+\bm{X}_n\cdot h(\bm{\chi}(\tau))}{C_n}\right) \right]$ of shares of asset $m$ to sell is at least the number of already sold shares $n\chi^m(\tau)$ \kosta{verstehe ich leider auch nicht}. Let us now translate this to the stochastic setting. Analogue to \eqref{eqn:fixed:point:sigma} we define the functions \kosta{koennen wir nicht einfach sagen, fuer jedes $n$ haben wir (4), jetzt vertuaschen wir die limits und kriegen (6)?}
%
\begin{equation}
f^m(\bm{\chi}) := \E\left[X^m\rho\left(\frac{L+\bm{X}\cdot h(\bm{\chi})}{C}\right) \right] - \chi^m,\quad m\in[M].\label{eqn:fire_sales:fm}
\end{equation}
Further, let 
\begin{equation}
\label{eq:S}
S:=\bigcap_{m\in[M]}\left\{\bm{\chi}\in\R_{+}^M\,:\,f^m(\bm{\chi})\geq0\right\}
\end{equation}
be the subset where all of the functions $f^m$, $m\in[M]$, are non-negative. 
%Then by the same heuristics as outlined for the deterministic case above, at no time $\tau$ the vector of sold shares $\bm{\chi}(\tau)$ will leave the set $S$ (clearly $\bm{\chi}(0)=\bm{0}\in S$) and the fire sales process will stop at a joint root $\hat{\bm{\chi}}$ of all the functions $f^m$. 
Let $\hat{\bm{\chi}}$ be the smallest joint root (which always exists by the Knaster-Tarski theorem) of the functions $f^m, m\in [M]$. While the necessity of Assumption \ref{ass:regularity:fire:sales} is obvious, it is surprising that in most cases it is actually also sufficient to ensure that 
\begin{equation}\label{convergence:chi}
\hat{\bm{\chi}}=\lim_{n\rightarrow \infty }\bm{\chi}_n .
\end{equation}
 Finally, in analogy to \eqref{eq:finalDefFixedSize} we can therefore expect for the ensemble that the final fraction of defaulted institutions in the auxiliary model is given by
\begin{equation}\label{eqn:theorem:heuristical}
\lim_{n\rightarrow \infty }n^{-1}\vert\mathcal{D}_n\vert=g(\hat{\bm{\chi}}),
\quad \text{where} \quad
%\end{equation}
%where
%\begin{equation}\label{eqn:def:g}
g(\bm{\chi}) := \P(L+\bm{X}\cdot h(\bm{\chi})\geq C).
\end{equation}
To illustrate under which conditions the previous statements are true and the intuition is right, let us look at an instructive example with one asset only ($M=1$). This is a carefully crafted `toy example' that is unlikely to model a real system, but it will reveal the important characteristics that determine the underlying behavior.
\begin{example}\label{example:1}
{\it ~Consider a system defined by
$$h(\chi)=\chi,~~\rho(y)=\1\{y\geq1\},~~X\sim\mathrm{Exp}(1),~~C\equiv c,~~\P(L=c)=0.15 \text{ and } \P(L=0)=0.85,$$
where $L$ is independent of $X$ and $c=0.75$. See Figure \ref{fig:one:joint:root:large} for an illustration of $f$, $S$ and $\hat{\bm{\chi}}$. There~$\hat{\bm{\chi}}$ is the only root of $f$ and the set $S$ is the interval $[0,\hat{\bm{\chi}}]$. The situation changes, however, as we vary the capital. See Figure \ref{fig:one:joint:root:small} for the case of $c=0.95$. Then $S$ consists of two disjoint intervals and~$f$ has three  roots. In fact, one can observe the splitting of $S$ and thus a  discontinuity of $\hat{\chi}$ at $c\approx0.91$, see Figure  \ref{fig:two:joint:roots}.
The situation in Figure  \ref{fig:two:joint:roots} presents a rather special and pathological case. In fact, only in this very special case above heuristics fail and the situation becomes more complex than suggested in \eqref{convergence:chi} and \eqref{eqn:theorem:heuristical}.
}
\end{example}

\noindent
In order to formalize the special situation encountered in the previous example, let us denote by $S_0$ the largest connected subset of $S$ containing $\bm{0}$ (clearly $\bm{0}\in S$). Further, let
\begin{equation}
\label{eq:defchistar}
 \bm{\chi}^*\in\R_{+}^M \quad\text{with}\quad (\chi^*)^m:=\sup_{\bm{\chi}\in S_0}\chi^m.
\end{equation}
\begin{lemma}\label{lem:existence:chi:hat}
There exists a smallest joint root $\hat{\bm{\chi}}$ of all functions %$\circFSuper{m}(\bm{\chi})$,%
$f^m(\bm{\chi})$, $m\in[M]$, with $\hat{\bm{\chi}}\in%\accentset{\circ}{S}_0
S_0$. Further, $\bm{\chi}^*$ as defined above is a joint root of the functions %\textcolor{red}{$\circFSuper{m}$ (anschaulich ja, aber stimmt das? Siehe Beweis)}, 
$f^m$, $m\in[M]$, and $\bm{\chi}^*\in S_0$.
\end{lemma}
The special case from Figure \ref{fig:two:joint:roots} is then described by $\hat{\bm{\chi}}\neq\bm{\chi}^*$, and in all other cases~$\hat{\bm{\chi}}=\bm{\chi}^*$. We shall henceforth term the former case as `pathological', since we can only obtain it if we maliciously fine-tune the parameters of the system to provoke such a behavior. We do not expect such situations to occur or to be particularly relevant, and in all settings that we study here it holds that $\hat{\bm{\chi}}=\bm{\chi}^*$. Note that in previous literature on fire sales the question of uniqueness of the fixed-point has been studied (\cite{Eisenberg2001,Rogers2013,RePEc:chf:rpseri:rp1520}). We do not require uniqueness of the fixed-point here and the situation $\hat{\bm{\chi}}=\bm{\chi}^*$ is only a stability property of the (component-wise) first fixed-point.

%To formalize this intuition about joint roots in $S_0$, consider the following lemma. Let
%\[
%\bm{\chi}^*\in\R_{+,0}^M \quad\text{with}\quad (\chi^*)^m:=\sup_{\bm{\chi}\in S_0}\chi^m.
%\]
%\begin{lemma}\label{lem:existence:chi:hat}
%There exists a smallest joint root $\hat{\bm{\chi}}$ of all functions %$\circFSuper{m}(\bm{\chi})$,%
%\daniel{$f^m(\bm{\chi})$}, $m\in[M]$, with $\hat{\bm{\chi}}\in%\accentset{\circ}{S}_0
%S_0$. Further, $\bm{\chi}^*$ as defined above is a joint root of the functions %\textcolor{red}{$\circFSuper{m}$ (anschaulich ja, aber stimmt das? Siehe Beweis)}, 
%$f^m$, $m\in[M]$, and $\bm{\chi}^*\in S_0$.
%\end{lemma}
%At the end of the section we will give a couple of examples illustrating the situation in concrete settings. Using the quantities $\hat{\bm{\chi}}$ and $\bm{\chi}^*$ as well as the functions $\circG$ and $g$, we can then describe the final state of the system after the fire sales process asymptotically as $n\to\infty$. 

Our first result provides in non-pathological cases (that is, when $\hat{\bm{\chi}} =\bm{\chi}^*$) the asymptotic number of finally sold shares for the auxiliary process and in all other cases upper and lower bounds. 
\begin{theorem}\label{thm:fire:sale:final:fraction}
Consider an $(\bm{X},C)$-system with initial shock $L$ satisfying Assumptions \ref{ass:regularity:fire:sales},~\ref{ass:main:part}. Then the number $\chi_n^m$ of finally sold shares of asset $m\in[M]$ divided by $n$ in the auxiliary process satisfies 
%\[ n^{-1}\vert\mathcal{D}_n\vert = g(\hat{\bm{\chi}}) + o(1), \qquad \chi_n^m = \hat{\chi}^m + o(1). \]
\[\begin{gathered}
%g (\hat{\bm{\chi}}) + o(1) \leq n^{-1}\vert\mathcal{D}_n\vert \leq g(\bm{\chi}^*) + o(1), \qquad
\hat{\chi}^m+o(1) \leq \chi_n^m \leq (\chi^*)^m+o(1).
\end{gathered}\]
In particular, for the final price impact $h^m(\bm{\chi}_n)$ on asset $m\in[M]$ it holds that  
\[
	h^m(\hat{\bm{\chi}})+o(1)
	\leq h^m(\bm{\chi}_n)
	\leq h^m(\bm{\chi}^*)+o(1).
\]
\end{theorem}
%Moreover, one easily observes that if $\P (X^m,L >0)>0$ for some $m\in [M]$, then $\hat{\bm{\chi}}\neq 0$. \kosta{Wozu das hier?}\nils{Kann von mir aus wieder raus. Ist vielleicht verwirrend weil klar sein sollte dass $\hat{\bm{\chi}}\neq 0$.}

\subsection{Resilience Criteria for the Auxiliary Process}
In the previous section we derived results that allow us to determine the final default fraction in $({\bf X},C)$-systems caused by  {the auxiliary process} and sparked by some exogenous shock $L$. In this section we go one step further and investigate whether a given  system in an \emph{initially} unshocked state is likely to be resilient to small shocks or susceptible to fire sales. 

Note that all information about an initial shock comes from the random variable $L$, whereas the system itself is specified by~$(\bm{X},C)$. So we can easily consider shocks of different magnitude~$L$ on the same a priori unshocked system. The unshocked system itself can then be described by the case $L\equiv0$. In the following, whenever we use the notation~$f^m$, $S$, $g$ and $\bm{\chi}^*$, we always mean the quantities~\eqref{eqn:fire_sales:fm}-\eqref{eq:defchistar} from the previous section for the $(\bm{X},C)$-system with initial shock $L\equiv0$, i.\,e.\, the unshocked system. 

\begin{remark}
For $L\equiv0$, in particular the smallest joint root is always given by $\hat{\bm{\chi}}=\bm{0}$, as $f^m(\bm{0})=0$. Instead (non-)resilience will be described in terms of $\bm{\chi}^*$ below. While in the previous subsection for $\P (L>0 )>0$ (shocked systems), we remarked that $\hat{\bm{\chi}}\neq\bm{\chi}^*$ only in pathological cases, we will see in this subsection for $L\equiv0$ that $\bm{\chi}^*\neq\bm{0}=\hat{\bm{\chi}}$ describes an important and non-pathological case. In fact, for $M=1$, if $f=f^1$ is such that $f' (0) > 0$, then always $\hat{\bm{\chi}}\neq\bm{\chi}^*$.
\end{remark}

We shall first provide resilience criteria for the auxiliary system. 
\begin{theorem}\label{thm:resilience}
For each $\epsilon>0$ there exists $\delta>0$ such that for all $L$ with $\E[L/C]<\delta$ the number $n\chi_{n,L}^m$ of finally sold shares of each asset $m\in[M]$ for the auxiliary process %and the final fraction of defaulted institutions $n^{-1}\vert\mathcal{D}_{n,L}\vert$ in the shocked system 
satisfy
\[\limsup_{n\to\infty} \chi_{n,L}^m \leq(\chi^*)^m+\epsilon, ~ m\in[M] .\]%\quad \text{and}\quad  \limsup_{n\to\infty}n^{-1}\vert\mathcal{D}_{n,L}\vert\leq g(\bm{\chi}^*)+\epsilon . \]
\end{theorem}
We immediately obtain the following handy resilience criterion.

\begin{corollary}[Resilience Criterion]\label{cor:resilience}
If $\bm{\chi}^*=0$, then the $(\bm{X},C)$-system is resilient under the auxiliary process.
\end{corollary}
%Note that $g(\bm{0})=0$ and hence there are only few defaults if the system is resilient (i.e.~$\bm{\chi}^*=\bm{0}$,  $S_0=\{\bm{0}\}$). It is possible, however, that $g(\bm{\chi}^*)=0$ while $\bm{\chi}^*\neq\bm{0}$ in which case a shock triggers a large fraction of asset sales but only very few defaults.
 
 We now turn to non-resilience criteria and study the case $\bm{\chi}^*>0$. Consider for example a system with one asset, assuming differentiability of $f=f^1$. Then $\bm{\chi}^*>0$ whenever $f '(0)>0$. (If on the contrary $f '(0)<0$, then $\bm{\chi}^*=0$ and the system is resilient by Corollary \ref{cor:resilience}.)

In fact, assuming that there is no additional zero between $\bm{0}$ and $\bm{\chi}^*>\bm{0}$, one can show that the first zero of any shocked system is lower bounded by $\bm{\chi}^*$ and thus Theorem~\ref{thm:fire:sale:final:fraction} guarantees that  $\chi_n^m \ge (\chi^*)^m - o(1)$ as $n\to\infty$, implying non-resilience of the system.

%Note that the considerations in the previous subsection assumed that $\P (L>0 )>0$ as otherwise the process would not start. This in particular implies that $f^m(0)>0$ for some $m\in [m]$ and we termed the situation $\hat{\bm{\chi}}\neq\bm{\chi}^*$ as pathological as it requires extensive fine tuning of the parameters. The situation is different for $L =0$ as in this case $f^m(0)=0, m\in [M]$ and therefore always $\hat{\bm{\chi}}=\bm{0}$. Whenever $\bm{\chi}^* = \bm{0} $ and thus $\hat{\bm{\chi}}=\bm{\chi}^*$, we saw above that the system is resilient to small shocks. In a system with one asset, assuming differentiability of $f=f^1$, this is the case in particular if $(f^1)'(0)<0$. If on the contrary $(f^1)'(0)>0$, then $\hat{\bm{\chi}}\neq\bm{\chi}^*$ and because of $L=0$ the separation of $\bm{\chi}^*$ from $\bm{0}$ withstands minor modifications to the system. If we now apply a shock $L$ to the system, the process actually gets started and $\bm{0}$ is not a joint root of the functions~\eqref{eqn:fire_sales:fm} anymore. Assuming that there is no additional zero between $\hat{\bm{\chi}}$ and $\bm{\chi}^*$, one can show that the first zero of the shocked system is lower bounded by $\bm{\chi}^*$ and thus Theorem~\ref{thm:fire:sale:final:fraction} then guarantees that  $\chi_n^m \ge (\chi^*)^m - o(1)$ as $n\to\infty$. More general, without any differentiability assumptions, the following theorem thus shows non-resilience when $\hat{\bm{\chi}}\neq \bm{\chi}^*$.

\begin{theorem}\label{thm:non-resilience}
Consider an $(\bm{X},C)$-system such that $\bm{\chi}^*>0$ and such that no $z\in\mathbb{R}^M \setminus\{\bm{0}, \bm{\chi}^*\}$ with $\bm{0} \le z \le \bm{\chi}^*$ (coordinate-wise) is a joint root of the functions $(f^m)_{m\in[M]}$. 
Then the system is non-resilient under the auxiliary process, more specifically
\[ %\circG
%-\epsilon
  \chi_{n,L}^m \geq (\chi^*)^m - o(1), \quad m\in[M], %\quad\text{and}\quad \liminf_{n\to\infty}n^{-1}\vert\mathcal{D}_{n,L}\vert \geq g (\bm{\chi}^*)%-\epsilon
  \]
  where $n\chi_{n,L}^m$ is the number of finally sold shares of asset $m$.
%  for all shocks $L$ such that $\sum_{m\in [M]} \P (L,X^m>0)>0$.
\end{theorem}

The assumption that there is no joint root between $\bm{0}$ and $\bm{\chi}^*$ can be weakened and the conclusion {of non-resilience} is for example also true if we request that there are only finitely many roots between $\bm{0}$ and $\bm{\chi}^*$, or that such roots are bounded away from $\bm{0}$, {an observation we shall regularly use in proofs}.
Constructing such highly artificial examples, where the infimum of the set of roots is $\bm{0}$, would require an enormous fine-tuning of the system parameters (and enough malicious energy).

Let us finally remark here that Theorem~\ref{thm:non-resilience} is not  true without the assumption that $\rho$ is strictly increasing. This becomes clear by looking at the particular example $\rho(u)=\1\{u\geq1\}$ of sales at default. Asset sales will only happen if some banks default initially, that is, $\P (L_n\geq C_n)>0$; a shock which is such that $\P (L_n \geq C_n)=0$ can not trigger any  sales.  In Theorem~\ref{thm:non-resilience:non:strictly:inc} we show, however, that for $\bm{\chi}^*>0$ the system is always weakly non-resilient. This completes our analysis of (non-)resilience under the auxiliary process. We close the section with an (again artificial) example demonstrating the qualitative difference between resilient and non-resilient systems.

\begin{figure}[t]
\centering
     \subfigure[]{\includegraphics[width=0.3\textwidth]{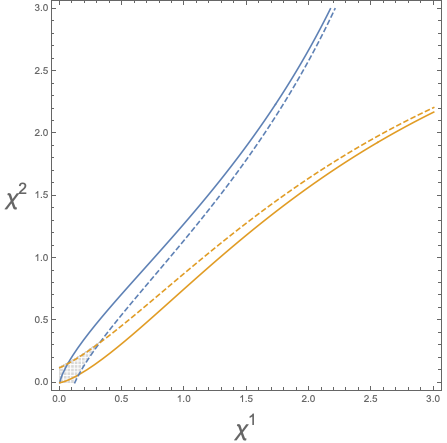}\label{fig:res}} $\qquad\qquad$\subfigure[]{\includegraphics[width=0.3\textwidth]{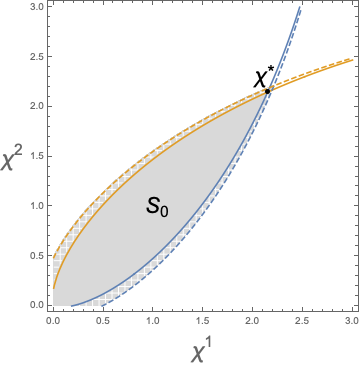}\label{fig:non-res}} 
    
\caption{Figure corresponding to Example \ref{example:2}. Plot of the zero lines for $f^1$ (blue) and $f^2$ (yellow) of the unshocked system (solid) and the shocked system (dashed). Left (a) depicts a resilient system  and right (b) a non-resilient system. }
\end{figure}

\begin{example}\label{example:2}
{\it ~Let $d(x) =   1.5 x^{-5/2}\mathbf{1}[x \ge 1]$ and let $X$ be a random variable with density $d$. Consider a system with two assets ($M = 2$) defined by
$$h(\chi)=\chi,~~\rho(u) = \min\{1,u^2\},~~X_1 = X_2=X,~~C= 10 \, X^\gamma$$% \text{ and } \P(L=0)=0.95,$$
for some $\gamma > 0$. Moreover, we consider a shock $L$ independent of $X$ with $\P(L=C)=0.03$. See Figure \ref{fig:res} for an illustration of $f^1, f^2$ when $\gamma = 0.6$ and~\ref{fig:non-res} for the case $\gamma = 0.4$. The solid yellow and blue lines mark the zero sets of the functions $f^1$ and $f^2$ for the \emph{unshocked} system. For $\gamma = 0.6$ (a) the system is resilient ($\bm{\chi}^* = 0$, $S_0 = \{0\}$) as the solid lines cross in $0$, for $\gamma = 0.4$ (b) the system is non-resilient and $S_0$ is the area bounded by the solid lines, which cross first in $\bm{\chi}^* \neq 0$. The dashed lines are the zero lines for the functions for the shocked system and the coordinates of their intersection mark the number of sold shares. With decreasing shock size, the dashed lines of the shocked system will approach the lines of the unshocked system. In the resilient case (a) this pushes the intersection of the dashed blue and yellow lines towards $0$ and only few shares are sold. In the non-resilient system (b), as the shock gets smaller, the intersection of the dashed lines is approaching $\bm{\chi}^* \neq 0$ and is thus lower bounded, no matter how small the shock is.
}
\end{example}

%\begin{corollary}[Non-Resilience Criterion]\label{cor:non:resilience}
%If $\bm{\chi}^*>0$, then the $(\bm{X},C)$-system is non-resilient under the auxiliary process.
%\end{corollary}

%As remarked earlier already, for most practical purposes it will hold that $\circG(\bm{\chi}^*)=g(\bm{\chi}^*)$ and the combination of Corollaries \ref{cor:resilience} and \ref{cor:non:resilience} hence fully describe stability of an $(\bm{X},C)$-system. Only if $\circG(\bm{\chi}^*)=0<g(\bm{\chi}^*)$ we cannot decide if an $(\bm{X},C)$-system is resilient or non-resilient. In this case depending on the exact convergence in Assumption \ref{ass:regularity:fire:sales} (a) the system can either be resilient or non-resilient.

\section{The Fire Sales Process}
\label{sec:fsp}

\subsection{Bounding the Fire Sales Process with the Auxiliary Process}\label{sec:real:process}

In this section we establish an explicit connection between the fire sales process~\eqref{eq:xik}--\eqref{eq:finalDefFixedSizeReal} and the auxiliary process described and studied in the previous section. In particular, we will use the final number of sold assets $\bm{\chi}_n = n^{-1}\lim_{k\to\infty}\bm{\sigma}_{(k)}$ in a specific auxiliary processes to bound the asset sales $\bm{\psi}_n = n^{-1}\lim_{k\rightarrow \infty}\bm{\tau}_{(k)}$ in the real process.

Let us start with a simple observation. For a given finite financial system $(\bm{X}_n,C_n)$ with shock~$L_n$ and specified functions $\rho$ and $h$, it easily follows from~\eqref{eq:lik} and~\eqref{eq:likbound} that the auxiliary process leads to more asset sales, to a higher price impact and as a result also to more defaults. Thus, the auxiliary process always serves as an upper bound for the fire sales process (regarding the actual asset sales and also the final default fraction).  This is summarized in the following theorem. In the rest of this section we (silently) consider a $(\bm{X},C)$-system with initial shock $L$ that satisfies Assumptions~\ref{ass:regularity:fire:sales} and~\ref{ass:main:part}.
\begin{theorem}\label{thm:fire:sale:upper:bound:real}
Let $\psi_n^m$ be the number of finally sold shares of asset $m\in[M]$ divided by $n$ in the fire sales  process. Let $\chi^*$ be as in~\eqref{eq:defchistar}. Then %the final default fraction $n^{-1}\vert\mathcal{D}_n\vert$ and 
\[ \limsup_{n\rightarrow \infty} \psi_n^m \leq (\chi^*)^m.\]
% \limsup_{n\rightarrow \infty} n^{-1}\vert\mathcal{D}_n\vert = g(\hat{\bm{\chi}}), \qquad
\end{theorem}
\proof
According to our assumptions, the functions $\rho:\R_{+}\to[0,1]$ and $h^m:\R_{+}^M\to[0,1]$ are non-decreasing and $\bm{\tau}_{(1)}= \bm{\sigma}_{(1)}=0$. {From~\eqref{eq:lik}  and~\eqref{eq:likbound} we obtain that}
\[\ell_{i,k} = \ell_i + \bm{x}_{i,k}h(\bm{\tau}_{(k)}/n) + \sum_{\ell=1}^{k-1} (\bm{x}_{i,\ell} - \bm{x}_{i,\ell+1})h(\bm{\tau}_{(\ell)}/n)  \le  \ell_i + \bm{x}_{i}h(\bm{\tau}_{(k)}/n) .\]
It follows that $\rho (\ell_{i,k})  \leq \rho (\ell_i + \bm{x}_{i}h(\bm{\tau}_{(k)}/n))$
and as a result $\bm{\tau}_{(k)} \leq  \bm{\sigma}_{(k)}$. Thus $$\bm{\psi}_n= \lim_{k\rightarrow \infty}\bm{\tau}_{(k)}n^{-1} \leq \lim_{k\rightarrow \infty} \bm{\sigma}_{(k)}n^{-1}=\bm{\chi}_n.$$ By Theorem~\ref{thm:fire:sale:final:fraction}, we know that for the auxiliary process,  $\limsup_{n\rightarrow \infty} \chi_n^m \leq  (\chi^*)^m$ and thus it follows that $\limsup_{n\rightarrow \infty} \psi_n^m \leq (\chi^*)^m$.
\hfill\hfill
\endproof
{More surprisingly, as we shall see below, we can adapt carefully the parameters of the given system so as to also obtain lower bounds from the auxiliary process.} The central idea is to define the auxiliary process on another system having smaller initial asset holdings and a sales function that is dominated point-wise by the original one.
%It allows us to obtain an auxiliary process that lower bounds the real process in the original system.
More specifically, let us fix $\varepsilon > 0$. In addition to $\rho$, define the sales function $\rho_{\varepsilon}:\R_{+}\to[0,1]$ by $\rho_{\varepsilon}(u) =\min \{ \rho (u), \varepsilon \}$.
Further, we consider a system that is derived from the $(\bm{X},C)$-system with initial shock $L$ but with asset holdings $(1-\varepsilon)\bm{X}$ instead.
In analogy to~\eqref{eqn:fire_sales:fm} we consider the functions
\begin{equation}
f_{\varepsilon}^m(\bm{\chi}) := \E\left[(1-\varepsilon)X^m\rho_{\varepsilon}\left(\frac{L+(1-\varepsilon)\bm{X}\cdot h(\bm{\chi})}{C}\right) \right] - \chi^m,\quad m\in[M].\label{eqn:fire_sales:fmeps}
\end{equation}
These functions describe a system where, in comparison to the original system, institutions hold fewer shares to start with and sell a smaller fraction of their assets as a reaction to a shock. 
Further, also in analogy to~\eqref{eq:S}, let 
\[ S_{\varepsilon}:=\bigcap_{m\in[M]}\left\{\bm{\chi}\in\R_{+}^M\,:\,f_{\varepsilon}^m(\bm{\chi})\geq0\right\} \]
and let $S_{\varepsilon,0}$ be the largest connected subset of $S_{\varepsilon}$ containing $\bm{0}$. Moreover, let $\hat{\bm{\chi}}_{\varepsilon}\in S_{\varepsilon,0}$ the smallest joint root of the functions $f_{\varepsilon}^m, m \in [M]$.
%\[ \bm{\chi}^{\varepsilon,*}\in\R_{+,0}^M \quad\text{with}\quad (\chi^{\varepsilon,*})^m:=\sup_{\bm{\chi}\in S_{\varepsilon,0}}\chi^m. \]
We obtain the following crucial lower bound:

\begin{theorem}\label{thm:fire:sale:lower:bound:real}
Let $\psi_n^m$ be the number of finally sold shares of asset $m\in[M]$ divided by $n$ in the fire sales  process. For every $\varepsilon > 0$ 
\[ \liminf_{n\rightarrow \infty} \psi_n^m \geq \hat{\chi}_{\varepsilon}^m. \]
\end{theorem}
\proof
{For every $n\in \mathbb{N}$ we consider a system $(\bm{X}_n,C_n,L_n)$ satisfying Assumptions~\ref{ass:regularity:fire:sales} and \ref{ass:main:part}. Then it is immediate that the sequence $((1-\varepsilon)\bm{X}_n,C_n,L_n)_{n\in \mathbb{N}}$ satisfies Assumption \ref{ass:regularity:fire:sales} with limiting distribution $((1-\varepsilon)\bm{X},C,L)$.} We now fix $n\in \mathbb{N}$. {As usual,} let $\bm{\tau}_{(k)}$ be the total number of assets sold in round $k$ in   $(\bm{X}_n,C_n,L_n)$ with  sales function $\rho$ under the fire sales process. Further, let $\bm{\sigma}_{(k)}$ be the total number of assets sold in round $k$ in   $((1-\varepsilon)\bm{X}_n,C_n,L_n)$ with   sales function $\rho_{\varepsilon}$ in the auxiliary process. We shall show that $\bm{\sigma}_{(k)} \leq  \bm{\tau}_{(k)}$, which implies that $\chi_n =\lim_{k\rightarrow \infty}\bm{\sigma}_{(k)} n^{-1} \leq  \lim_{k\rightarrow \infty}\bm{\tau}_{(k)} n^{-1}=\psi_n$ and since by Theorem~\ref{thm:fire:sale:final:fraction}, $\liminf_{n\rightarrow \infty} \chi^m_n \geq \hat{\chi}_{\varepsilon}^m$, the claim follows.
In order to show that $\bm{\sigma}_{(k)} \leq  \bm{\tau}_{(k)}$ we use induction. By definition  $\bm{\sigma}_{(1)} =  \bm{\tau}_{(1)}=0$. So, lets assume that $\bm{\sigma}_{(1)} \leq   \bm{\tau}_{(1)},\dots,  \bm{\sigma}_{(k)} \leq   \bm{\tau}_{(k)}$. We determine the number of assets bank $i$ sells in round $k$. 
First consider the case that $\ell_{i,k} /c_i\leq \delta$, that is, the loss of bank $i$ at the beginning of round $k$ is bounded by $\delta$ and thus the fraction of shares sold so far is less than $\varepsilon$. As a result  $\bm{x}_{i,k}\geq (1-\varepsilon)\bm{x}_{i}$ and thus
\begin{eqnarray}
&&\bm{x}_i (1-\varepsilon ) \rho_{\varepsilon} \left(\frac{\ell_i + (1-\varepsilon) \bm{x}_i\cdot h(\bm{\sigma}_{(k)} n^{-1})}{c_i} \right)\leq \bm{x}_i \rho \left(\frac{\ell_i + (1-\varepsilon) \bm{x}_i\cdot h(\bm{\sigma}_{(k)} n^{-1})}{c_i} \right)\\
&\leq &\bm{x}_i \rho \left(\frac{\ell_i + \bm{x}_{i,k} \cdot h(\bm{\sigma}_{(k)} n^{-1})}{c_i} \right) \leq \bm{x}_i \rho \left(\frac{\ell_i + \bm{x}_{i,k} \cdot h(\bm{\tau}_{(k)} n^{-1})}{c_i} \right) \leq \bm{x}_i \rho \left(\frac{\ell_{i,k}}{c_i} \right).
\end{eqnarray}
If, on the contrary, $\ell_{i,k} /c_i> \delta$, then 
\[ \bm{x}_i (1-\varepsilon ) \rho_{\varepsilon} \left(\frac{\ell_i + (1-\varepsilon) \bm{x}_i\cdot h(\bm{\sigma}_{(k)} n^{-1})}{c_i} \right) \leq \bm{x}_i (1-\varepsilon ) \varepsilon \leq \bm{x}_i \rho \left(\frac{\ell_{i,k}}{c_i} \right) \] 
and again bank $i$ has sold more shares  in step $k$ of the {fire sales process} in $(\bm{X}_n,C_n,L_n)$ with sales function $\rho$ than in the auxiliary process in   $((1-\varepsilon)\bm{X}_n,C_n,L_n)$ with sales function $\rho_{\varepsilon}$. We infer
\[\bm{\sigma}_{(k+1)} =  \sum_{i \in [n]} \bm{x}_i (1-\varepsilon ) \rho_{\varepsilon} \left(\frac{\ell_i + (1-\varepsilon) \bm{x}_i\cdot h(\bm{\sigma}_{(k)} n^{-1})}{c_i} \right) \leq \sum_{i \in [n]}  \bm{x}_i \rho \left(\frac{\ell_{i,k}}{c_i} \right) =  \bm{\tau}_{(k+1)}  \] 
and the induction step is completed.
\hfill\hfill
\endproof

\subsection{(Non-)Resilience Criteria for the Fire Sales Process}\label{ssec:resilience:real}

{As considered before for the auxiliary process, we now investigate resilience properties of the fire sales process}.  We denote by $\bm{\chi}^{*,\varepsilon}$ the smallest joint root of the functions $f_{\varepsilon}^m, m \in [M]$ as defined in  Section~\ref{sec:real:process} with $L=0$. Due to the conservative character of the auxiliary process, it is immediate that resilience of the auxiliary process for system $(\bm{X},C)$ implies resilience of the fire sales process for the system. Similarly, non-resilience under the fire sales  process for $(\bm{X},C)$ implies non-resilience also under the auxiliary process for $(\bm{X},C)$.

\begin{corollary}
\label{cor:nonresfiresales}
If $\bm{\chi}^*=\bm{0}$, then the $(\bm{X},C)$-system is resilient under the fire sales process. If there exists $\varepsilon > 0$ such that $\bm{\chi}^{*,\varepsilon} >\bm{0}$ and such that no $z\in\mathbb{R}^M \setminus\{0, \bm{\chi}^{*,\varepsilon} \}$ with $0 \le z \le \bm{\chi}^{*,\varepsilon}$ (coordinate-wise) is a joint root of the functions $(f^m_{\varepsilon})_{m\in[M]}$, then the $(\bm{X},C)$-system is non-resilient under the fire sales process.
\end{corollary}
\proof
The first claim about resilience is a straightforward consequence of Theorem~\ref{thm:fire:sale:upper:bound:real}. For the second part note that if there exist $\varepsilon > 0$ such that $\bm{\chi}^{*,\varepsilon} >\bm{0}$, then the system $((1-\varepsilon)\bm{X},C)$ with sales function $\rho_{\varepsilon}$ is non-resilient for the auxiliary process by Theorem~\ref{thm:non-resilience} (and its extension in \ref{ec:non:resilience}) and for every shock $L$, the number of sold assets over $n$, $\bm{\chi}_n$ it holds that $\liminf_{n\rightarrow \infty}  \bm{\chi}^m_n \geq \bm{\chi}^{*,\varepsilon} >\bm{0}$. By Theorem~\ref{thm:fire:sale:lower:bound:real} it follows that under the real process in the system $(\bm{X},C)$ with sales function $\rho$ the number $n\psi_n^m$ of assets $m\in [M]$ sold fulfills $\liminf_{n\rightarrow \infty} \psi_n^m \geq (\bm{\chi}^{*,\varepsilon} )^m>\bm{0}$ for every shock $L$, which implies non-resilience.\hfill\hfill
\endproof
We will now present several more explicit (non-)resilience criteria for the fire sales process; in particular, we will show that in many natural situations the fire sales and the auxiliary processes behave \emph{the same} with respect to their resilience properties. A starting point is the following observation, which states that in many cases an analysis of the derivatives at $0$ of the functionals $f^m, m\in [M]$ describing the auxiliary system can be used to study (non-)resilience.
\begin{theorem}\label{thm:non:resilience:real}
Assume that $\rho$ is differentiable in $0$, and that there exists $m\in[M]$ such that all  partial right-derivatives $\frac{\partial}{\partial \chi^m}h^\ell (\bm{0}),\ell \in [M]$ exist (possibly with value $\infty$) and
\begin{equation}\label{assump:aux:process:deriv}
\E\left[X^m \rho ' (0) \left( \sum_{\ell\in [M]} X^\ell \frac{\partial}{\partial \chi^m}h^\ell (\bm{0}) \right)/C \right] >1.
\end{equation}
Then the $(\bm{X},C)$-system with sales function $\rho$ is non-resilient under the fire sales process.
\end{theorem}
\proof
Let $\bm{\chi}^m_0$ be the $M$ dimensional vector which has value $\chi$ in the $m$-th entry and zero in all other entries. It is sufficient to show that there exists $\varepsilon > 0$ such that $\liminf_{\chi \rightarrow 0} f_{\varepsilon}^m (\bm{\chi}^m_0) /\chi >0$ since $f_{\varepsilon}^m$ is monotone increasing in the arguments $[M] \setminus \{ m\}$. Using the lemma of Fatou and that $\rho(u)=\rho_{\varepsilon} (u)$ for small $u$, we obtain
\begin{eqnarray}
 \liminf_{\chi \rightarrow 0} f_{\varepsilon}^m (\bm{\chi}^m_0) /\chi & =&\liminf_{\chi \rightarrow 0}  \E\left[(1-\varepsilon)X^m\rho_{\varepsilon}\left(\frac{(1-\varepsilon)\bm{X}\cdot h(\bm{\chi}^m_0)}{C}\right)/\chi \right] - 1 \nonumber\\
 & \geq & \E\left[\liminf_{\chi \rightarrow 0}(1-\varepsilon)X^m\rho_{\varepsilon}\left(\frac{(1-\varepsilon)\bm{X}\cdot h(\bm{\chi}^m_0)}{C}\right)/\chi \right] - 1 \nonumber \\
  & = & \E\left[(1-\varepsilon)X^m \liminf_{\chi \rightarrow 0}\rho \left(\frac{(1-\varepsilon)\bm{X}\cdot h(\bm{\chi}^m_0)}{C}\right)/\chi \right] - 1\nonumber \\
 &=& (1-\varepsilon)^2 \E\left[X^m \rho ' (0) \left( \sum_{\ell\in [M]} X^\ell \frac{\partial}{\partial \chi^m}h^\ell (\bm{0}) \right)/C \right] -1 \nonumber,
\end{eqnarray}
and by (\ref{assump:aux:process:deriv}) this expression is larger than zero for small $\varepsilon> 0$. Thus non-resilience follows for the fire sales process.\hfill\hfill
\endproof
The previous theorem allows us to show that in many relevant situations actually the auxiliary and the real process are equivalent in terms of resilience. Let us  consider the one dimensional system (i.e., $m=1$) in the important case of linear price impact (i.e., $h(\chi) = \chi$). The following corollary shows that  in most cases the resilience properties of the processes coincide.
\begin{corollary}\label{cor:lin:price}
Let $h(\chi) = \chi$ and let $\rho$ be differentiable in $0$. Then the  following holds.
\begin{enumerate}
\item If $\E [X^2 /C] = \infty$ and $\rho'(0)>0$: the system is non-resilient under both (original and auxiliary) fire sales processes. 
\item If $\E [X^2 /C] < \infty$: the systems are resilient under both processes if $\E [(X^2 /C) \rho'(0)]< 1$ and non-resilient if $\E [(X^2 /C) \rho'(0)]> 1$.
%\item If $\E [X^2 /C] < \infty$ and $\rho'(0)=0$: the systems are resilient under both processes.
%\item For the case $\E [X^2/C]=\infty $ und $\rho ' (0) =0$ the resilience question cannot be answered without specifying $\rho$ more precisely. Above in (\ref{assump:aux:process:deriv}) we always get 0 as $\E [(X^2/C) 0]=0$ and based on Theorem~\ref{thm:non:resilience:real}  we cannot proof non-resilience of the real system in this case. However, we provide more conditions under specific assumptions on $\rho$ which allow us to even make statements in this case. know that the auxiliary system is non-resilient for $C=X^{\gamma}$ with $\gamma < 1- (\beta - 2)$. In this case then $\E [X^2/C] > \E[X^{3-(\beta -2)} ] = \E[X^{5-\beta } ]=\infty $  for $\beta \in (2,3)$ and $\rho ' (0) =0$.%\nils{What are we actually doing in this case? Can we derive our capital requirements directly for the real system?}
\end{enumerate}
\end{corollary}
\proof
For 1. non-resilience follows from Theorem~\ref{thm:non:resilience:real} as the expectation in (\ref{assump:aux:process:deriv}) is infinite. For 2., if $\E [(X^2 /C) \rho'(0)]> 1$ non-resilience follows again directly from Theorem~\ref{thm:non:resilience:real}. Let $f=f^1$. To show resilience for $\E [(X^2 /C) \rho'(0)]< 1$, note that by the Dominated Convergence Theorem
$$f' (0) = \E\left[X \rho ' (0) 
\left( X \frac{\partial}{\partial \chi}h  (0) \right)/C \right] -1 =\E[X^{2} \rho ' (0)/C ]-1<0$$
and thus $S_0=\{ 0\}$ and $\bm{\chi}^*=0$, which implies with  Corollary~\ref{cor:resilience} that the system is resilient under the auxiliary process, and so also under the fire sales process.
\hfill\hfill
\endproof
The conditions provided in Corollary~\ref{cor:lin:price} highlight two characteristics that fuel fire sales: 
\begin{enumerate}
\item[a)] The capital is too small in comparison to the asset holdings.
\item[b)] Pronounced selling as a response to a drop in asset prices. 
\end{enumerate}
For a system with very low capitalization ($\E [X^2 /C] = \infty$), Corollary~\ref{cor:lin:price} 1. shows that the system is non-resilient if market participants immediately react to price drops by selling assets ($\rho'(0)>0$). For better capitalized systems ($\E [X^2 /C] <\infty$), the picture is more fine grained and Corollary~\ref{cor:lin:price} 2. shows that the system is only non-resilient if the reaction to a price drop is prompt and strong ($\E [(X^2 /C) \rho'(0)]> 1$). It also follows that if $\E [X^2 /C] < \infty$ and $\rho'(0)=0$ the system is resilient under both processes. However, it misses the case $\E [X^2/C]=\infty $ und $\rho ' (0) =0$. This case represents a scenario of a low capitalized, heterogeneous system with a rather low market reaction to stress, and is thus of particular interest. In this case the expectation in (\ref{assump:aux:process:deriv}) equals 0 and based on Theorem~\ref{thm:non:resilience:real} we cannot prove non-resilience of the real system. Moreover, we cannot use dominated convergence to derive the derivative of $f=f^1$ and conclude (non-)resilience.

In order to address this case we must make more refined assumptions. In particular, we assume that $h(\chi) = \chi^{\nu}$ for some $\nu>0$ and $\rho (u) =u^q \wedge 1$ for $q \in (0,\infty ]$ where $q = \infty $ describes sales at default (i.e. $\rho (u) =\1_{\{u\geq 1\}}$). The next result, whose proof is presented in Section~\ref{ssec:proofs:resilience}, shows that (non-)resilience is a property that depends on the interplay of $\rho$ and $h$.
\begin{theorem}\label{res:power:h_and_rho}
Let $m=1$ and $\P (X>0,C> 0)=1$. Then the following holds.
\begin{enumerate}
\item $1-\nu q > 0$: the system is  non-resilient.
\item $1-\nu q = 0$: the system is resilient if $\E [ X (X/C )^{1/\nu} ] < 1$ and non-resilient if $\E [ X (X/C )^{1/\nu} ] > 1$.
\item $1-\nu q < 0$ or $q=\infty$: Let $\alpha^* =\sup \{ \alpha \in \mathbb{R}: \E [X^{1+\alpha}/C^\alpha ] < \infty  \}$. If $\alpha^* > 1/\nu$, then the system is resilient. If $\alpha^* < 1/\nu$ and $f' (0)$ exists, then the system is non-resilient.
\end{enumerate}
\end{theorem}
The last theorem adds another dimension to the picture and that is the interplay of $\rho$ and $h$: Strong sales activity as a reaction to price drops needs to be compensated by low price impact to make the system resilient. If strong price impact comes together with strong sales as a reaction to price decline ($1-\nu q > 0$), then the system is non-resilient, independent of the capital. For all other cases ($1-\nu q \leq 0$) the relation of asset and capital is crucial. Informally, the stronger the price impact, the more capital is required in relation to the asset holdings.

%\nils{Mir ist noch nicht ganz klar was die beste multidim erweiterung ist. Waere dafuer das dann nur zu diskutieren. Kann ja auch aufgenommen werden in eine section for "Further investigations" Zunaechst dachte ich das setting $h^m(\bm{\chi}) = (\chi^m)^{\nu_m}$ with $\nu=(\nu_1,...,\nu_m)$. Einfach ist dann zu zeigen, dass wenn $1-\min_{m\in [M]} \{ \nu_m \} q>0 $ dann das system non-resilient ist. Probleme macht der Fall $1-\min_{m\in [M]} \{ \nu_m \} q\leq 0$ und die Sache wuerde auch nicht viel leichter wenn man sich nur $1-\min_{m\in [M]} \{ \nu_m \} q < 0$ anschaut. Ohne Probleme kann es erweitert werden wenn man $\nu=\nu_1=\dots \nu_M$ annimmt aehnlich wie wir mal bei power law gemacht hatten. Dann ist die resilience Bedingung im Fall  $1-\nu q = 0$ gegeben durch 
%\[ \max_{m\in [M]} \E [  X^m \left( \sum_{l\in [M]} X^l /C\right)^{1/\nu}  ]  <1\]
%und non-resilient fuer $>1$. Im Fall $1-\nu q < 0$ haetten wir resilience auf jeden Fall fuer 
%\begin{equation}\label{tmp1}
%\alpha^{*,m} =\sup \{ \alpha \in \mathbb{R}: \E \left[X^m \left(\sum_{l\in [M]} X^l/C\right)^\alpha \right] < \infty  \} > 1+1/\nu 
%\end{equation}
%  fuer all $m\in [M]$ (Richtungsableitung negative in Richtung der Diagonale $(1,..,1)$) und non-resilience fuer 
%\[ \alpha^{*,m} =\sup \{ \alpha \in \mathbb{R}: \E \left[X^m \left(X^m /C\right)^\alpha \right] < \infty  \} < 1+1/\nu \] 
%fuer ein $m\in [M]$, wobei ich denk man kann das noch verschaerfen zu (\ref{tmp1}) mit $< 1+1/\nu$.}
The previous results are restricted to systems with one asset only. They can be generalized to systems with multiple assets in various forms. For example, we can assume that for each asset $m\in[M]$ with $M \ge 2$ we have a specific price impact function $h^m(\bm{\chi}) = (\chi^m)^{\nu_m}$. Then an argument as before shows that if $1 - q\min_{m\in [M]}\nu_m > 0$, then the system is non-resilient. Moreover, if
\[
	1 - q\min_{m\in [M]}\nu_m = 0
	\quad \text{ and }	\quad
	\max_{m\in [M]} \E\left [  X^m \Big( \sum_{\ell\in [M]} X^\ell /C\Big)^{1/\nu}\right] > 1,
\]
the system is non-resilient (corresponding to a generalization of 2.~in Theorem~\ref{res:power:h_and_rho}). Resilience is more difficult to characterize, as it may depend on the interplay among different assets. In these cases, Theorem~\ref{thm:non:resilience:real} and Corollary~\ref{cor:nonresfiresales} may be used to study resilience properties for the system at hand. We leave it as a research problem to extract further and more explicit (non-)resilience conditions from our main results.

\section{Applications: Minimal Capital Requirements \& The Effect of the Portfolio Structure}
\label{sec:applications}
 
In this section we apply the theory developed in previous sections to investigate which structures or properties of systems hinder or promote the emergence and spread of fire sales. First, in Section~\ref{ssec:capreq} we address the important question of formulating adequate capital requirements: how much capital is necessary and sufficient for each bank, so that the system is resilient? We give an answer to this question in the specific setting where the asset holdings follow a power-law distribution, which is the most typical case to consider. Then, 
in Section~\ref{ssec:diversification:similarity} we consider systems parametrized by two orthogonal characteristic quantities: \emph{portfolio diversification} and \emph{portfolio similarity}. We first study their effect analytically based on our previous results, and we also verify our findings with simulations for finite systems of reasonable size. Our examples demonstrate the effects of portfolio diversification and similarity on resilience in various sensible settings.

\subsection{Minimal Capital Requirements}
\label{ssec:capreq}

While the results in the previous section might allow regulators to test the resilience of financial systems with respect to fire sales, a natural question from a regulatory perspective is to determine capital requirements  that actually ensure resilience. This question is addressed now. In this section we have mainly leveraged institutions (for example banks) in mind.

Historically, the first approach to systemic risk was to apply a monetary risk measure to some aggregated, system-wide risk factor, see \cite{ChenIyengarMoallemi,KromerOverbeckZilch,HoffmannMeyer-BrandisSvindland} for an axiomatic characterization of this family of risk  measures. %Capital requirements for individual institutions are  then specified by some rule to allocate the total systemic risk capital.
Another approach is to determine the total  risk by specifying capital requirements on an institutional level before aggregating to a system-wide risk factor, see \cite{RePEc:hal:wpaper:hal-01764398,doi:10.1111/mafi.12170,doi:10.1137/16M1066087}. An important question is then whether from an institutional viewpoint the requirements correspond to fair shares of the systems risk, see e.g.~\cite{doi:10.1111/mafi.12170}. In particular, a major problem in the implementation of the methodologies proposed in the aforementioned papers is that a given individual capital requirement in general depends on the configuration of the whole system. As a consequence, one institution's capital requirement may be manipulated by other institutions' behaviour, or the entrance of a new institution into the system would potentially alter the capital requirements of all other institutions. An important contribution of the  methodology proposed here is that, besides certain global parameters that need to be determined by a regulating institution, the implied capital requirement for a given institution $i\in[n]$ depends on its local parameters, i.e., its capital and asset holdings, only.

Abstractly, a capital requirement is a function $b: \mathbb{R}_+^M \rightarrow \mathbb{R}_+$ that assigns to any bank $i\in [n]$ with asset holdings $x_i^1,\dots , x_i^M$ a (minimal) required capital $c_i= b(x^1_i,\dots, x^M_i)$. For our ensemble this implies that the random variable $C$ is of the form $C=b(X^1, \dots ,X^M)$ and we would like to give conditions on $b$ that ensure resilience. We provide only stylized examples here for which analytic derivations are possible to illustrate the power of our asymptotic approach. For  general system configurations one can always resort to numerical calculations. 

Let us first restrict to one asset $X=X^1$ and assume that the distribution $F_X$ of asset holdings has power law tail in the sense that there exist constants $B_1,B_2\in(0,\infty)$ such that for $x$ large enough
\begin{equation}
\label{eq:Xpowerlaw}
B_1 x^{1-\beta}\leq 1-F_X(x) \leq B_2 x^{1-\beta},
\end{equation}
for some $\beta > 2$. There is empirical evidence for power laws in investment volumes, see e.g.~\cite{Garlaschelli2005}. Moreover, we assume the setting of Section~\ref{ssec:resilience:real}, that is, we assume that $h(\chi) = \chi^{\nu}$ for some $\nu$ and $\rho (u) =u^q \wedge 1$ for $q \in (0,\infty]$.

One natural  choice for the capitals $c_i$ as a function of the asset holdings is of power form. We immediately get the following corollary, whose proof is a direct consequence of Theorem~\ref{res:power:h_and_rho}, that determines sharply the magnitude of the minimal capital requirements to ensure resilience.
\begin{corollary}
\label{cor:gamma*}
Let $1-\nu q < 0$ or $q=\infty$ and $c_i = \alpha x_i^\gamma$ for $\alpha >0$ and $\gamma\in\R_{+}$. Then,
\begin{enumerate}
\item if $\gamma>1-\nu(\beta-2)$, then the system is resilient,
%\item if $\gamma=1-\nu(\beta-2)$ and $\alpha>\mu_2 \left(B_2\frac{\beta-1}{\beta-2}\right)^\nu$, then the system is resilient.
%\item if $\gamma=1-\nu(\beta-2)$ and $\alpha<\mu_1\left(B_1\frac{\beta-1}{\beta-2}\right)^\nu$, then the system is non-resilient.
\item if $\gamma<1-\nu(\beta-2)$, then the system is non-resilient.
\end{enumerate}
\end{corollary}
\proof
We verify the conditions of Theorem~\ref{res:power:h_and_rho}, i.e., we show that $\alpha^*>1/\nu$ (resp.~$<1/\nu$) where $\alpha^*= \sup \{ \alpha \in \mathbb{R}: \E [X^{1+\alpha}/C^\alpha ] < \infty  \}$. By the choice of capital $C=X^\gamma$ we get that $\E [X^{1+\alpha}/C^\alpha ] = \E [X^{1+\alpha (1- \gamma ) } ] $
which is $<\infty $ (resp.~$\infty$) for $\alpha <(\beta -2)/(1- \gamma )$ (resp.~$\alpha >(\beta -2)/(1- \gamma ))$. Under Assumption 1. then $\alpha^*>1/\nu$ while under Assumption 2. $\alpha^*<1/\nu$.
\hfill
\endproof

Additionally we can derive sufficient capital requirements in the multidimensional case. In this setting we assume that there exist constants $B_{1,m},B_{2,m}\in(0,\infty)$ and $\beta_m$ for $m\in [M]$ such that for $x$ large enough
\begin{equation}
B_{1,m} x^{1-\beta_m}\leq 1-F_{X^m}(x) \leq B_{2,m} x^{1-\beta_m}.
\end{equation}
\begin{corollary}
Let $\rho (u) =u^q \wedge 1$ and $h^m(\bm{\chi})=(\chi^m)^\nu$ for $m\in [M]$. Let $1-\nu q < 0$ or $q=\infty$ and $c_i = (\sum_{m\in [M]} \alpha_m x^m_i)^\gamma$ for $\alpha_m >0, m\in [M]$ and $\gamma\in\R_{+}$. Let $\beta_{\min}:= \min \{ \beta_1,\dots , \beta_m \}$. Then,
\begin{enumerate}
\item if $\gamma>1-\nu(\beta_{\min}-2)$, the system is resilient.
%\item if $\gamma=1-\nu(\beta-2)$ and $\alpha>\mu_2 \left(B_2\frac{\beta-1}{\beta-2}\right)^\nu$, then the system is resilient.
%\item if $\gamma=1-\nu(\beta-2)$ and $\alpha<\mu_1\left(B_1\frac{\beta-1}{\beta-2}\right)^\nu$, then the system is non-resilient.
\item if $\gamma<1-\nu(\beta_{\min}-2)$, the system is non-resilient.
\end{enumerate}
\end{corollary}
\proof
{The proof is immediate from Lemma~\ref{sum:of:powers} below, which allows us to reduce to the one dimensional situation.}
\hfill\hfill
\endproof
The following simple result might be known, but we could not find it in the literature. We provide a proof for completeness.
\begin{lemma}\label{sum:of:powers}
Let $X^m, m\in [M]$ be random variables with power law tails as above and let $\beta_{\min}:= \min \{ \beta_1,\dots , \beta_m \}$. Then there exists $B_1,B_2 > 0$ such that 
\begin{equation}
B_{1} x^{1-\beta_{\min}}\leq 1-F_{X^1 + \dots X^M}(x) \leq B_{2} x^{1-\beta_{\min}}.
\end{equation}
\end{lemma}
\proof
{We assume that $M=2$; the result for general $M$ follows by induction. Let us further assume without loss of generality that $F_{X^1} (x)=F_{X^2} (x)=0$ for $x<1$ and $\beta_{\min} =\beta_1\leq \beta_2$. 
Clearly, 
\[1-F_{X^1 + X^2}(x) =\P (X_1+X_2>x)\geq \P (X_1 >x) =1-F_{X^1}(x)\geq B_{1,1} x^{1-\beta_{1}} ,\]
which provides the lower bound. 
By the union bound
\begin{eqnarray}
1-F_{X^1 + X^2}(x) &=&\P (X_1+X_2>x)  \leq \P (X_1>x/2) + \P (X_2>x/2) \nonumber \\
& \leq & 2-F_{X^1}(x/2) -F_{X^2}(x/2) \nonumber \\ 
&\leq & B_{2,1} (x/2)^{1-\beta_{1}} + B_{2,2} (x/2)^{1-\beta_{2}} \nonumber \\ 
&\leq & (B_{2,1} 2^{\beta_1-1}+ B_{2,2} 2^{\beta_{2}-1}) x^{1-\beta_{1}}\nonumber
\end{eqnarray}
and we obtain the upper bound.}
%Since $X = X^1$ and $Y = X^2$ are independent we infer from Fubini's theorem\begin{eqnarray}
%\mathbb{P} (X+Y > t )&=& \int_{\mathbb{R}_+^2} \1 \{ x+ y \geq t \} d F_{X,Y} (x,y) \nonumber \\
%%&=& \int_1^{\infty} \left(  \int_1^{\infty} \1 \{ x+y \geq t \}  dF_{X} (x) \right) dF_{Y}(y) \nonumber \\ 
%&=& \int_1^{\infty }  \left( \int_{(t-y)\wedge 1}^{\infty }  dF_X (x) \right) dF_{Y} (y) \nonumber \\ 
%&=& \int_1^{\infty} \mathbb{P} (X\geq (t-y)\wedge 1) dF_Y (y) \nonumber \\
%&= &\int_{t-1}^{\infty}  d F_{Y} (y) + \int_1^{t-1} \mathbb{P} (X\ge t-y) dF_{Y}(y)\nonumber \\ 
%&=& \mathbb{P} (Y\geq t-1) + \int_1^{t/2} (t-y)^{-\beta_1+1} d  F_{Y} (y) + \int_{t/2}^{t-1} (t-y)^{-\beta_1+1} dF_Y (y) \nonumber \\ 
%&=&  \Theta (1) t^{-\beta_2+1 } + \Theta (1) t^{-\beta_1+1} \mathbb{P} (Y \in (1,t/2]) + \Theta (1) t^{-\beta_2+1 } =  \Theta (1) t^{-\beta_1+1 } \nonumber,
%\end{eqnarray}
%where in the last step we used that $\lim_{t\rightarrow \infty } \mathbb{P} (Y \in (1,t/2]) =1$. \kosta{waive independence??}
\hfill\hfill
\endproof

\subsection{The Effect of Portfolio Diversification and Similarity}\label{ssec:diversification:similarity}

For simplicity, throughout this section we assume that the limiting total asset holdings $X^\text{tot}=X^1+\ldots+X^M$ are Pareto distributed with density $d_{X^\text{tot}}(x)=(\beta-1)x^{-\beta}\1\{x\geq1\}$ for some exponent $\beta>2$. One can generalize the example also to more general distributions. Further, we make the assumptions that $\rho(u)=\1\{u\geq1\}$  and $h^m(\bm{\chi})=1-e^{-\chi^m}$ to simplify calculations, but also for other sensible choices our observations below are applicable.

In a first setting we consider a system of institutions whose investment in each asset $m\in[M]$ makes up a fraction $\lambda^m >0$ of their total asset holdings, where $\sum_{m \in [M]}\lambda^m=1$. 
\begin{example}\label{ex:diversification}
For a system as described above, the functions $f^m(\bm{\chi})$ are given by
\[ f^m(\bm{\chi}) = \lambda^m\E\left[X^\text{tot}\1\left\{X^\text{tot}\sum_{\ell=1}^M\lambda^\ell\left(1-e^{-\chi^\ell}\right)\geq C\right\}\right] - \chi^m,\quad m\in[M].
\]
Let us write $t = \sum_{1\le \ell \le M}\lambda^\ell(1-e^{-\chi^\ell})$ for short. Now assume similar to Corollary \ref{cor:gamma*} that $C=\alpha(X^\text{tot})^\gamma$ for some constants $\alpha,\gamma\in\R_{+}$. Then
\begin{align*}
f^m(\bm{\chi})
& = \lambda^m\E\left[X^\text{tot}\1\left\{X^\text{tot}\geq\left({\alpha}/t\right)^\frac{1}{1-\gamma}\right\}\right] - \chi^m\\
& = \lambda^m \int_{\max\left\{1 , \left({\alpha}/t\right)^{{1}/({1-\gamma})}\right\}}^\infty (\beta-1)x^{1-\beta}\,\dd x - \chi^m
=  \lambda^m \frac{\beta-1}{\beta-2} \min\left\{1, (t\alpha^{-1})^\frac{\beta-2}{1-\gamma}\right\} - \chi^m.
\end{align*}
Motivated by the symmetry of the functions, we consider $f^m(\bm{\chi})$ along direction $\bm{v}\in\R_+^M$, with $v^m=(\lambda^m)^{-1}$. Then 
\[
\frac{f^m(\chi\bm{v})}{\lambda^m} = \frac{\beta-1}{\beta-2}\left(\alpha^{-1}\sum_{\ell=1}^M \lambda^\ell \left(1-e^{-\chi/\lambda^\ell}\right)\right)^\frac{\beta-2}{1-\gamma} - \frac{\chi}{(\lambda^m)^2}.
\]
Let $\gamma_c:=3-\beta$ and $\alpha_c:=\sum_{m\in[M]}(\lambda^m)^2(\beta-1)/(\beta-2)$. We infer that if $\chi\in\R_{+}$ is small enough and $\gamma>\gamma_c$ or $\gamma=\gamma_c$ and $\alpha>\alpha_c$, then $\frac{\dd}{\dd \chi}f^m(\chi\bm{v})<0$ for all $m\in[M]$. That is, $\bm{\chi}^*=\bm{0}$ and the auxiliary system is resilient by Corollary~\ref{cor:resilience}. On the other hand, if either $\gamma<\gamma_c$ or $\gamma=\gamma_c$ and $\alpha<\alpha_c$, then $\frac{\dd}{\dd \chi}f^m(\chi\bm{v})>0$ for all $m\in[M]$ and the system is non-resilient by Theorem \ref{thm:non-resilience}. Since $\gamma_c$ does not depend on the choice of $\{\lambda^m\}_{m\in[M]}$, it makes sense to consider $\alpha_c$ as a measure for stability of the system (the smaller $\alpha_c$, the more stable the system). Clearly, $\alpha_c$ becomes minimized for $\lambda^m=M^{-1}$ for all $m\in[M]$ and hence a perfectly diversified system requires least capital and can be seen as the most stable. This somewhat surprising observation is discussed after Example~\ref{ex:diversification:similarity} in more detail. 
\hfill\hfill
\end{example} 

\noindent Next, we consider a financial system that comprises of $S\in\N$ subsystems of equal size $n/S$. For each subsystem $s\in[S]$ there is  a set of $D\in\N$ specialized assets that can only be invested in by institutions from  $s$. In addition to these $S\cdot D$ specialized assets, there is a set of $J\in\N$ joint assets that can be invested in by any institution. Thus, each institution can choose from $\Delta:=D+J$ different assets to invest in. We call $\Delta$ the \emph{diversification} of the system. Further, let $\Sigma:=J/\Delta$ be the \emph{(portfolio) similarity} in the system. Then, as in Example \ref{ex:diversification}, one possible route to take is to determine the optimal investments for each institution (that is shifted towards investing in the specialized assets to avoid overlap with other subsystems). Instead we assume in the following example that each institution still perfectly diversifies its investment over the $\Delta = D+J$ assets available to it. %This is reasonable if the single institutions do not have a perfect overview of the whole financial system. %The effect of diversification $\Delta$ and similarity $\Sigma$ is similar for the two different allocations.
\begin{example}\label{ex:diversification:similarity}
Consider a system as described above consisting of $S$ subsystems and allowing each institution to invest in $D$ specialized assets and in $J$ joint assets in equal shares. Then the system is described by the following functions:
\begin{align*}
f^j(\bm{\chi}) &:= S^{-1}\sum_{s=1}^S\E\left[\frac{X^\text{tot}}{D+J}\1\left\{\frac{X^\text{tot}}{D+J}\left(\sum_{k=1}^J\left(1-e^{-\chi^k}\right)+\sum_{d=1}^D\left(1-e^{-\chi^{s,d}}\right)\right)\geq C\right\}\right] - \chi^j,\\
f^{s,d}(\bm{\chi}) &:= S^{-1}\E\left[\frac{X^\text{tot}}{D+J}\1\left\{\frac{X^\text{tot}}{D+J}\left(\sum_{j=1}^J\left(1-e^{-\chi^j}\right)+\sum_{e=1}^D\left(1-e^{-\chi^{s,e}}\right)\right)\geq C\right\}\right] - \chi^{s,d},
\end{align*}
%Now use symmetry in $s$ to reduce above $J+DS$ functions to the following $J+D$ functions:
%\begin{align*}
%f^j(\bm{\chi}) &= \E\left[\frac{X^\text{tot}}{D+J}\1\left\{\frac{X^\text{tot}}{D+J}\left(\sum_{k=1}^J\chi^k+\sum_{d=1}^D\chi^{1,d}\right)\geq C\right\}\right] - \chi^j,\\
%f^{1,d}(\bm{\chi}) &= S^{-1}\E\left[\frac{X^\text{tot}}{D+J}\1\left\{\frac{X^\text{tot}}{D+J}\left(\sum_{j=1}^J\chi^j+\sum_{e=1}^D\chi^{1,e}\right)\geq C\right\}\right] - \chi^{1,d}
%\end{align*}
%Next, we can consider $(D+J)^{-1}\left(\sum_{j=1}^Jf^j(\bm{\chi})+\sum_{d=1}^Df^{1,d}(\bm{\chi})\right)$ which reduces the system to the following single function:
%\[ \tilde{f}(\chi) := \frac{J+\frac{D}{S}}{(D+J)^2}\E\left[X^\text{tot}\1\{X^\text{tot}\chi\geq C\}\right] - \chi \]
where $j\in[J]$, $s\in[S]$, $d\in[D]$ and $\bm{\chi}=(\chi^1,\ldots,\chi^J,\chi^{1,1},\ldots,\chi^{S,D})\in\R_{+}^{J+SD}$ with small misuse of notation. Similar as in Example \ref{ex:diversification} we derive that
\[ \gamma_c=3-\beta\quad \text{and}\quad \alpha_c=\frac{J+\frac{D}{S}}{(D+J)^2}\frac{\beta-1}{\beta-2}=\frac{1+(S-1)\Sigma}{\Delta S}\frac{\beta-1}{\beta-2}. \]
From the formula it is obvious that $\alpha_c$ decreases (i.\,e.~capital required to ensure resilience) as $\Delta$ increases or $\Sigma$ decreases. 
\hfill\hfill
\end{example}
\begin{figure}[t]
    \hfill\subfigure[]{\includegraphics[width=0.45\textwidth]{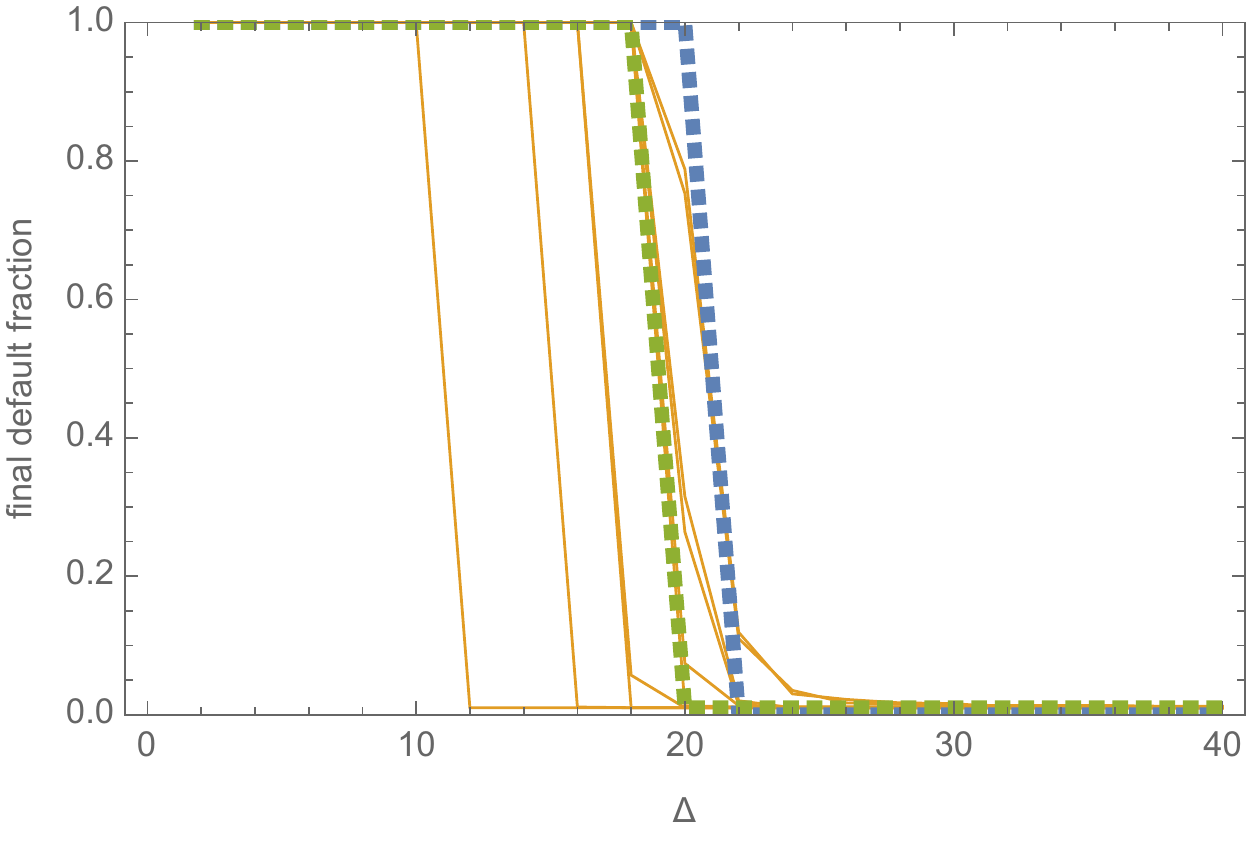}\label{fig:varying:diversification}}
    \hfill\subfigure[]{\includegraphics[width=0.45\textwidth]{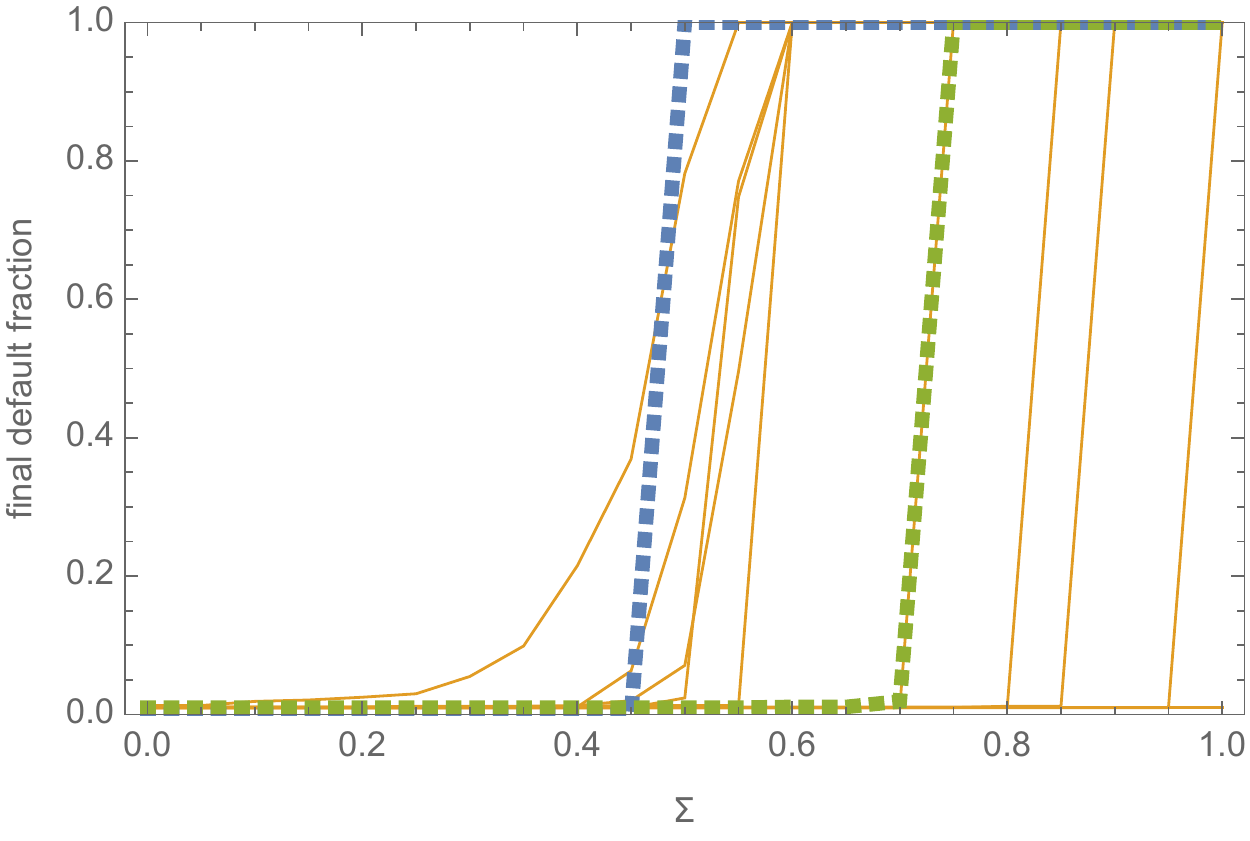}\label{fig:varying:similarity}}\hfill
\caption{Figure corresponding to the simulation study. (a)~The effect of varying portfolio diversification $\bm{\Delta}$ as $\bm{\Sigma=0.5}$ is fixed. (b)~The effect of varying portfolio similarity as $\bm{\Delta=20}$ is fixed. In blue: the theoretical final default fraction. In orange: $\bm{10}$ exemplary simulations. In green: the median over $\bm{10^3}$ simulations.}
\end{figure}
For this particular setup it thus turns out that diversification reduces the capital necessary to ensure resilience whereas stronger similarity between the institutions' portfolios increases it. The fact that diversification is favorable might seem at odds with other studies in the literature (see for example \cite{Ibragimov2011,Beale2011,Wagner2010,Capponi}), which find that diversification, when it comes with higher portfolio similarity, can increase externalities and is less favorable from a systemic risk perspective. In the second example this discrepancy is not surprising, as increased diversity arises due to an increased number of assets. In the first example it is due to the fact that capital requirements ensure resilience and that amplification effects are small. The sales pressure arising from the shock to some banks is then distributed among more assets, each of them being less impacted. Spillover effects to other institutions are contained due to the nature of the capital requirements based on $X^\text{tot}$. In summary, in a well-capitalized system, diversification might actually be good to reduce the impact of small shocks. However, we want to stress that the results obtained in these two examples should not be considered as universal statements. They only show that our mathematical framework allows one to investigate these questions in quite some detail for flexible specifications of the system.

\noindent
{\bf Simulation Study} Note that all previous conclusions build on the (asymptotic) theory developed in the previous sections. To verify and back up the result for finite systems, however, we also provide a simulation based verification for a series of moderate size  ($n=10^4$) systems. In the setting of the last example our choices of the parameters are as follows. We set $\beta=3$ and $S=2$ and $D=J=10$. Then we obtain $\Delta=20$, $\Sigma=0.5$, $\gamma_c=0$ and $\alpha_c=0.075$. We assign to each institution the capital $c_i=\alpha_c$. Further, we draw the total asset holdings $x_i^\text{tot}$ for each institution $i\in[n]$ as random numbers according to the above described Pareto distribution. Finally, we randomly choose a set of initially defaulted institutions of size $0.01n$ and equally distribute it across the $S$ subsystems.

To see the effect of diversification, we first fix $\Sigma=0.5$ and let $D=J$ vary from $1$ to $20$ (i.e.~$\Delta\in[40]$). The results are plotted in Figure \ref{fig:varying:diversification}. Since we calibrated the capitals $c_i=\alpha_c$ to the values $\Delta=20$ and $\Sigma=0.5$, the theoretical (asymptotic) final default fraction is $1$ for $\Delta\leq20$ and $0$ otherwise. This curve is shown in blue. In orange we illustrate $10$ of the $10^3$ simulations. One can see that in each simulation the final default fraction rapidly decreases at a certain value for $\Delta$ close to the theoretical value of $20$. In green finally, we plot the median over all $10^3$ simulations which is very close to the theoretical curve despite the finite system size and hence verifies that systems become more resilient as $\Delta$ increases. Deviations from the theoretical curve %given by $\1\{\Delta\leq100\}$ %are due to the finite system size and 
become smaller as $n$ increases. %(and the initial infection probability decreases).

Furthermore, in the same setting we conducted simulations for systems of fixed diversification $\Delta=20$ and varying similarity $\Sigma$ between $0$ and $1$ ($J\in[0,20]$ and $D=20-J$). The results are shown in Figure \ref{fig:varying:similarity}. Again, in blue we plot the theoretically predicted curve which is $0$ for $\Sigma<0.5$ and $1$ otherwise. In orange $10$ exemplary simulations are shown. For these, one can see that either there exists an individual threshold for $\Sigma$ close to $0.5$ at which the final default fraction rapidly increases or the final fraction stays constant at $1\%$. The median over the $10^3$ simulations for each $\Sigma$ can be seen in green %in Figure \ref{fig:varying:similarity}.
and it verifies that the system becomes less resilient as the similarity $\Sigma$ increases. Again deviations from the theoretical curve %given by $\1\{\Sigma\geq0.5\}$ \marginpar{\tiny warum ist das die theoretical curve} 
are due to finite size effects and become smaller as $n$ increases.

\section{Conclusions \& Future Research Directions}

In this paper we introduced and studied a mathematical framework for fire sales. Our model describes the state of a financial system by specifying for each participant  the capital and the amount of $M$ distinct assets that she possesses. All this information is concisely represented by the empirical distribution function of the previously mentioned parameters. Then this system gets initially shocked -- the participants suffer losses -- and, as a consequence, assets are sold. This causes the prices to go down, and an iterated process of selling and price decline starts rolling. We studied the final state of the system, in particular the number of sold assets and the final default fraction. Our main driving question was to understand which structural characteristics promote or hinder the outbreak of fire sales, and thus our approach is fundamentally asymptotic in nature: we `gathered' all systems that are similar -- in terms of the empirical distribution -- and studied, as the system size gets large, the effect of various statistical parameters of the (limiting) distribution function.  Our results provided a characterization for any system as either resilient or non-resilient. Moreover, it allowed us to formulate explicit capital requirements and to study the effect of portfolio correlations.

There is a whole bunch of questions that may be investigated within the framework developed here.

\begin{itemize}
	\item {\bf Finite Systems} ~ Our results about resilience are completely explicit for an ensemble of systems and thus, \emph{a priori}, apply to large systems only. Our simulation studies show however that they are already good approximations for moderate-sized systems. A challenging direction for future research is to quantify precisely the \emph{finite-size effect}: how do our results qualitatively carry over to systems of a given size $n$? How can we adapt the notions of (non-)resilience, and what effect does it have, for example on the capital requirements that we develop?
	\item {\bf Refined Models} ~Our model describes the process of (distressed) selling of the assets. However, there is no explicit modeling of market participants who might start buying when the price drops. These buyers might thus slow down the price decline. This effect is captured here in an abstract way by the price impact function $h$. More refined models representing the statistical properties of further market participants and actions would constitute  significant extensions of our model.
	\item {\bf Capital Allocation Strategies} ~ The results presented here allow us to formulate minimal capital requirements that make the system resilient to fire sales. Thus, they constitute a building block of a more global capital allocation strategy that should be capable of making complex systems (including other types of interactions as default or liquidity contagion) resilient to initial stress. For example, our capital requirements can be combined with classical value-at-risk requirements used in the Basel III framework for leveraged institutions such as banks. Developing such global risk management strategies and studying the interplay of various contagion channels is an interesting research topic.
	\item {\bf Portfolio Optimization} ~ Our results might not only be applicable from the view of regulators; they might also be utilized from market participants to optimize their portfolio with respect to systemic risk considerations. Developing adequate investment strategies by taking into consideration the effect of fire sales is a challenging open problem.
\end{itemize}

\section{Generalized Statements and Proofs}\label{sec:proofs}

\subsection{Description and Analysis of the Fire Sales Process for a Discontinuous Sales Function $\rho$}\label{discont:description}

To focus on the underlying phenomena of fire sales, rather than being restrained by technical details, in the main part of this article we considered the special case of a continuous sales functions $\rho$. As already the simple example of sales at default $\rho(u)=\1\{u\geq1\}$ shows, however, it is not untypical for thresholds to exist at which institutions sell a positive fraction of their assets and thus $\rho$ may be discontinuous. By the explanation via thresholds, right-continuous sales functions are more natural than sales functions with discontinuities from the right, and we thus assume $\rho$ to be right-continuous in the sequel and denote by $\circRho(u):=\lim_{\epsilon\to0+}\rho((1-\epsilon)u)$ its left-continuous modification. Replacing $\rho$ by its right-continuous modification $\overline{\rho}(u):=\lim_{\epsilon\to0+}\rho((1+\epsilon)u)$ in the following, however, really makes our results applicable to any non-decreasing sales function $\rho$. In addition to the results presented in the main body of the paper, the results presented here  quantify, besides the final number of sold shares, also the final number of defaulted banks.

\subsubsection{The Deterministic Model}
As for the continuous case in the main part of this article, we start by considering the auxiliary process for a given system of size $n$ first.  For a general non-decreasing sales function $\rho$ the equation
\begin{equation}\label{eqn:fixed:point:sigma:copied}
\E\left[\bm{X}_n\rho\left(\frac{L_n+\bm{X}_n\cdot h(\bm{\chi})}{C_n}\right)\right] - \bm{\chi} = 0
\end{equation}
(compare \eqref{eqn:fixed:point:sigma}) has a smallest solution ${\overline{\bm{\chi}}_n}$ by the Knaster-Tarski theorem, and as in~\eqref{eqn:induction:sigma} it holds $n^{-1}\lim_{k\to\infty}\bm{\sigma}_{(k)}\leq{\overline{\bm{\chi}}_n}$. For left-continuous $\rho$ in fact we would derive equality but for right-continuous $\rho$ it is in general possible that $n^{-1}\lim_{k\to\infty}\bm{\sigma}_{(k)}<{\overline{\bm{\chi}}_n}$. This is the case if $\lim_{k\to\infty}\bm{\sigma}_{(k)}$ sold shares would be enough to start a new round of fire sales but this quantity is actually never reached in finitely many rounds. Then the following holds and extends Proposition~\ref{Prop:det:setting}; the proof is straight-forward by bounding $\rho$ from below with its left-continuous modification $\accentset{\circ}{\rho}$.
\begin{proposition}
\label{prop:upperlowerfixedsize}
Consider the {auxiliary process} with a right-continuous sales function $\rho$ and the corresponding left-continuous modification $\accentset{\circ}{\rho}$. Let {$\overline{\bm{\chi}}_n\in\R_{+}^M$} denote the smallest solution of \eqref{eqn:fixed:point:sigma:copied}. Moreover, let $\hat{\bm{\chi}}_n\in\R_{+}^M$ be the smallest solution of 
\[ {\E\left[\bm{X}_n\accentset{\circ}{\rho}\left(\frac{L_n+\bm{X}_n\cdot h(\bm{\chi})}{C_n}\right)\right] - \bm{\chi} = 0.} \]
Then the number of sold shares divided by $n$ at the end of the fire sales process satisfies 
\begin{equation}\label{eqn:bounds:chi:n}
\hat{\bm{\chi}}_n \leq n^{-1}\lim_{k\to\infty}\bm{\sigma}_{(k)}\leq{\overline{\bm{\chi}}_n}.
\end{equation}
\end{proposition}
The equilibrium vector $n\overline{\bm{\chi}}_n$ (in the sense of \eqref{eqn:fixed:point:sigma:copied}) is thus a conservative bound on the final number of sold shares $n\bm{\chi}_n=\lim_{k\to\infty}\bm{\sigma}_{(k)}$. However, as discussed above, the convergence of the fire sales process to a non-equilibrium heavily relies on the assumption of \emph{arbitrarily} small sale sizes towards the end of the process. For real  systems this is obviously not realistic since the least possible number of shares sold by an institution is lower bounded by $1$. For all practical purposes it will therefore hold that $\bm{\chi}_n=\overline{\bm{\chi}}_n$ and fire sales stop at an equilibrium state.

\subsubsection{Large systems, $n\rightarrow \infty$}\label{ssec:stochastic:model:generalized}
We now consider a sequence of financial systems of increasing size $n\in\N$ as before satisfying Assumption \ref{ass:regularity:fire:sales}. As in Section~\ref{ssec:aux:stoch} let $f^m,g:\R_{+}^M\to\R$, $m\in [M]$ be given by
\begin{align*}
	f^m(\bm{\chi}) := \E\left[X^m\rho\left(\frac{L+\bm{X}\cdot h(\bm{\chi})}{C}\right) \right] - \chi^m,~m\in[M],
	\quad
	g(\bm{\chi}) := \P(L+\bm{X}\cdot h(\bm{\chi})\geq C).
%\circG(\bm{\chi}) &:= \P(L+\bm{X}\cdot h(\bm{\chi})>C).
\end{align*}
Moreover, motivated by the lower bound in Proposition \ref{prop:upperlowerfixedsize} define the lower semi-continuous functions
\begin{equation}
\circFSuper{m}(\bm{\chi}) := \E\left[X^m\accentset{\circ}{\rho}\left(\frac{L+\bm{X}\cdot h(\bm{\chi})}{C}\right) \right] - \chi^m,\quad m\in[M],
\end{equation}
and let
\[ \circG(\bm{\chi}) := \P(L+\bm{X}\cdot h(\bm{\chi})>C). \]
In addition to the sets $S$ and $S_0$ from Section~\ref{ssec:aux:stoch}, now also define
\[ \accentset{\circ}{S}:=\bigcap_{m\in[M]}\left\{\bm{\chi}\in\R_{+}^M\,:\,\circFSuper{m}(\bm{\chi})\geq0\right\} \]
and denote by $\accentset{\circ}{S}_0$ the largest connected subset of $\accentset{\circ}{S}$ containing $\bm{0}$ (clearly $\circFSuper{m}(\bm{0})\geq0$ for all $m\in[M]$ and thus $\bm{0}\in \accentset{\circ}{S}$). Note that still $S$ and $S_0$ are closed sets as $f^m$, $m\in[M]$, are upper semi-continuous for the choice of right-continuous sales function $\rho$. Finally, as in Section~\ref{ssec:aux:stoch} set 
\[ \bm{\chi}^*\in\R_{+}^M \quad\text{with}\quad (\chi^*)^m:=\sup_{\bm{\chi}\in S_0}\chi^m. \]
We can then state the following generalization of Lemma \ref{lem:existence:chi:hat}.
\begin{lemma}\label{lem:existence:chi:hat:generalized}
There exists a smallest joint root $\hat{\bm{\chi}}$ of all functions $\circFSuper{m}(\bm{\chi})$, $m\in[M]$, with $\hat{\bm{\chi}}\in%\accentset{\circ}{S}_0
S_0$. Further, $\bm{\chi}^*$ as defined above is a joint root of the functions %\textcolor{red}{$\circFSuper{m}$ (anschaulich ja, aber stimmt das? Siehe Beweis)}, 
$f^m$, $m\in[M]$, and $\bm{\chi}^*\in S_0$.
\end{lemma}
In complete analogy to Theorem \ref{thm:fire:sale:final:fraction}, but now without imposing Assumption~\ref{ass:main:part}, we can describe the extent and impact of the {auxiliary process}  together with bounds for the number of defaulted institutions.
\begin{theorem}\label{thm:fire:sale:final:fraction:generalized}
{
Consider a $(\bm{X},C)$-system with initial shock $L$ satisfying Assumption \ref{ass:regularity:fire:sales}. Assume that $\rho$ is right-continuous.
Then for  $\chi_n^m$, the number of finally sold shares of asset $m\in[M]$ divided by $n$ in the auxiliary process,  and for the final default fraction $n^{-1}\vert\mathcal{D}_n\vert$,}
\[\begin{gathered}
\hat{\chi}^m+o(1) \leq \chi_n^m \leq (\chi^*)^m+o(1) , \qquad \circG(\hat{\bm{\chi}}) + o(1) \leq n^{-1}\vert\mathcal{D}_n\vert \leq g(\bm{\chi}^*) + o(1) .
\end{gathered}\]
In particular, for the final price impact $h^m(\bm{\chi}_n)$ on asset $m\in[M]$ 
\[ h^m(\hat{\bm{\chi}})+o(1) \leq h^m(\bm{\chi}_n) \leq h^m(\bm{\chi}^*)+o(1). \]
\end{theorem}
In addition to instability of the joint root $\hat{\bm{\chi}}$ (as is explained in Subsection \ref{ssec:aux:stoch}) another explanation for $\hat{\bm{\chi}}\neq\bm{\chi}^*$ could be that $\hat{\bm{\chi}}$ is a point of discontinuity for some $f^m$
%(cf.~Figure \ref{fig:example:non:continuous})
but this again is a rather pathological case. It thus usually holds $\circG(\hat{\bm{\chi}})=g(\bm{\chi}^*)$ and Theorem \ref{thm:fire:sale:final:fraction:generalized} determines the limits of $\bm{\chi}_n$ and $n^{-1}\vert\mathcal{D}_n\vert$ as $n\to\infty$.

\subsubsection{(Non-)Resilience}\label{ec:non:resilience}

{
In this section we study (non-)resilience properties of $(\bm{X},C)$-systems in the general setting; note that these notions did not depend on our previous Assumption~\ref{ass:main:part}. The following statement thus generalizes Theorem~\ref{thm:resilience} and also includes a statement about the default fraction.}
\begin{theorem}\label{thm:resilience:with:defaults}
Consider a $(\bm{X},C)$-system satisfying Assumption 1. Assume that $\rho$ is right-continuous. Then for each $\epsilon>0$ there exists $\delta>0$ such that for all initial shocks $L$ with $\E[L/C]<\delta$, the number $n\chi_{n,L}^m$ of finally sold shares of  asset $m\in[M]$ and the final set of defaulted institutions $\mathcal{D}_{n,L}$ in the shocked system satisfy
\[  \limsup_{n\to\infty} \chi_{n,L}^m \leq(\chi^*)^m+\epsilon, ~~ m\in[M] \quad \text{and}\quad \limsup_{n\to\infty}n^{-1}\vert\mathcal{D}_{n,L}\vert\leq g(\bm{\chi}^*)+\epsilon. \]
\end{theorem}
Using the notations of Section~\ref{ssec:stochastic:model:generalized}, also the wording of Corollary~\ref{cor:resilience} remains completely unchanged;  that is, if $\bm{\chi}^* =0$, then the  $(\bm{X},C)$-system satisfying Assumption 1 with right-continuous $\rho$ is resilient. 

We now consider $\bm{\chi}^* >0$. Theorem~\ref{thm:non-resilience} remains unchanged in case that $\rho$ is strictly increasing in a neighborhood of $0$. If $\rho$ is not strictly increasing, we distinguish between non-resilience and weak non-resilience.

\begin{theorem}\label{thm:non-resilience:non:strictly:inc}
%If the initial shock $L$ satisfies above described properties, that is $\P(L=2C)>0$ and $L$ independent of $(\bm{X},C)$, then
Consider an $(\bm{X},C)$-system satisfying Assumption 1. Assume that $\rho$ is right-continuous.
Let $\epsilon>0$. For every $\delta>0$ there exists a shock $L$ with $\E [L/C]<\delta $ and 
\[  \liminf_{n\to\infty} \chi_{n,L}^m > (\chi^*)^m-\epsilon \quad \text{and} \quad \liminf_{n\to\infty}n^{-1}\vert\mathcal{D}_{n,L}\vert > \circG(\bm{\chi}^*)-\epsilon . \]
\end{theorem}
In particular, from Theorem~\ref{thm:non-resilience:non:strictly:inc} we immediately get the following result.
\begin{corollary}[Non-resilience Criterion]\label{cor:weak:resilience}
If $\bm{\chi}^*>0$, then the $(\bm{X},C)$-system is weakly non-resilient under the auxiliary process.
\end{corollary} 

%Corollary~\ref{cor:non:resilience} remains unchanged in wording and holds in both case ($\rho$ strictly increasing or above assumptions on $L$).

\subsection{Proofs of Statements}\label{proofs:general:case}

\subsubsection{Proofs for Section~\ref{sec:aux:model} and Section~\ref{discont:description} (auxiliary process) }\label{ssec:proofs:2}

\proof[Proof of Lemma \ref{lem:existence:chi:hat:generalized}.]
Existence of $\hat{\bm{\chi}}$ follows from the Knaster-Tarski theorem. We now construct a joint root $\accentset{\circ}{S}_0\ni\bar{\bm{\chi}}\leq\hat{\bm{\chi}}$ such that we can conclude $\hat{\bm{\chi}}=\bar{\bm{\chi}}\in \accentset{\circ}{S}_0$.
 It holds $\circFSuper{m}(\hat{\bm{\chi}})=0$ for all $m\in[M]$ and thus (for any fixed $m\in[M]$) $\circFSuper{m}(\bm{\chi})\leq0$ for all $\hat{\bm{\chi}}\geq\bm{\chi}\in\R_{+}^M$ such that $\chi^m=\hat{\chi}^m$ by monotonicity of $\circFSuper{m}$. %Now let $D_0:=\{\bm{\chi}\in S_0\,:\,\exists m\in[M]:\circFSuper{m}(\bm{\chi})=0\}$. 
Consider then the following sequence $(\bm{\chi}_{(k)})_{k\in\N}\subset\R_{+}^M$:
\begin{itemize}
\item $\bm{\chi}_{(0)}=\bm{0}\in \accentset{\circ}{S}_0$
\item $\bm{\chi}_{(1)}=(\chi_{(1)}^1,0,\ldots,0)$, where $0\leq\chi_{(1)}^1\leq\hat{\chi}^1$ is the smallest possible value such that $\circFSuper{1}(\bm{\chi}_{(1)})=0$. It is possible to find such $\chi_{(1)}^1$ since $\circFSuper{1}(\bm{\chi})+\chi^1$ is monotonically increasing in $\chi^1$, $\circFSuper{1}(\bm{0})\geq 0$ and $\circFSuper{1}(\hat{\chi}^1,0,\ldots,0)\leq0$. By monotonicity of $\circFSuper{m}$ with respect to $\chi^1$ for all $m\in[M]\backslash\{1\}$, it then holds $\circFSuper{m}(\bm{\chi}_{(1)})\geq \circFSuper{m}(\bm{0})\geq0$ for all $1\neq m\in [M]$ and in particular $\bm{\chi}_{(1)}\in \accentset{\circ}{S}_0$.
\item $\bm{\chi}_{(2)}=\bm{\chi}_{(1)}+(0,\chi_{(2)}^2,0,\ldots,0)$, where $0\leq\chi_{(2)}^2\leq\hat{\chi}^2$ is the smallest value such that $\circFSuper{2}(\bm{\chi}_{(2)})=0$. Again it is possible to find such $\chi_{(2)}^2$ since $\circFSuper{2}(\bm{\chi})+\chi^2$ is monotonically increasing in $\chi^2$, $\circFSuper{2}(\bm{\chi}_{(1)})\geq0$ and $\circFSuper{2}(\bm{\chi}_{(1)}+(0,\hat{\chi}^2,0,\ldots,0))\leq0$. By monotonicity of $\circFSuper{m}$ with respect to $\chi^2$ for all $m\in[M]\backslash\{2\}$, it then holds $\circFSuper{m}(\bm{\chi}_{(2)})\geq \circFSuper{m}(\bm{\chi}_{(1)})\geq0$ for all $2\neq m\in [M]$ and in particular $\bm{\chi}_{(2)}\in \accentset{\circ}{S}_0$.
\item $\bm{\chi}_{(i)}$, $i\in\{3,\ldots,M\}$, are found analogously, changing only the corresponding coordinate.
\item $\bm{\chi}_{(M+1)}=\bm{\chi}_{(M)}+(\chi_{(M+1)}^1-\chi_{(M)}^1,0,\ldots,0)$, where $\chi_{(M)}^1\leq\chi_{(M+1)}^1\leq\hat{\chi}^1$ is the smallest value such that $\circFSuper{1}(\bm{\chi}_{(M+1)})=0$, which is again possible by monotonicity of $\circFSuper{1}(\bm{\chi})+\chi^1$% with respect to $\chi^1$
, $\circFSuper{1}(\bm{\chi}_{(M)})\geq0$ and $\circFSuper{1}(\bm{\chi}_{(M)}+(\hat{\chi}^1-\chi_{(M)}^1,0,\ldots,0))\leq0$. Further, it still holds $\bm{\chi}_{(M+1)}\in \accentset{\circ}{S}_0$.
\item Continue for $\bm{\chi}_{i}$, $i\geq M+2$.
\end{itemize}
The sequence $(\bm{\chi}_{(k)})_{k\in\N}$ constructed this way has the following properties: it is non-decreasing in each coordinate and %$(\bm{\chi}_{(k)})_{k\in\N}\subset D_0$. Further, it is 
bounded inside $[\bm{0},\hat{\bm{\chi}}]$. Hence by monotone convergence, each coordinate of $\bm{\chi}_{(k)}$ converges and so $\bar{\bm{\chi}}=\lim_{k\to\infty}\bm{\chi}_{(k)}$ exists. Since the convergence is from below, 
\begin{align*}
\circFSuper{m}(\bar{\bm{\chi}}) &= \E\left[X^m\circRho\left(\frac{L+\bm{X}\cdot h(\lim_{k\to\infty}\bm{\chi}_{(k)})}{C}\right)\right] - \lim_{k\to\infty}\chi_{(k)}^m\\
&= \lim_{k\to\infty}\E\left[X^m\circRho\left(\frac{L+\bm{X}\cdot h(\bm{\chi}_{(k)})}{C}\right)\right]-\chi_{(k)}^m = \lim_{k\to\infty}\circFSuper{m}(\bm{\chi}_{(k)}) \geq 0
\end{align*}
and thus $\bar{\bm{\chi}}\in\accentset{\circ}{S}_0$. %$\bar{\bm{\chi}}\in\overline\circS_0\cap \circS$
Now suppose there is $m\in[M]$ such that $\circFSuper{m}(\bar{\bm{\chi}})>0$. By lower semi-continuity of $\circFSuper{m}$ then also $\circFSuper{m}(\bm{\chi}_{(k)})>\epsilon$ for some $\epsilon>0$ and $k$ large enough. This, however, is a contradiction to the construction of the sequence $(\bm{\chi}_{(k)})_{k\in\N}$ since $\circFSuper{m}(\bm{\chi}_{(k)})=0$ in every $M$-th step. Hence $\circFSuper{m}(\bar{\bm{\chi}})= 0$ for all $m\in[M]$ and $\bar{\bm{\chi}}$ is a joint root of all functions $\circFSuper{m}$, $m\in[M]$.

Now turn to the proof that $\bm{\chi}^*\in S_0$.
% and it is a joint root of all functions $\textcolor{red}{\circFSuper{m}}$, $f^m$, $m\in[M]$
We first consider the case that $\rho$ is continuous (Lemma~\ref{lem:existence:chi:hat}). We approximate $\bm{\chi}^*\in S_0$ by the sequence $(\hat{\bm{\chi}}(\epsilon))_{\epsilon>0}$ of smallest fixed-points for the functions $f^m(\bm{\chi})+ \epsilon$. This allows us to apply the Knaster-Tarski Theorem and the monotonicity properties of $f^m+ \epsilon$  as above. Simple topological arguments will then allow us to conclude that $\bm{\chi}^*\in S_0$. Let for $\epsilon>0$
\[ S(\epsilon) := \bigcap_{m\in[M]}\left\{\bm{\chi}\in\R_{+}^M\,:\,f^m(\bm{\chi})\geq -\epsilon\right\} \]
and denote by $S_0(\epsilon)$ the connected component of $\bm{0}$ in $S(\epsilon)$. By the same procedure as for $\hat{\bm{\chi}}$ above, we now derive that there exists a smallest (componentwise) point $\hat{\bm{\chi}}(\epsilon)\in S_0(\epsilon)$ such that $f^m(\hat{\bm{\chi}}(\epsilon))=-\epsilon$ for all $m\in[M]$. Clearly, $\hat{\bm{\chi}}(\epsilon)$ is non-decreasing (componentwise) in $\epsilon$ and hence $\tilde{\bm{\chi}}:=\lim_{\epsilon\to0+}\hat{\bm{\chi}}(\epsilon)$ exists (we will show that $\tilde{\bm{\chi}}=\bm{\chi}^*$ in fact).

Now by monotonicity of $S_0(\epsilon)$, we derive that $\hat{\bm{\chi}}(\delta)\in S_0(\delta)\subseteq S_0(\epsilon)$ for all $\delta\leq \epsilon$. Since $S_0(\epsilon)$ is a closed set, it must hold that also $\tilde{\bm{\chi}}=\lim_{\delta\to0+}\hat{\bm{\chi}}(\delta)\in S_0(\epsilon)$ for all $\epsilon>0$ and in particular, $\tilde{\bm{\chi}}\in\bigcap_{\epsilon>0}S_0(\epsilon)$. Further, we derive that $\bigcap_{\epsilon>0}S_0(\epsilon)\subseteq\bigcap_{\epsilon>0}S(\epsilon)\subseteq S$. Moreover, $\bigcap_{\epsilon>0}S_0(\epsilon)$ is the intersection of a chain of connected, compact sets in the Hausdorff space $\R^M$ and it is hence a connected, compact set itself. Since it further contains $\bm{0}$, we can then conclude that $\bigcap_{\epsilon>0}S_0(\epsilon)\subseteq S_0$ and thus $\tilde{\bm{\chi}}\in S_0$.

Consider now an arbitrary $\bm{\chi}\in S_0$. We want to show that $\bm{\chi}\leq\tilde{\bm{\chi}}$ componentwise and thus $\tilde{\bm{\chi}}=\bm{\chi}^*$. It suffices to show that $S_0\subset[\bm{0},\hat{\bm{\chi}}(\epsilon)]$  for all $\epsilon$. Then $\bm{\chi}\leq\hat{\bm{\chi}}(\epsilon)$ and $\bm{\chi}\leq\lim_{\epsilon\to0+}\hat{\bm{\chi}}(\epsilon)=\tilde{\bm{\chi}}$. Hence assume that $S_0\not\subset[\bm{0},\hat{\bm{\chi}}(\epsilon)]$. By connectedness of $S_0$ we find $\bar{\bm{\chi}}\in S_0$ with $\bar{\chi}^m\leq\hat{\chi}^m(\epsilon)$ for all $m\in[M]$ and equality for at least one coordinate (otherwise $S_0\cap\partial[\bm{0},\hat{\bm{\chi}}(\epsilon)]=\emptyset$ and $S_0=\left(S_0\cap\left(\R_{+}^M\backslash[\bm{0},\hat{\bm{\chi}}(\epsilon)]\right)\right)\cup\left(S_0\cap[\bm{0},\hat{\bm{\chi}}(\epsilon))\right)$ is the union of two open non-empty sets and hence not connected). W.\,l.\,o.\,g.~let this coordinate be $\bar{\chi}^1$. By monotonicity of $f^1$ with respect to $\chi^m$ for every $1\neq m\in[M]$, we thus derive that $f^1(\bar{\bm{\chi}})\leq f^1(\hat{\bm{\chi}}(\epsilon))=-\epsilon<0$ which is a contradiction to $\bar{\bm{\chi}}\in S_0$.

Now consider the general case that $\rho$ is right-continuous and let $(\rho_r(u))_{r\in\N}$ a sequence of continuous sales functions approximating $\rho$ from above. Denoting by $S^r$ the analogue of $S$ for the sales function $\rho_r$, we derive that $S=\bigcap_{r\in\N}S^r$ since clearly $S^r\supseteq S$ for all $r\in\N$ and further by dominated convergence for every $\bm{\chi}\in\bigcap_{r\in\N}S^r$,
\[ \chi^m \leq \E\left[X^m\rho_r\left(\frac{L+\bm{X}\cdot h(\bm{\chi})}{C}\right)\right] \to \E\left[X^m\rho\left(\frac{L+\bm{X}\cdot h(\bm{\chi})}{C}\right)\right],\quad\text{as }r\to\infty, \]
so that $\bigcap_{r\in\N}S^r\subseteq S$. If we further let $S_0^r$ denote the largest connected subset of $S^r$ containing $\bm{0}$, then $S_0^r$ is compact and connected for every $r\in\N$ and hence so is $\bigcap_{r\in\N}S_0^r$. Since further $\bm{0}\in\bigcap_{r\in\N}S_0^r$, we derive that $\bigcap_{r\in\N}S_0^r=S_0$. Let now $\bm{\chi}_r^*$ denote the analogue of $\bm{\chi}^*$ for the sales function $\rho_r$. Then $\lim_{r\to\infty}\bm{\chi}_r^*\in S_0^R$ for all $R\in\N$ and hence $\lim_{r\to\infty}\bm{\chi}_r^*\in\bigcap_{R\in\N}S_0^R=S_0$. Now suppose there existed a vector $\bm{\chi}\in S_0$ and $m\in[M]$ such that $\chi^m>\lim_{r\to\infty}(\chi_r^*)^m$. Then also for $R$ large enough, $\chi^m>(\chi_R^*)^m$ and hence $\bm{\chi}\not\in S_0^R$. This, however, contradicts the assumption that $\bm{\chi}\in S_0=\bigcap_{R\in\N}S_0^R$. Hence there exists no such $\bm{\chi}\in S_0$ and $\bm{\chi}^*=\lim_{r\to\infty}\bm{\chi}_r^*\in S_0$.

Finally, we show that $\bm{\chi}^*$ is a joint root of $f^m$, $m\in[M]$: Since $\bm{\chi}^*\in S_0$, it holds that $f^m(\bm{\chi}^*)\geq0$ for all $m\in[M]$. Assume now that $f^m(\bm{\chi}^*)>0$ for some $m\in[M]$. We can then gradually increase the $m$-coordinate of $\bm{\chi}^*$ (until $f^m(\bm{\chi}^*)=0$). By monotonicity of $f^k(\bm{\chi})$ with respect to $\chi^m$ for every $m\neq k\in[M]$, however, we can be sure that we do not leave the set $S_0$ by this procedure which is a contradiction to the definition of $\bm{\chi}^*$. Hence $\bm{\chi}^*$ is a joint root of $f^m$, $m\in[M]$. %The same argument shows that $\circFSuper{m}(\bm{\chi}^*)\leq0$ for all $m\in[M]$ since $f^m(\bm{\chi})\geq\circFSuper{m}(\bm{\chi})$. Finally, by definition of $\bm{\chi}^*$ there exists a sequence $(\bm{\chi}_{(k)})_{k\in\N}\subset S_0$ such that $\bm{\chi}_{(k)}\to\bm{\chi}^*$ from below. Hence $\circFSuper{m}(\bm{\chi}^*) = \lim_{k\to\infty}f^m(\bm{\chi}_{(k)}) \geq 0$. \textcolor{red}{If $\bm{\chi}_{(k)}^m=(\bm{\chi}^*)^m$ for some $m\in[M]$ it is possible that $\circFSuper{m}(\bm{\chi}^*) < \lim_{k\to\infty}f^m(\bm{\chi}_{(k)})$}
\hfill
\endproof

\begin{remark}\label{rem:sequence:chi:*}
In the proof of Lemma~\ref{lem:existence:chi:hat:generalized}, for the case that $\rho$ is continuous, we constructed $\bm{\chi}^*$ as the limit of a sequence $(\hat{\bm{\chi}}(\epsilon))_{\epsilon>0}$ such that $f^m(\hat{\bm{\chi}}(\epsilon))=-\epsilon$ for all $m\in[M]$. For non-continuous $\rho$ by the Knaster-Tarski theorem we still know that there exists a smallest vector $\hat{\bm{\chi}}(\epsilon)$ such that $f^m(\hat{\bm{\chi}}(\epsilon))=-\epsilon$, but the construction of $\hat{\bm{\chi}}(\epsilon)$ as for $\hat{\bm{\chi}}$ in the proof of Lemma \ref{lem:existence:chi:hat} fails and we can not be sure a priori that $\hat{\bm{\chi}}(\epsilon)\in S_0(\epsilon)$. Hence let further $\tilde{\bm{\chi}}(\epsilon)$ be defined as the smallest vector in $S_0(\epsilon)$ such that $f^m(\tilde{\bm{\chi}}(\epsilon))=-\epsilon$. This vector exists again by the Knaster-Tarski theorem noting that analogously to Lemma \ref{lem:existence:chi:hat} $S_0(\epsilon)$ contains its componentwise supremum $\bm{\chi}^*(\epsilon)$. %Then by definition $\hat{\bm{\chi}}(\epsilon)\leq\tilde{\bm{\chi}}(\epsilon)$ but $\hat{\bm{\chi}}(\epsilon)\geq \tilde{\bm{\chi}}(\delta)$ for every $\delta<\epsilon$. %$S_0(\delta)\subsetneq[\mathb{0},\hat{\bm{\chi}}(\epsilon)]$ by monotonicity
%Now $\tilde{\bm{\chi}}(\epsilon)$ is a strictly increasing lower bounded sequence and it can thus only have at most countably many points of discontinuities as $\epsilon\to0$. We can hence find arbitrarily small $\epsilon>0$ such that $\lim_{\delta\to\epsilon-}\tilde{\bm{\chi}}=\tilde{\bm{\chi}}(\epsilon)$ and in particular, $\hat{\bm{\chi}}(\epsilon)=\tilde{\bm{\chi}}(\epsilon)$.
Then by the same means as above, we derive that $\bm{\chi}^*=\lim_{\epsilon\to0}\tilde{\bm{\chi}}(\epsilon)$.
\hfill
\end{remark}
The rest of the section is devoted to the proof of Theorem \ref{thm:fire:sale:final:fraction:generalized} (and Theorem~\ref{thm:fire:sale:final:fraction}) which deal with a sequence of financial systems. The following technical lemma about the convergence of the smallest joint roots will be quite handy.
\begin{lemma}\label{lem:convergence:chi:hat}
Let a sequence (for $r\in\N$) of financial systems be described by functions $\circFSuperSub{m}{r}$, $m\in[M]$, with smallest joint root $\hat{\bm{\chi}}_r$. If $\liminf_{r\to\infty}\circFSuperSub{m}{r}(\bm{\chi})\geq\circFSuper{m}(\bm{\chi})$ pointwise for every $m\in[M]$, then $\liminf_{r\to\infty}\hat{\bm{\chi}}_r\geq\hat{\bm{\chi}}$, where $\hat{\bm{\chi}}$ denotes the smallest joint root of the functions $\circFSuper{m}$, $m\in[M]$.
\end{lemma}
\proof
The main difficulty in showing the claim is that $\liminf_{r\to\infty}\circFSuperSub{m}{r}(\bm{\chi})\geq\circFSuper{m}(\bm{\chi})$ only pointwise but not uniformly in $\bm{\chi}$. A further difficulty is the multidimensionality. The main idea is to construct a path in analogy to the construction in Lemma \ref{lem:existence:chi:hat:generalized} that leads to a point $\tilde{\bm{\chi}}(\epsilon)$ smaller but close to $\hat{\bm{\chi}}$. On this path the functions $\circFSuperSub{m}{r}$, $m\in[M]$ are all positive for $r$ large. It can then be compared componentwise with a path leading to $\hat{\bm{\chi}}_r$.

For this consider the construction of $\hat{\bm{\chi}}$ in Lemma \ref{lem:existence:chi:hat:generalized} and change it in such a way that in each step $k=LM+m$ (where $L\in\N_0$ and $m\in[M]$) a point $\bm{\chi}_{(k)}(\epsilon)$ is chosen such that $\circFSuper{m}(\bm{\chi}_{(k)}(\epsilon))\leq\epsilon$ for some fixed $\epsilon>0$ (choose $\chi_{(k)}^m(\epsilon)\geq\chi_{(k-1)}^m(\epsilon)$ as the smallest possible value such that this inequality holds; it will then either be $\circFSuper{m}(\bm{\chi}_{(k)}(\epsilon))=\epsilon$ or $\bm{\chi}_{(k)}(\epsilon)=\bm{\chi}_{(k-1)}(\epsilon)$). Note that $\circFSuper{m}(\bm{\chi}_{(k)}(\epsilon))<\epsilon$ can only happen if $\circFSuper{m}(\bm{0}) <\epsilon$ in which case there exists $k_0\in \N_{\infty}$ such that  $\chi^m_{(k)}=0$ and $\circFSuper{m}(\bm{\chi}_{(k)}) <\epsilon$ for all $k\leq k_0$ but $\circFSuper{m}(\bm{\chi}_{(k)}) \geq \epsilon$ and $\chi^m_{(k)}>0$ for $k>k_0$. Then $(\bm{\chi}_{(k)}(\epsilon))_{k\in\N}$ is a non-decreasing (componentwise) sequence bounded by $\hat{\bm{\chi}}$ and hence $\tilde{\bm{\chi}}(\epsilon)=\lim_{k\to\infty}\bm{\chi}_{(k)}(\epsilon)$ exists. Further, it holds that $\circFSuper{m}(\tilde{\bm{\chi}}(\epsilon)) \leq \liminf_{k\to\infty}\circFSuper{m}(\bm{\chi}_{(k)}(\epsilon)) \leq \epsilon$. Finally, $\tilde{\bm{\chi}}(\epsilon)$ is non-increasing componentwise in $\epsilon$ and bounded inside $[\bm{0},\hat{\bm{\chi}}]$ and thus $\tilde{\bm{\chi}}=\lim_{\epsilon\to0+}\tilde{\bm{\chi}}(\epsilon)$ exists. Moreover, $\circFSuper{m}(\tilde{\bm{\chi}}) \leq \liminf_{\epsilon\to0+}\circFSuper{m}(\tilde{\bm{\chi}}(\epsilon)) \leq \liminf_{\epsilon\to0+}\epsilon = 0$ and hence in particular $\tilde{\bm{\chi}}=\hat{\bm{\chi}}$.

Fix now $\delta>0$ and choose $\epsilon>0$ small enough such that $\tilde{\chi}^m(\epsilon)>\hat{\chi}^m(1-\delta)^{1/2}$ for all $m\in[M]$. Further, choose $K=K(\epsilon)\in\N$ large enough such that $\chi_{(K)}^m(\epsilon)>\tilde{\chi}^m(\epsilon)(1-\delta)^{1/2}$ for all $m\in[M]$. In particular, $\chi_{(K)}^m(\epsilon)>\hat{\chi}^m(1-\delta)$. Now note that $\hat{\bm{\chi}}_r$ can be constructed by a sequence $(\bm{\chi}_{(k,r)})_{k\in\N}$ analogue to $\hat{\bm{\chi}}$ in the proof of Lemma \ref{lem:existence:chi:hat} as well. We can then in each step $k\in\N$ cap the element of the constructing sequence $\bm{\chi}_{(k,r)}$ at $\bm{\chi}_{(k)}(\epsilon)$, which clearly does not increase the limit of the sequence. We want to make sure that in fact the cap is used in every step $k\leq K$ if we only choose $r$ large enough. Then we can conclude that $\hat{\bm{\chi}}_r \geq \bm{\chi}_{(K)}(\epsilon) \geq \hat{\bm{\chi}}(1-\delta)$ and hence letting $\delta\to0$, $\liminf_{r\to\infty}\hat{\bm{\chi}}_r \geq \hat{\bm{\chi}}$.

We now show that the cap is applied in every step $k\leq K$ for $r$ large enough by an induction argument. For $k=0$, clearly $\bm{\chi}_{(k,r)}=\bm{\chi}_{(k)}(\epsilon)=\bm{0}$ and the cap is applied. Now let us assume it holds for $k\leq k_0<K$. If $\bm{\chi}_{(k_0+1)}(\epsilon)=\bm{\chi}_{(k_0)}(\epsilon)$, then of course the cap is also applied in step $k_0+1$ as the sequence $\bm{\chi}_{(k,r)}$ is increasing. Otherwise, note that by definition of $\bm{\chi}_{(k_0+1)}(\epsilon)$, it holds $\circFSuper{m}(\bm{\chi})\geq\epsilon$ for all $\bm{\chi}\in\R_{+}^M$ such that $\chi^m\in[\chi_{(k_0)}^m(\epsilon),\chi_{(k_0+1)}^m(\epsilon)]$ and $\chi^\ell=\chi_{(k_0)}^\ell(\epsilon)=\chi_{(k_0+1)}^\ell(\epsilon)$ for all $\ell\in[M]\backslash\{m\}$. Now choose a discretization $\{\chi_j\}_{0\leq j\leq J}$ of $[\chi_{(k_0)}^m(\epsilon),\chi_{(k_0+1)}^m(\epsilon)]$ for $J<\infty$ such that $\chi_0 = \chi_{(k_0)}^m(\epsilon)$, $\chi_J=\chi_{(k_0+1)}^m(\epsilon)$ and $\chi_{j-1}<\chi_j<\chi_{j-1}+\epsilon/3$ for all $j\in[J]$. We now use the assumption that $\liminf_{r\to\infty}\circFSuperSub{m}{r}(\bm{\chi}_j)\geq\circFSuper{m}(\bm{\chi}_j)$ for every $0\leq j\leq J$, where $\chi^m_j=\chi_j$ and $\chi^\ell_j = \chi_{(k_0)}^\ell(\epsilon)$ for $\ell\in[M]\backslash\{m\}$. Then for $r$ large enough, $\circFSuperSub{m}{r}(\bm{\chi}_j)\geq\circFSuper{m}(\bm{\chi}_j)-\epsilon/3\geq 2\epsilon/3$. Finally, for any linear interpolation $\bm{\chi}=\alpha\bm{\chi}_{j-1}+(1-\alpha)\bm{\chi}_j$ between $\bm{\chi}_{j-1}$ and $\bm{\chi}_j$ ($\alpha\in[0,1]$),  
\[ \circFSuperSub{m}{r}(\bm{\chi}) \geq \circFSuperSub{m}{r}(\bm{\chi}_{j-1}) + \chi_{j-1}^m-\chi_j^m \geq 2\epsilon/3 - \epsilon/3 = \epsilon/3. \]
%However, as the derivative of $\circFSuperSub{m}{r}$ with respect to its $m-$th argument \daniel{doesn't necessarily exist} is lower bounded by $-1$, it follows that $\circFSuperSub{m}{r}(\bm{\chi})\geq \epsilon/3$ for any linear interpolation between $\bm{\chi}_{j-1}$ and $\bm{\chi}_j$ because the distances between $\bm{\chi}_{j-1}$ and $\bm{\chi}_j$ were chosen such that $\abs{\bm{\chi}_{j}^m-\bm{\chi}_{j-1}^m}\leq \epsilon/3$. 
Hence the cap is applied in step $k_0+1$. As there are only finitely many steps $k\leq K$, this finishes the proof. \hfill
\endproof

\proof[Proof of Theorem \ref{thm:fire:sale:final:fraction:generalized}.]
We start by proving the lower bound. Recall from Proposition \ref{prop:upperlowerfixedsize} that $\bm{\chi}_n\geq\hat{\bm{\chi}}_n$. Using weak convergence of the random vector $(\bm{X}_n,C_n,L_n)$ and approximating $\circRho$ from below by a sequence of continuous sales functions $(\rho_r)_{r\in\N}$, we derive for $U>0$ that pointwise
\begin{align*}
\liminf_{n\to\infty}\E\left[X_n^m\circRho\left(\frac{L_n+\bm{X}_n\cdot h(\bm{\chi}))}{C_n}\right)\right] &\geq \lim_{n\to\infty}\E\left[\left(X_n^m\wedge U\right)\rho_r\left(\frac{L_n+\bm{X}_n\cdot h(\bm{\chi}))}{C_n}\right)\right]\\
&= \E\left[\left(X^m\wedge U\right)\rho_r\left(\frac{L+\bm{X}\cdot h(\bm{\chi}))}{C}\right)\right].
\end{align*}
Hence as $U\to\infty$ and $r\to\infty$ by monotone convergence,
\begin{equation}\label{eqn:convergence:finite:f}
\liminf_{n\to\infty}\E\left[X_n^m\circRho\left(\frac{L_n+\bm{X}_n\cdot h(\bm{\chi})}{C_n}\right)\right] - \chi^m \geq \circFSuper{m}(\bm{\chi}))
\end{equation}
and we can use Lemma \ref{lem:convergence:chi:hat} to derive that $\liminf_{n\to\infty}\bm{\chi}_n \geq \liminf_{n\to\infty}\hat{\bm{\chi}}_n\geq\hat{\bm{\chi}}$.

We now want to show the lower bound on the final default fraction. Fix some $\delta>0$ and choose $n$ large enough such that $\bm{\chi}_n\geq\hat{\bm{\chi}}_n\geq(1-\delta)\hat{\bm{\chi}}$ componentwise. Then
\[ n^{-1}\vert\mathcal{D}_n\vert = \P\left(L_n+\bm{X}_n\cdot h(\bm{\chi}_n)\geq C_n\right) \geq \P\left(L_n+\bm{X}_n\cdot h((1-\delta)\hat{\bm{\chi}})> C_n\right). \]
However, using weak convergence of $(\bm{X}_n,C_n,L_n)$ and approximating the indicator function $\1\{y>1\}$ from below by continuous functions $(\phi_t)_{t\in\N}$, we derive that
\[ \liminf_{n\to\infty}n^{-1}\vert\mathcal{D}_n\vert \geq \lim_{n\to\infty}\E\left[\phi_t\left(\frac{L_n+\bm{X}_n\cdot h((1-\delta)\hat{\bm{\chi}})}{C_n}\right)\right] = \E\left[\phi_t\left(\frac{L+\bm{X}\cdot h((1-\delta)\hat{\bm{\chi}})}{C}\right)\right] \]
and as $t\to\infty$,
\[ \liminf_{n\to\infty}n^{-1}\vert\mathcal{D}_n\vert \geq \P\left(L+\bm{X}\cdot h((1-\delta)\hat{\bm{\chi}})>C\right) = \accentset{\circ}{g}((1-\delta)\hat{\bm{\chi}}). \]
This quantity now tends to $\accentset{\circ}{g}(\hat{\bm{\chi}})$ as $\delta\to0$ by lower semi-continuity of $\accentset{\circ}{g}$.

Now we approach the second part of the theorem. Recall from Proposition \ref{prop:upperlowerfixedsize} that $\bm{\chi}_n\leq\overline{\bm{\chi}}_n$. By the construction of $\bm{\chi}^*$ in the proof of Lemma \ref{lem:existence:chi:hat:generalized}
, we have a non-increasing (in $\epsilon$) sequence $(\hat{\bm{\chi}}(\epsilon))_{\epsilon>0}$  such that $\lim_{\epsilon\to0+}\hat{\bm{\chi}}(\epsilon)=\bm{\chi}^*$. (See Remark \ref{rem:sequence:chi:*} for non-continuous $\rho$.) In particular, $\bm{\chi}^*\leq\hat{\bm{\chi}}(\epsilon)$ for every $\epsilon>0$ and $f^m(\hat{\bm{\chi}}(\epsilon))=-\epsilon$. Using weak convergence of $(\bm{X}_n,C_n,L_n)$ we derive for $U>0$ and $(\rho_s)_{s\in\N}$ an approximation of $\rho$ from above by continuous sales functions that
\begin{align*}
&\limsup_{n\to\infty}\E\bigg[X_n^m\rho\bigg(\frac{L_n+\bm{X}_n\cdot h(\hat{\bm{\chi}}(\epsilon))}{C_n}\bigg)\bigg] = \E[X^m] - \liminf_{n\to\infty}\E\bigg[X_n^m\bigg(1-\rho\bigg(\frac{L_n+\bm{X}_n\cdot h(\hat{\bm{\chi}}(\epsilon))}{C_n}\bigg)\bigg)\bigg]\\
&\hspace{1cm}\leq \E[X^m] - \liminf_{n\to\infty}\E\left[(X_n^m\wedge U)\left(1-\rho_s\left(\frac{L_n+\bm{X}_n\cdot h(\hat{\bm{\chi}}(\epsilon))}{C_n}\right)\right)\right]\\
&\hspace{1cm}= \E[X^m] - \E\left[(X^m\wedge U)\left(1-\rho_s\left(\frac{L+\bm{X}\cdot h(\hat{\bm{\chi}}(\epsilon))}{C}\right)\right)\right]
\end{align*}
and as $U\to\infty$, $s\to\infty$, by monotone convergence
\[ \limsup_{n\to\infty}\E\left[X_n^m\rho\left(\frac{L_n+\bm{X}_n\cdot h(\hat{\bm{\chi}}(\epsilon))}{C_n}\right)\right] \leq f^m(\hat{\bm{\chi}}(\epsilon)) + \hat{\chi}^m(\epsilon) = \hat{\chi}^m(\epsilon) - \epsilon. \]
Hence for $n$ large enough it holds
\[ \E\left[X_n^m\rho\left(\frac{L_n+\bm{X}_n\cdot h(\hat{\bm{\chi}}(\epsilon))}{C_n}\right)\right] - \hat{\chi}^m(\epsilon) \leq -\epsilon/2 <0 \]
for all $m\in[M]$. In particular, we know that $\overline{\bm{\chi}}_n\leq\hat{\bm{\chi}}(\epsilon)$.  Letting $\epsilon\to0$, this shows that $\limsup_{n\to\infty}\chi_n^m\leq\limsup_{n\to\infty}\overline{\chi}_n^m\leq(\chi^*)^m$ for all $m\in[M]$ and hence completes the proof of the upper bound on finally sold assets.

For the upper bound on the final default fraction $n^{-1}\vert\mathcal{D}_n\vert=\P(L_n+\bm{X}_n\cdot h(\bm{\chi}_n)\geq C_n)$, approximate the indicator function $\1\{y\geq1\}$ from above by continuous functions $(\psi_t)_{t\in\N}$ and use weak convergence of $(\bm{X}_n,C_n,L_n)$ to derive
\[ \limsup_{n\to\infty}n^{-1}\vert\mathcal{D}_n\vert \leq \lim_{n\to\infty}\E\left[\psi_t\left(\frac{L_n+\bm{X}_n\cdot h(\hat{\bm{\chi}}(\epsilon))}{C_n}\right)\right] = \E\left[\psi_t\left(\frac{L+\bm{X}\cdot h(\hat{\bm{\chi}}(\epsilon))}{C}\right)\right] \]
and as $t\to\infty$, $\limsup_{n\to\infty}n^{-1}\vert\mathcal{D}_n\vert\leq g(\hat{\bm{\chi}}(\epsilon))$. Letting $\epsilon\to0$, thus shows the second part of the theorem by upper semi-continuity of $g$.

%
%Finally, consider the general case of right-continuous $\rho$ and choose an approximation $(\rho_r)_{r\in\N}$ from above by continuous sale functions $\rho_r$. For $\bm{\chi}_r^*$, the analogue of $\bm{\chi}^*$ for $\rho_r$, it clearly holds $\bm{\chi}_r^*\geq\bm{\chi}^*$ and hence $n^{-1}\vert\mathcal{D}_n\vert \leq g(\bm{\chi}_r^*) + o(1)$ (note that $g$ does not depend on $\rho$). Further, by the proof of Lemma \ref{lem:existence:chi:hat} we know that $\bm{\chi}_r^*\to\bm{\chi}^*$ as $r\to\infty$ and hence $n^{-1}\vert\mathcal{D}_n\vert \leq g(\bm{\chi}^*) + o(1)$. Finally, for $\bm{\chi}_{n,r}$ the vector of finally sold shares in the financial system with sale function $\rho_r$,
%\[ \chi_n^m \leq \chi_{n,r}^m \leq (\chi_r^*)^m + \frac{\epsilon}{2} \leq (\chi^*)^m +\epsilon \]
%for $\epsilon>0$ arbitrary, $r=r(\epsilon)$ large enough such that $(\chi_r^*)^m\leq(\chi^*)^m+\epsilon/2$ and $n=n(\epsilon,r)$ large enough such that $\chi_{n,r}^m \leq (\chi_r^*)^m+\epsilon/2$. In particular, $\limsup_{n\to\infty}\chi_n^m\leq(\chi^*)^m$.
\hfill
\endproof

In the following two proofs we use the notations $g$, $f^m$, $\circG$, $\circFSuper{m}$, $\hat{\bm{\chi}}$ and $\bm{\chi}^*$ as introduced in \ref{ssec:stochastic:model:generalized} but for an unshocked $(\bm{X},C)$-system. If instead we index these quantities by $\cdot_L$, we mean the system shocked by $L$.
\proof[Proof of Theorem \ref{thm:resilience}.]
By %construction of $(z^*,\bm{\chi}^*)$ in Lemma \ref{lem:existence:z:chi:hat},
Remark \ref{rem:sequence:chi:*}, there exists a sequence of vectors $\tilde{\bm{\chi}}(\gamma)\in\R_{+}^M$ such that $f^m(\tilde{\bm{\chi}}(\gamma))=-\gamma$ for all $m\in[M]$ and arbitrary $\gamma>0$. Now for arbitrary $\alpha\in\R_+$ it holds that
\begin{align*}
f_L^m(\bm{\chi}) &= \E\left[X^m\rho\left(\frac{L+\bm{X}\cdot h(\bm{\chi})}{C}\right)\right] - \chi^m\\
&\leq \E\left[X^m\1\left\{\frac{L}{C}\geq\alpha\right\}\right] + \E\left[X^m\rho\left(\frac{\alpha C+\bm{X}\cdot h(\bm{\chi})}{C}\right)\right] - \chi^m.
\end{align*}
Since $\E[L/C]<\delta$, by Markov's inequality it holds that $\P(L/C\geq\alpha)\leq\delta/\alpha$ and hence for $\delta>0$ small enough, we have $\E[X^m\1\{L/C\geq\alpha\}]\leq\gamma/3$ (recall that $\E[X^m]<\infty$). By dominated convergence and right-continuity of $\rho$, it thus holds that $f_L^m(\bm{\chi})\leq f^m(\bm{\chi})+2\gamma/3$ for $\alpha>0$ small enough such that
\[ \E\left[X^m\rho\left(\frac{\alpha C+\bm{X}\cdot h(\bm{\chi})}{C}\right)\right] \leq \E\left[X^m\rho\left(\frac{\bm{X}\cdot h(\bm{\chi})}{C}\right)\right] + \gamma/3. \]
In particular, $f_L^m(\tilde{\bm{\chi}}(\gamma)) \leq -\gamma/3<0$ and hence $\bm{\chi}_L^*<\tilde{\bm{\chi}}(\gamma)$ for $\delta$ small enough. By similar means, we further derive that for $\delta$ small enough it holds $g_L(\tilde{\bm{\chi}}(\gamma))\leq g(\tilde{\bm{\chi}}(\gamma))+\epsilon/3$. Together with Theorem \ref{thm:fire:sale:final:fraction}, we thus derive that
\[ \limsup_{n\to\infty}n^{-1}\vert\mathcal{D}_{n,L}\vert \leq g_L(\bm{\chi}_L^*) + \epsilon/3 \leq g_L(\tilde{\bm{\chi}}(\gamma)) + \epsilon/3 \leq g(\tilde{\bm{\chi}}(\gamma)) + 2\epsilon/3. \]
Now since $\tilde{\bm{\chi}}(\gamma)\to\bm{\chi}^*$ and by upper semi-continuity of $g$, we can choose $\gamma>0$ small enough such that $g(\tilde{\bm{\chi}}(\gamma))\leq g(\bm{\chi}^*)+\epsilon/3$ and conclude that $\limsup_{n\to\infty}n^{-1}\vert\mathcal{D}_{n,L}\vert \leq g(\bm{\chi}^*)+\epsilon$.

For the bound on $(\chi_{n,L}^m)$ choose $\gamma$ and $\delta$ small enough such that $(\chi_L^*)^m \leq \tilde{\chi}^m(\gamma)+\epsilon/3 \leq (\chi^*)^m+2\epsilon/3$ and conclude by Theorem \ref{thm:fire:sale:final:fraction} that
\[ \limsup_{n\to\infty}(\chi_{n,L}^m) \leq (\chi_L^*)^m + \epsilon/3 \leq (\chi^*)^m+\epsilon. %\qedhere
\]\hfill
\endproof
\proof[Proof of Theorem \ref{thm:non-resilience}.]
For some $l\in [M]$ it must hold that $ \P (L, X^{l} >0)>0$ and thus there exists $\delta > 0$ and $\mathcal{A}\subset \Omega$ such that $L,X^{l}\geq \delta$ on $\mathcal{A}$. Then we can find $\tilde{\mathcal{A}}\subset  \mathcal{A}$ and $x_0$ large enough such that $C\leq x_0$ on $\tilde{\mathcal{A}}$  and therefore $L/C\geq \delta/x_0$ on $\tilde{\mathcal{A}}$. Since $\rho$ is assumed to be strictly increasing we can conclude that $\varepsilon:=\rho (\delta/x_0)>0 $. We obtain that
\[ \E\left[X^l\rho\left(\frac{L+\bm{X}\cdot h(0)}{C}\right)\right] \geq  \varepsilon \delta \P (\tilde{\mathcal{A}}) >0\]
and thus $\circFSuper{l}_L (\bm{0})>0$. Therefore, $\bm{0}$ is not a joint root of the functions $\circFSuper{m}_L (\bm{0}), m\in [M]$ and in fact the auxiliary  process starts. Because of $\circFSuper{m}_L (\bm{\chi}) \geq \circFSuper{m} (\bm{\chi}) $ it follows that $\hat{\bm{\chi}}_L \geq \bm{\chi^*}$ and thus Theorem~\ref{thm:resilience:with:defaults} completes the proof.
\endproof

\proof[Proof of Theorem \ref{thm:non-resilience:non:strictly:inc}.]
For a given $\delta>0$ we consider a shock which is such that $\P (L=2C)=\delta /3$ and $\P (L=0)=1-\delta/3$. Such shock can be chosen to be  independent of $\bm{X}$. Then clearly $\E [L/C]=(2/3)\delta<\delta$. Moreover since $\rho(1)=1$ it follows that $\circFSuper{m}_L (\bm{0})=\E[X^m](2/3)\delta$ for $ m\in [M]$. Since $\E[X^l]>0$ for some $l\in [M]$, it follows that $\circFSuper{l}_L (\bm{0})>0$ for some $l\in [M]$. Now the same arguments as in the proof of Theorem \ref{thm:non-resilience} can be applied.

\hfill\hfill
\endproof

\subsubsection{Proofs for Section \ref{sec:fsp} (Fire Sales Process)}
\label{ssec:proofs:resilience}

\proof[Proof of Theorem~\ref{res:power:h_and_rho}.]
We start with the case $1-\nu q > 0$. Using Theorem~\ref {thm:fire:sale:lower:bound:real} it is enough to show for some fixed $\varepsilon \in(0,1)$ that $\liminf_{n \rightarrow \infty} \E [X \rho_{\varepsilon} \left( \frac{X h(a_n)}{C}\right) ]/a_n > 1$ for any sequence $(a_n)_{n\geq 1}$ with $\lim_{n\rightarrow \infty} a_n = 0$. Using Fatou's lemma we get that
\begin{eqnarray}
\liminf_{n \rightarrow \infty} \E \left[X \rho_{\varepsilon} \left( \frac{X h(a_n)}{C}\right)/a_n \right]  
&\geq & \E \left[\liminf_{n \rightarrow \infty}X \rho_{\varepsilon} \left( \frac{X h(a_n)}{C}\right) /a_n\right] \nonumber\\
& = & \E \left[\liminf_{n \rightarrow \infty}X \left( \frac{X a_n^{\nu}}{C}\right)^q /a_n\right] \nonumber\\
& = & \E \left[\liminf_{n \rightarrow \infty} a_n^{\nu q -1 } X \left( \frac{X}{C}\right)^q \right] = \E \left[ \infty \cdot \1_{\{X \left( {X}/{C}\right)^q \neq 0\}} \right]=\infty, \nonumber
\end{eqnarray}
and thus the real system is non-resilient.

We next look at the case $1-\nu q = 0$. Let $\E [ X( X/C )^q ] < 1$. We show resilience of the auxiliary process from which resilience for the real process follows by Theorem~\ref{thm:fire:sale:upper:bound:real}. As before, let $(a_n)_{n\geq 1}$ be a sequence with $\lim_{n\rightarrow \infty} a_n = 0$. We need to show that $\limsup_{n \rightarrow \infty} \E [X \rho \left( {X h(a_n)}/{C}\right) ]/a_n < 1$. Define the event $\mathcal{A}_n:= \{  (X/C)^{q} \leq 1/a_n \}$. First observe that since $q=1/\nu$,
\begin{eqnarray}
X a_n^{-1} \rho \left( \frac{X h(a_n)}{C}\right) &=& X a_n^{-1}  \left( \frac{X h(a_n)}{C}\right)^q \1_{\mathcal{A}_n}+  X a_n^{-1}  \1_{\mathcal{A}^c_n} \nonumber \\
&=& X(X/C)^q \1_{\mathcal{A}_n} +  X a_n^{-1}  \1_{\mathcal{A}^c_n} \nonumber \\
&\leq & X(X/C)^q \1_{\mathcal{A}_n} +  X (X/C)^q  \1_{\mathcal{A}^c_n} = X(X/C)^q. \nonumber
\end{eqnarray} It follows that 
\[ \lim_{n\rightarrow \infty } \E \left[X a_n^{-1} \rho \left( \frac{X h(a_n)}{C}\right)\right]  \leq \lim_{n\rightarrow \infty } \E \left[ X(X/C)^q \right]  =  \E \left[ X(X/C)^q \right]<1, \]
and thus the system is resilient.
On the contrary if $\E [ X( X/C )^q ] > 1$, then for fixed $\varepsilon >0 $ define $\mathcal{A}_n:= \{  (X/C)^{q} \leq (1/a_n) \varepsilon\}$ and observe that $\mathcal{A}_n \nearrow \Omega$. It follows for $\varepsilon>0$ small enough that 
\[ \lim_{n\rightarrow \infty } \E \left[(1-\varepsilon)X a_n^{-1} \rho_{\varepsilon} \left( \frac{X h(a_n)}{C}\right)\right]  \geq \lim_{n\rightarrow \infty }  \E [ (1-\varepsilon) X(X/C)^q \1_{\mathcal{A}_n} ] >1,\]
 and thus the real system is non-resilient.

Now consider the remaining case $1-\nu q < 0$. First assume that $\alpha^* > 1/\nu$, where we claim resilience of the auxiliary process. Choose $1<\delta $ such that $\delta ( 1/\nu) <  \min \{ \alpha^*, q\}$ and set $\mathcal{A}_n:= \{  (X/C)^{1/\nu} \leq 1/a_n \}$. Then 
\begin{eqnarray}
\E \left[ X a_n^{-1} \rho \left( \frac{X h(a_n)}{C}\right) \right]&=& \E \left[ X a_n^{-1}  \left( \frac{X h(a_n)}{C}\right)^q \1_{\mathcal{A}_n} \right] +  \E \left[  X a_n^{-1}   \1_{\mathcal{A}_n^c} \right] \nonumber \\
 &=&a_n^{\nu q-1} \E \left[ X   \left( \frac{X}{C}\right)^q \1_{\mathcal{A}_n} \right] +  \E \left[  X a_n^{-1}  \1_{\mathcal{A}_n^c} \right]. \label{expt:terms1}
\end{eqnarray}
For the first term observe that
\begin{eqnarray}
a_n^{\nu q-1} \E \left[ X   \left( \frac{X}{C}\right)^q \1_{\mathcal{A}_n} \right] &=& a_n^{\nu q-1} \E \left[ X   \left( \frac{X}{C}\right)^{\delta/\nu} \left(\frac{X}{C}\right)^{q-\delta/\nu} \1_{\mathcal{A}_n} \right] \nonumber\\
&=& a_n^{\nu q-1} \E \left[ X   \left( \frac{X}{C}\right)^{\delta/\nu}  \left(\frac{X}{C}\right)^{q-\delta/\nu}  \1_{ \{  (X/C)^{q-\delta/\nu} \leq (1/a_n)^{\nu q - \delta} \}} \right]\nonumber\\
&\leq & a_n^{\nu q-1} \E \left[ X   \left( \frac{X}{C}\right)^{\delta/\nu} a_n^{-\nu q + \delta} \1_{ \{  (X/C)^{q-\delta/\nu} \leq a_n^{-\nu q+ \delta} \}} \right] 
\le  a_n^{\delta-1} \E \left[ X   \left( \frac{X}{C}\right)^{\delta/\nu} \right], \nonumber
\end{eqnarray}
and $\lim_{n\rightarrow \infty }a_n^{\delta-1} \E \left[ X ( X/C)^{\delta/\nu} \right] =0$ by the choice of $\delta$ and our assumption $\E \left[ X   ({X}/{C})^{{\delta}/{\nu}} \right]<\infty$. For the second term in~\eqref{expt:terms1} we obtain by Markov's inequality that 
\[ \E \left[  X a_n^{-1}  \1_{\mathcal{A}_n^c} \right] = a_n^{-1}  \E \left[  X  \1_{  \{  (X/C)^{\delta/\nu } > (1/a_n)^\delta \}  } \right] \leq  (1/a_n)^{1-\delta} \E \left[  X (X/C)^{\delta/\nu} \right]  \]
and $ \lim_{n\rightarrow \infty }(1/a_n)^{1-\delta} \E \left[  X (X/C)^{\delta/\nu} \right] =0$ by the choice of $\delta $. 

To show non-resilience for $\alpha^* < 1/\nu$, note that we assumed that the right derivative of $f(\cdot)$ in $0$ exists (possibly with value $\infty$). It is thus sufficient to find some $\varepsilon>0$ and a particular sequence $(a_n)_{n\geq 1}$ with $\lim_{n\rightarrow \infty} a_n = 0$ such that 
\[ \lim_{n\rightarrow \infty } \E \left[ X (1-\varepsilon) a_n^{-1} \rho_{\varepsilon} \left( \frac{X h(a_n)}{C}\right) \right]> 1.\] 
Define the event $\mathcal{A}_n:= \{  (X/C)^{1/\nu} \varepsilon^{-{1}/{q\nu}} \geq 1/a_n \}$. Since 
\[ \E \left[ X (1-\varepsilon)a_n^{-1} \rho_{\varepsilon} \left( \frac{X h(a_n)}{C}\right) \right] \geq\E \left[ X (1-\varepsilon) a_n^{-1} \rho_{\varepsilon} \left( \frac{X h(a_n)}{C}\right)\1_{\mathcal{A}_n}  \right] \geq a_n^{-1}\E \left[ X (1-\varepsilon)  \varepsilon\1_{\mathcal{A}_n}  \right], \]
it is sufficient to find $(a_n)_{n\geq 1}$  such that $\E \left[ X \1_{\mathcal{A}_n}  \right] \gg a_n$.
For this, choose $\delta <1 $ such that $\delta/\nu> \alpha^*$. 
Using Fubini's theorem we obtain that
\begin{eqnarray}
&&\E \left[  X\left( \frac{X}{C} \right)^{\delta/\nu}   \right] = \E \left[ X \int_0^{(X/C)^{\delta/\nu}} 1 \dd t  \right]  = \E \left[ X \int_0^{\infty} \1_{\{t\leq (X/C)^{\delta/\nu} \}} \dd t  \right] \nonumber \\
&=& \E \left[ X \int_0^{\infty} \1_{\{t\leq (X/C)^{\delta/\nu} \}} \dd t  \right] =  \E \left[ X \int_0^{\infty} \1_{\{t^{1/\delta} \varepsilon^{-1/q\nu} \leq (X/C)^{1/\nu}\varepsilon^{-1/q\nu} \}} \dd t  \right] \nonumber \\
&=& \int_0^{\infty} \E \left[ X  \1_{\{t^{1/\delta} \varepsilon^{-1/q\nu} \leq (X/C)^{1/\nu}\varepsilon^{-1/q\nu} \}} \right] \dd t  = \int_0^{\infty}  \E \left[ X \1_{\{ u\leq (X/C)^{1/\nu}\varepsilon^{-1/q\nu} \}} u^{\delta -1} \varepsilon^{-1/q\nu} \right]  \dd u. \nonumber
\end{eqnarray}
Since $\E [  X\left( {X}/{C} \right)^{\delta/\nu}  ] =\infty$ it follows that  $\E \left[ X \1_{\{ u\leq (X/C)^{1/\nu}\varepsilon^{-1/q\nu} \}} u^{\delta -1} \varepsilon^{-1/q\nu} \right] $ can not be dominated by the function $C{u^{-\alpha}}$ for any $\alpha >1$ and $C >0$. Choose now $\alpha$ such that $\delta + \alpha <2$, $C=1$ and a sequence $(u_n)_{n\geq 1}$ such that
\[ \E \left[ X \1_{\{ u\leq (X/C)^{1/\nu}\varepsilon^{-1/q\nu} \}} u^{\delta -1} \varepsilon^{-1/q\nu} \right] \geq u_n^{-\alpha}. \]
Set now $a_n=1/u_n$. It follows by the choice of $\alpha$   and assuming that $(1-\varepsilon)\varepsilon \ge 1/2$
\[
	\lim_{n\rightarrow \infty}a_n^{-1}\E \left[ X (1-\varepsilon)  \varepsilon\1_{\mathcal{A}_n}  \right]
	\ge \lim_{n\rightarrow \infty}u_n \E \left[ X \1_{ \{ u_n  \leq (X/C)^{1/\nu} \varepsilon^{-{1}/{q\nu}}  \}}  \right] /2
	=  \lim_{n\rightarrow \infty}u_n^{2-\alpha-\delta} /2 =\infty.
\]
\hfill\hfill
\endproof

\subsection{The Choice of the Sales Function}
\label{eq:salesfctex}
Some concrete examples for sales functions $\rho$ are as follows:
\begin{itemize}
\itemsep-0.1em
\item The perhaps simplest non-trivial example is $\rho(u)=\1\{u\geq1\}$. It describes complete liquidation of the portfolio at default (if the institution is leveraged) resp.~dissolution.
\item A more involved example can be derived from a leverage constraint that prohibits an institution from investing more money into risky assets than a certain multiple $\lambda_\text{max}\geq 1$ of its capital/equity. In the one-asset case this means that $xp/c=:\lambda\leq\lambda_\text{max}$, where $x$ denotes the number of shares held, $p$ is the price per share and $c$ denotes the institution's capital. Assume now that while the asset price $p$ stays constant, the institution suffers an exogenous shock $\ell$ and $c$ is reduced to $\tilde{c}=c-\ell$. 
\begin{itemize}
\itemsep-0.2em
\item If $\ell\leq (1-\lambda/\lambda_\text{max})c$, then the leverage constraint $xp/\tilde{c}\leq\lambda_\text{max}$ is satisfied and no reaction is required by the institution.
\item However, if $\ell> (1-\lambda/\lambda_\text{max})c$, then the institution must divest some shares; suppose that it sells $\delta x$ of them, for some $0<\delta\le 1$. In order for the leverage constraint $(1-\delta)xp/\tilde{c}\leq\lambda_\text{max}$ to hold, it is easy to verify that $\delta\geq 1-(1-\frac\ell{c})\frac{\lambda_\text{max}}\lambda$.
\end{itemize}
The relative asset sales are hence given by $\rho(\ell/c)$, where $\rho(u):=(1-(1-u)\lambda_\text{max}/\lambda)^+$ for $u\in[0,1]$; this amounts to linear sales (with respect to the losses) once the threshold $1-\lambda\lambda_\text{max}^{-1}$ is reached.
\item Taking an alternative route in the previous example, suppose that the loss of the institution stems only from  a price change $p\to \tilde{p}<p$, which reduces the capital to $\tilde{c} = c - x(p-\tilde{p})$. If $\tilde{p}\geq p(1-\frac1\lambda)/(1-\frac1{\lambda_\text{max}})$, then no action is required  to comply with the leverage constraint. In the remaining cases the institution must sell a fraction of $1-\lambda_\text{max}+\lambda_\text{max}\frac{p}{\tilde{p}}(1-\frac{1}{\lambda})$ of their assets, and we obtain  $\rho(u)=(1-\lambda_\text{max}(1-u)/(\lambda-u))^+$ for $u\in[0,1]$. 
\item Finally, it can be shown that for price changes combined with exogenous losses the sales function is bounded from above and below by the two previous cases. Leverage constraints hence imply a sales function which is $0$ below a certain threshold and then grows linearly. %in good approximation.
\end{itemize}

\bibliography{finance}
\bibliographystyle{abbrv}
%\end{multicols}}
\end{document}

%% file: definitions.tex
\usepackage{mathrsfs}

\newcommand{\E}{{\mathbb{E}}}
\providecommand{\R}{{\mathbb{R}}}

\newcommand{\dd}{{\rm d}}

\providecommand{\N}{{\mathbb N}}

\newcommand{\1}{\ensuremath{\mathbf{1}}}

\ifx\prop\undefined
\newtheorem{prop}{Proposition}[section]
\fi

\newtheorem{remark}[theorem]{Remark}

\def\P{{\mathbb P}}